\documentclass[lineno]{JFM-FLM_Au}

\usepackage{orcidlink}
\usepackage{enumitem} 
\usepackage{bm}
\usepackage{subcaption} 
\usepackage{graphicx}
\usepackage[figuresright]{rotating}

\lefttitle{A.~I.~El-Nadi, R.~Vinuesa and S.~T.~M.~Dawson}

\title{Low-dimensional Galerkin projection models for predicting turbulent secondary mean flow in a square duct}

\author{Ahmed I. El-Nadi\,\orcidlink{0009-0006-3083-6222}\aff{1}, Ricardo Vinuesa\aff{2} and Scott T.~M.~Dawson\aff{1}}

\affiliation{
\aff{1}Mechanical, Materials \& Aerospace Engineering Department, Illinois Institute of Technology, IL 60616, USA
\aff{2}Department of Aerospace Engineering, University of Michigan, MI 48109, USA

\par\vspace{2pt}
\textbf{First author}:Ahmed I. El-Nadi \email{aelnadi@hawk.illinoistech.edu}
}

\corresau{Scott T.~M.~Dawson, \email{sdawson5@illinoistech.edu}}
\begin{document}
\nolinenumbers
\maketitle

\begin{abstract}
\begin{center}
    \textbf{Abstract}
\end{center}
\vspace{7pt}
The presence of sidewalls in turbulent duct flows leads to the emergence of secondary flow structures in the form of counter-rotating streamwise vortices near the corners (Prandtl's secondary flow of the second kind). This work develops reduced-order Galerkin projection models that can predict the emergence and structure of these turbulent mean secondary flows in a square duct geometry, without requiring any prior knowledge of the turbulent statistics. The models are obtained by projecting the Navier--Stokes equations onto eigenmodes of the linearised system, with the resulting systems of ordinary differential equations simulated with the addition of zero-mean forcing. We show that most models obtained using leading streamwise-constant eigenmodes predict the correct shape and direction of the secondary mean, with the minimal such model requiring only two modes. In these models, the secondary mean contribution arises due to the nonzero average coefficient of an eigenmode that possesses all of the symmetry properties expected of a secondary mean. The models are sufficiently simple such that the relationship between this mean coefficient and the joint second moment of other mode coefficients can be computed analytically. We confirm that running streamwise-constant direct numerical simulations (DNS) with the same forcing structure as used in the reduced-order models produces similar secondary-flow structures. We additionally demonstrate that our models produce qualitatively similar Reynolds stress distributions to fully resolved direct numerical simulations, with improved agreement as model dimension increases.
\end{abstract}


\section{Introduction}
\label{Introduction}

Turbulent flows in ducts and non-circular geometries are common in nature and industrial systems, including rivers, canals, and heat exchangers. Therefore, understanding the associated flow physics is essential for accurately predicting species~\citep{wang2024}, energy-transfer mechanisms~\citep{modesti2022,karabulut2024}, and wall shear stresses~\citep{doehring2024,santese2024}. A characteristic feature of such flows is the presence of secondary mean velocity components normal to the streamwise direction, typically generated by turbulent stress imbalances and/or duct curvature. For ducts with both circular and non-circular cross sections, the centrifugal force caused by this curvature induces a secondary mean flow classified as Prandtl's secondary flow of the first kind. This secondary flow exists in both laminar and turbulent flows, with a secondary mean velocity magnitude of approximately $20$--$30\%$ of the streamwise laminar velocity~\citep{demuren1984}. In contrast, Prandtl's secondary flows of the second kind, which are the focus of this work, are observed in non-circular straight ducts exhibiting transitional or fully developed turbulence~\citep{prandtl1934}. The secondary mean velocity magnitude in this flow is about $1$--$3\%$ of the streamwise turbulent mean velocity. Despite being small in magnitude, this type of secondary motion plays a major role in the convection of momentum and energy from the mean flow towards the corners. This momentum transport contributes to wall shear stresses near the corners \citep{modesti2018}, affecting quantities such as erosion~\citep{winkler2006} and heat transfer~\citep{fuzhang2021}. These secondary flows have been a subject of investigation for more than a century, starting with early experimental visualizations and evolving into detailed numerical studies.

The first documented visualizations of the isotachs (lines of constant velocity magnitude) of the streamwise mean velocity distortions in the corner regions were made by~\cite{nikuradse1926} in experimental investigations of turbulent flow in rectangular and triangular cross-section ducts. Subsequently, \citet{prandtl1927} initiated an effort to understand the secondary flow characteristics and generation mechanisms, proposing that isotach distortions arise from the momentum transport from the main flow towards the corners. Later,~\citet{prandtl1952} formulated an explanation based on the surface friction stress balance. His theoretical explanation was confirmed experimentally~\citep{hoagland1962} where the surface friction change is consistent with the secondary mean flow direction. Other researchers used turbulent kinetic energy balance~\citep{hinze1967} to explain the generation and motion of the secondary flows by quantifying the difference between the TKE and viscous dissipation. However, at that time, it was not yet clear whether secondary flows arise in laminar flows or turbulent flows, nor was the role of Reynolds stresses in their generation fully understood. A subsequent mathematical analysis by \cite{moissis1957} demonstrated that secondary flows cannot exist in fully developed laminar flow through a straight duct, while the study by \cite{einstein1958} was the first to clearly link secondary flow generation to the vorticity production from Reynolds stresses. Researchers have presented varying perspectives on the role of individual Reynolds production terms in the generation of secondary flows~\citep{brundrett1964}. These differences were generally attributed to the limitations in experimental measurements~\citep{perkins1970}. Further discussion on the streamwise vorticity equation and the secondary flow production will be provided in \S\ref{SecFlwProd}.

The emergence of Direct Numerical Simulation (DNS) has further improved our understanding of secondary flow dynamics in ducts~\citep{gavrilakis1992,huser1993}, enabling the accurate computation of all flow quantities over the entire duct cross-section. For square ducts, simulations have been performed over a wide range of Reynolds numbers in smooth~\citep{zhang2015direct,pirozzoli2018,modesti2018,modesti2022,xiang2023,doehring2024} and ribbed ducts~\citep{santese2024}, enabling the investigation of turbulence statistics and Reynolds number dependence of the secondary flow. Other studies focused on marginally turbulent duct flows at lower Reynolds numbers, where  the secondary flow streamlines become increasingly aligned with the mean streamwise vorticity \citep{pinelli2010}, which in turn emerges due 
to the preferential emergence of high-speed streaks in the corner region~\citep{uhlmann2007}. It has further been demonstrated by \citet{uhlmann2010traveling} that the square duct geometry permits travelling wave solutions that result in a secondary mean consistent with that observed in DNS. The duct aspect ratio effect on the secondary flow characteristics has also been numerically \citep{vinuesa2014,vinuesa2015,vinuesa2018secondary} and experimentally~\citep{hoagland1962} investigated. Similar to the square duct, in rectangular ducts a pair of counter-rotating vortices exists at each corner, with small and large vortices present along the short and long sides of the duct, respectively. This is in contrast to Prandtl's original prediction of multiple vortices along the long side of the duct~\citep{prandtl1925}.

While DNS can accurately capture the characteristics and statistics of turbulent flows in square ducts and other corner geometries, the associated computational cost and scaling with increasing Reynolds number~\citep{choi2012} motivates the use and development of closure models capable of capturing features such as the secondary mean. Standard linear eddy viscosity models, based on the Boussinesq relation for the Reynolds stress tensor, have been found to be inadequate for capturing secondary mean flows~\citep{mani2013}. More sophisticated Reynolds stress models (RSM) that solve several transport equations for the components of the Reynolds stress tensor can capture anisotropy of the Reynolds stresses and secondary flow production, though at an increased computational cost. For example, the Speziale-Sarkar-Gatski (SSG) model~\citep{speziale1991} has demonstrated success in predicting mean quantities in rectangular~\citep{de2009} and trapezoidal~\citep{ansari2011} duct flows. An alternative and less computationally expensive approach to improve upon linear eddy viscosity models involves their generalisation to nonlinear eddy-viscosity formulations, such as the Spalart-Quadratic-Constitutive (QCR) model initially developed by~\cite{spalart2000}. This and several extensions of the QCR framework~\citep{rumsey2020,sabnis2021,tamaki2024} have been shown to predict the secondary flow in square ducts with accuracy comparable to an RSM model~\citep{prudenzano2025}. See~\citep{modesti2020} for a comprehensive comparison of the performance of different eddy viscosity models in capturing the anisotropic Reynolds stress tensor in square duct flows.

An alternative approach to simplifying the NSE involves the use of restricted nonlinear (RNL) models based on statistical state dynamics (SSD) \citep{farrell1993,farrell2017}. In this approach, the dynamics of the system are decomposed into those for the streamwise-averaged and fluctuating components, with the nonlinear fluctuation-fluctuation interactions neglected or otherwise modelled. This approach has been applied to predict the presence and features of turbulent flow in a range of configurations, including wall-bounded geometries~\citep{farrell2012,farrell2016,gayme2019coherent}. It has also been demonstrated that this RNL approach can accurately predict mean secondary flow components in turbulent flow over riblets \citep{zhu2025restricted}.

The preceding paragraphs have discussed general methods that approximate the solutions to the NSE at reduced computational cost. For a given flow, it is natural to also seek the simplest model capable of reproducing certain pertinent features. For example, various low-dimensional models with sinusoidal basis functions have been proposed to capture the interaction of streamwise streaks and vortices in shear-driven turbulent flows \citep{waleffe1997self,moehlis2004,cavalieri2021structure}. Such reduced-order models (ROMs) are obtained via Galerkin projection, a method whose origins come from~\citet{galerkin1915} and are derived from a variational principle belonging to the Rayleigh-Ritz class of problems seeking a simplified system capable of capturing the original system dynamics. The method was originally applied to mechanical systems~\citep{duncan1937,batdorf1947}, then extended to fluid mechanics~\citep{murphy1973}. 
An advantage of Galerkin projection is that it can be used with any set of linearly independent (ideally orthogonal) basis functions, as long as they span the subspace in which the system attractor is approximately resolved. The introduction of proper orthogonal decomposition (POD)~\citep{lumley1967,holmes2012} to the fluid mechanics community opened the door for the extensive use of such data-driven empirical basis functions with Galerkin projection across a broad range of flow configurations~\citep{moehlis2002,rempfer1994,aubry1988,deane1991,rowley2009,schlegel2009}. The trade-off when using a truncated set of basis functions identified from data is that typically the resulting models are highly tailored to a specific flow configuration and are not robust to parameter variation~\citep{noack2011,noack2003,jorgensen2003}. We are not aware of any prior works seeking Galerkin projection models for modelling turbulent secondary flow in square ducts, though \citet{wedin2008} considered a low-dimensional model for self-sustaining processes in a square duct geometry, analogous to the work of \citet{waleffe1997self} for a wall-bounded shear flow without sidewalls. POD has also been performed in a square duct geometry for the purposes of identifying the shape and dynamics of coherent structures present in such flows \citep{matin2018coherent,khan2020dynamics,lopez2024linear}. 
Since our goal in the present work is to predict properties of turbulent duct flow without any prior data, here rather than using POD modes to obtain a subspace for Galerkin projection, we instead use eigenmodes associated with the underlying linearised dynamics. This choice is motivated in part by the fact that linear mechanisms are known to play an important role in the formation of coherent structures that exist within shear-driven turbulent flows \citep{lee1990,kim2000,hwang2010,mckeon2010,jimenez2013linear}. 

This work seeks to find the simplest model capable of predicting the existence and qualitative structure of the secondary mean present in turbulent flows through a square duct. The specific approach taken is motivated in part by the findings of the prior work mentioned above. We form Galerkin projection models arising from the interaction of a small number of modes, motivated by the fact that similar low-dimensional models have been found to capture pertinent dynamics in wall-bounded turbulent flows, such as the near-wall cycle \citep{waleffe1997self}. Furthermore, we focus on streamwise-constant modes, motivated in part by the fact that streamwise-elongated streaks are prevalent in the corner region of turbulent duct flows and are important for the emergence of secondary mean flow \citep{pinelli2010,vinuesa2016convergence,atzori2021intense}. 
As our objective is to find the simplest ROMs capable of capturing the features of secondary mean flows, additional justification of this streamwise-constant assumption will come from the results obtained, which will demonstrate that streamwise-constant fluctuations are sufficient for this purpose. For our models to yield non-trivial dynamics, we find it necessary to apply an external input to the equations, analogous to the stochastic closure terms used in SSD models for turbulent phenomena \citep{farrell2012}.

The paper is structured as follows: an overview of the governing equations and the geometrical configuration is outlined in \S\ref{Governing equations}. The linearisation of the NSE along with the associated eigenvalue computational approach and symmetry properties are discussed in \S\ref{Eigenfunctions and eigenvalues}. In \S\ref{Galerkin  projection}, we describe the Galerkin projection method used to obtain ROMs. \S\ref{results} shows the numerical predictions from running these ROMs, alongside theoretical results obtained directly by analysing their structure. To demonstrate how these ROMs relate to the full system, forced two-dimensional DNS simulation results are  presented in \S\ref{sec:dns}, and  ROM predictions of secondary flow production through Reynolds stress quantities are discussed in \S\ref{SecFlwProd}. Additional discussion and conclusions are provided in \S\ref{sec:conc}.

\section{Problem setup and governing equations}
\label{Governing equations}

This section describes the duct flow configuration and the basic properties of the corresponding incompressible NSE.
\begin{figure}
  \centerline{\includegraphics[width=0.6\textwidth]{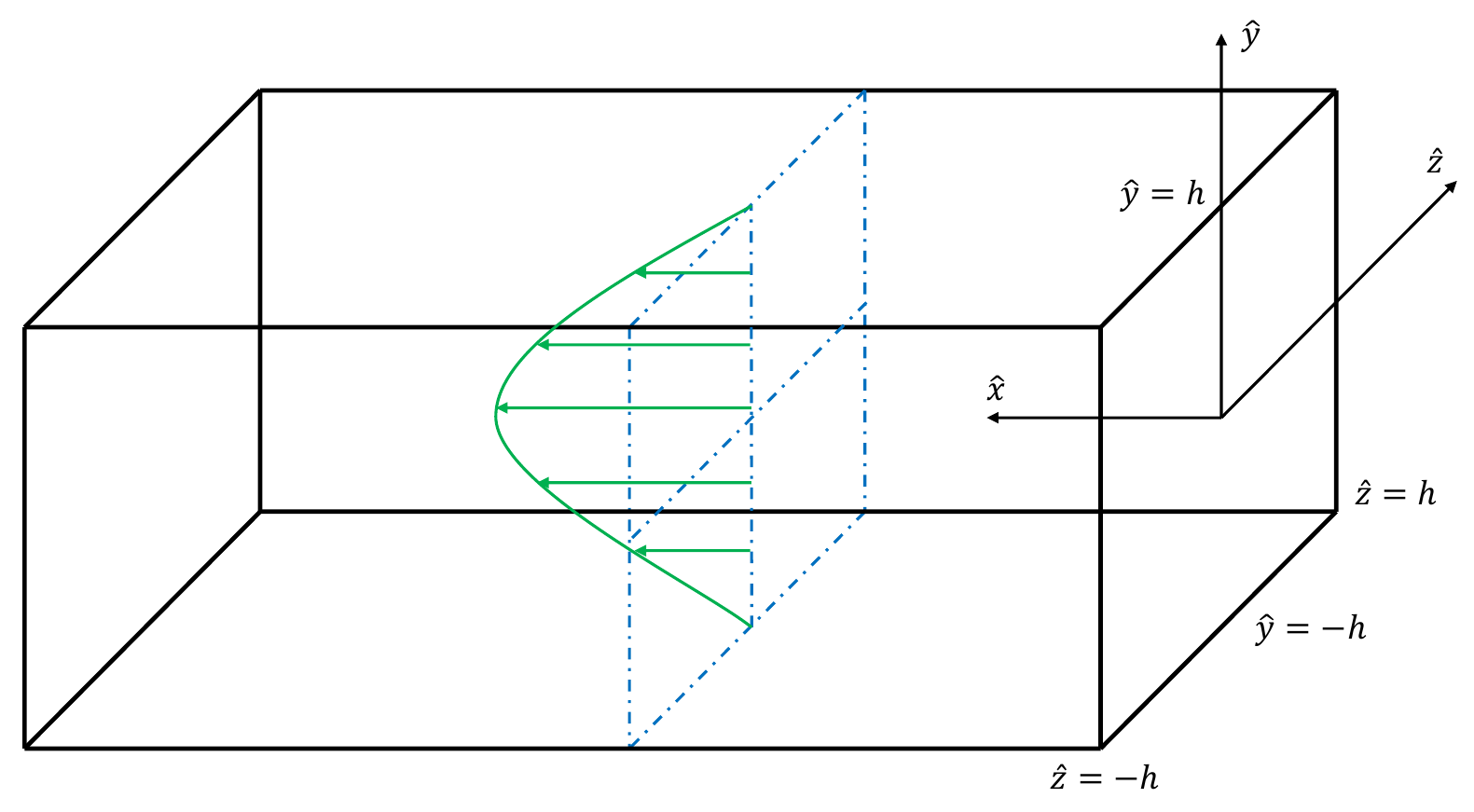}}
  \caption{Schematic drawing of the square duct geometry and the associated coordinate system. The duct half height is $h$, and the streamwise flow is in the positive $x$-direction, as indicated by the green arrows and curve. While the laminar streamwise flow is plotted as a function of $y$ at $z=0$, the full 2D profile is a function of both $y$ and $z$ (as depicted in figure \ref{LaminarProfile}).}
\label{fig:schematic}
\end{figure}
Figure~\ref{fig:schematic} shows a schematic drawing of the square duct along with the non-dimensional coordinates $x$, $y$, and $z$ denoting the streamwise and cross-stream directions, respectively. The associated velocity vector is defined as $\hat{\bsu}=u \textbf{i}+v \textbf{j} +w\textbf{k}$ where $\textbf{i},\;\textbf{j}\; \mathrm{and}\; \textbf{k}$ are the Cartesian unit vectors. 
We choose $h,\; \hat{U}_c, \;h/\hat{U}_c\; \mathrm{and}\; \rho \hat{U}_c^2$ as length, velocity, time, and pressure scales, respectively, where $(\hat{\boldsymbol{\cdot}})$ denotes the dimensional quantities. The velocity $\hat{U}_c$ is the centreline laminar velocity and $h$ is the duct half height. The associated non-dimensional NSE and continuity equations are given by: 
\begin{subequations}
    \begin{equation}
 \partial_t \bsu + \bsu \cdot \nabla \bsu =-\nabla p +\Rey^{-1} \;\nabla^2 \bsu,
 \label{eqn3.1}    
 \end{equation}
 \begin{equation}
\nabla \cdot \bsu =0,
    \end{equation}
\label{TotNavStokes}
\end{subequations}
with no-slip boundary conditions
\begin{equation}
\bsu(x,y = \pm 1,z) = \bsu(x,y, z = \pm 1)=0,
\label{eq:BCs}
\end{equation}
where the Reynolds number is defined as $\Rey=\hat{U}_c h/\nu$. We decompose the flow field variables into a base state and 
time-dependent fluctuations as
\begin{subequations}
\begin{equation}
\bsu (\bsx,t)=\bsU(\bsx)+  \bsu'(\bsx,t),
\end{equation}   
\begin{equation}
p (\bsx,t)=P_o+\frac{dP(x)}{dx}x+\; p'(\bsx,t),
\end{equation}
\label{ReDecomp}
\end{subequations}
where $\bsU(\bsx)=U(y,z)\textbf{i}+0\textbf{j}+0\textbf{k}$ is the base (laminar) velocity profile, $\bsu'=u'\textbf{i}+v'\textbf{j}+w'\textbf{k}$, 
 and $P_o$ is an integration constant. While we consider a general 3-component fluctuating velocity for the purposes of this derivation, the ROMs that we identify will only consider cross-stream ($v',w'$) fluctuations, which themselves will be streamwise-constant. Both $\bsU$ and $\bsu'$ satisfy the same boundary conditions as the total velocity indicated in \eqref{eq:BCs}. Setting $\bsu'=0$ in~\eqref{ReDecomp} and substituting the resulting
decomposition into~\eqref{TotNavStokes} gives the steady laminar
base-flow equations
\begin{subequations}
    \begin{equation}
     \bsU \cdot \nabla \bsU =-\nabla P +\Rey^{-1} \nabla^2 \bsU   , \end{equation}
     \begin{equation}
       \nabla \cdot \bsU =0.  
    \end{equation}
     \label{MeanNavier}
\end{subequations}

The linearised equations governing the fluctuating components are obtained by subtracting~\eqref{MeanNavier} from~\eqref{TotNavStokes} while utilising the decomposition~\eqref{ReDecomp} and neglecting the nonlinear $\bsu'\cdot \nabla\bsu'$ term, giving
\begin{subequations}
    \begin{equation}
    \partial_t \bsu' +\bsU\cdot \nabla\bsu' +\bsu' \cdot \nabla \bsU + \nabla p' -\Rey^{-1} \nabla^2 \bsu' = 0 ,
    \end{equation}
    \begin{equation}
\nabla \cdot \bsu'= 0    .
\end{equation}
    \label{PerturbSystem}
\end{subequations}

\begin{figure}
\begin{subfigure}{0.5\textwidth} 
  \centerline{\includegraphics[width=\textwidth]{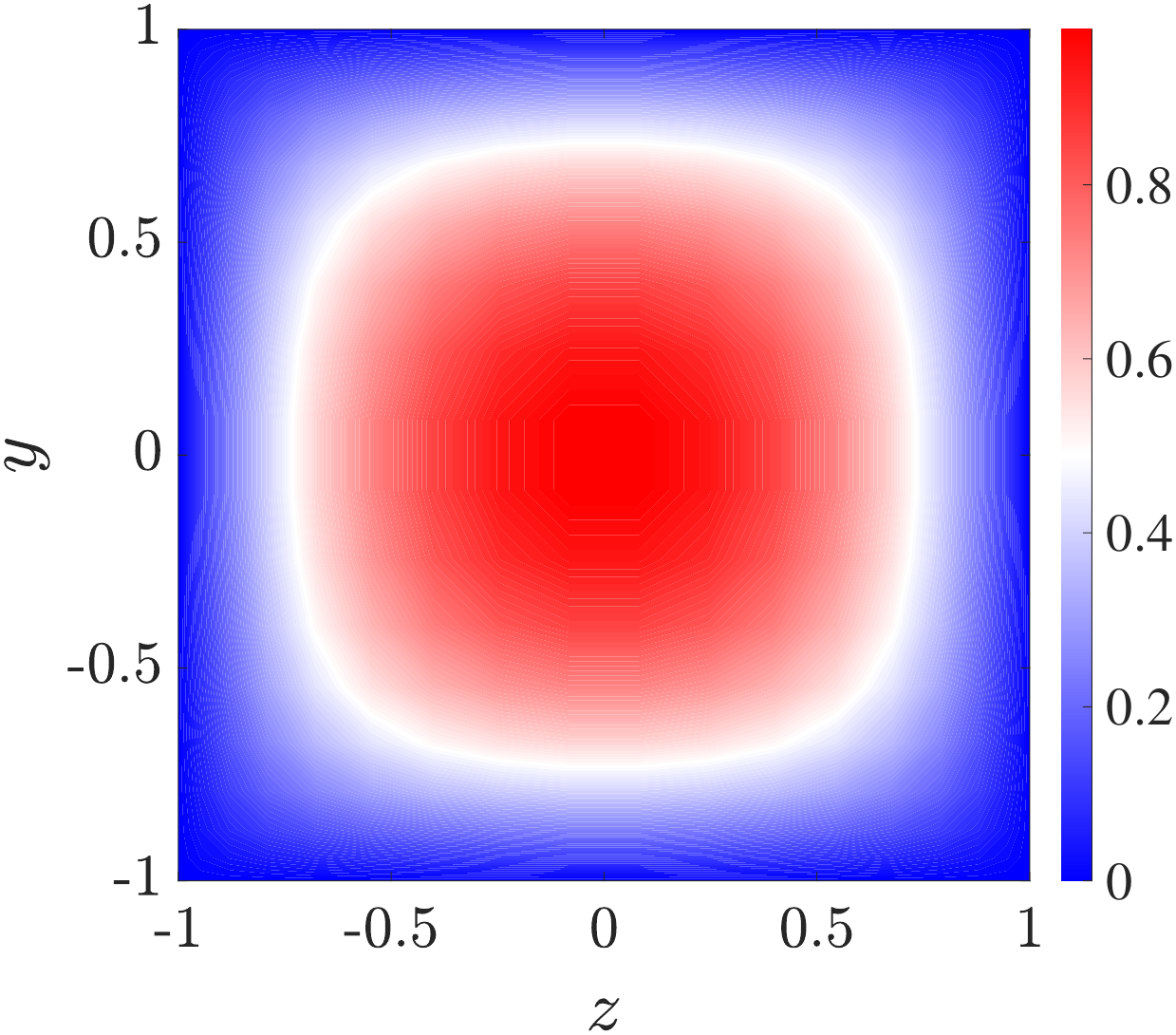}}
   \caption{}\label{LaminarProfile}
\end{subfigure}
\begin{subfigure}{0.5\textwidth} 
        \centerline{\includegraphics[width=\textwidth]{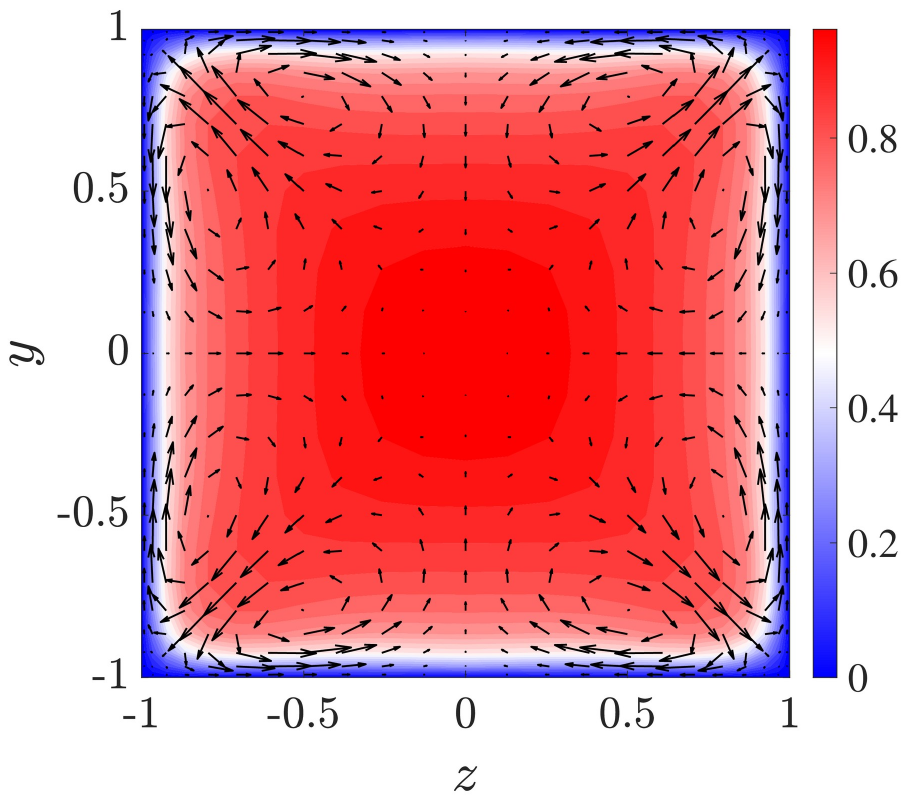}}
\caption{}\label{TurbulentProfile}
    \end{subfigure}
  \caption{ (a) Laminar  and (b) turbulent mean flow from DNS simulations at $\Rey=2796$~\citep{vinuesa2014}. The coloured contours show the streamwise velocity distribution, and the vectors show the secondary turbulent mean velocity. Both contour plots are normalised by the corresponding maximum streamwise velocity.}
    \label{LamAndTurb}
\end{figure}

The base laminar solution to~\eqref{MeanNavier} is obtained by solving the 2D Poisson equation
    \begin{equation}
\left(\partial_{yy}+\partial_{zz}\right)U(y,z)=-\Rey \; \frac{d P(x)} {d x} =C_1,
\label{Poission}
    \end{equation}
again with the no-slip boundary conditions $U(y,z=\pm 1) = U(y=\pm 1,z)=0$. Setting $C_1=-3.393449$~\citep{wedin2008} gives a dimensionless centreline velocity $U_c=1$. The solution of~\eqref{Poission} can be expressed as a series solution \citep{panton2024}
\begin{equation}
    U(y,z)= C_2\left( 1 -y^2\right)+ C_3  \sum_{n=1}^{\infty} \frac{(-1)^n}{\alpha_n ^2} \cos{(\alpha_n y)}  \frac{\cosh {(\alpha_n z)} }{\cosh {(\alpha_n )}},
    \label{LamProf}
\end{equation}
 where $\alpha_n=(2n-1)\pi/2$, $C_2 = 1.69672$, and $C_3 = 6.78689 $. Figure~\ref{LaminarProfile} shows contours of this 2D laminar flow (which is independent of Reynolds number), compared to a turbulent mean flow (figure \ref{TurbulentProfile}) from DNS simulations~\citep{vinuesa2014} at  a Reynolds number $\Rey = 2796$ (based on the centreline velocity). As shown, the streamwise velocity is flattened across the duct cross-section towards the walls and deformed near the corner due to the secondary flow. 

\section{Basis function computation and properties}\label{Eigenfunctions and eigenvalues}
Using Galerkin projection models requires the selection of basis functions upon which we project the governing equations. Here, we describe how these basis functions are chosen and selected. Since we do not assume any prior knowledge of the turbulent flow field, we cannot use POD modes as often when identifying low-dimensional models via Galerkin projection. Instead, we choose to use eigenmodes of the linearised NSE for this system as basis functions. We describe these linearised equations for the square duct in \S\ref{sec:lin}, before discussing the numerical methods by which these eigenmodes are computed in \S\ref{sec:computation}. The properties of these eigenmodes, particularly the geometrical symmetries that they can possess, are discussed in \S\ref{sec:symmetries}.

\subsection{ Linearised equations and their eigendecomposition}
\label{sec:lin}

The eigenvalue problem of the linearised system~\eqref{PerturbSystem} around a streamwise laminar base flow is obtained by assuming a fully developed flow that is homogeneous in the streamwise direction. In particular, we consider a fluctuating velocity field of the form
\begin{equation}
\bsu'(\bsx,t)=
\tilde{\bsu}(k_x,\lambda;y,z)\;\exp{(\text{i} k_x x+\lambda t)},
\label{Temoralgrowth}
\end{equation}
where $\lambda$ determines the temporal evolution of this fluctuation, and $k_x$ is the streamwise wave number. Note that while $\lambda$ is generally complex (and indeed is often expressed in terms of a complex frequency), it is real for the streamwise-constant eigenproblems considered in the present work. Substituting~\eqref{Temoralgrowth} into~\eqref{PerturbSystem}, we get a linearised eigensystem decomposition of the form
\begin{equation}
    \lambda \tilde{\bsu} =-\bsU \cdot \tilde{\nabla} \tilde{\bsu} -\tilde{\bsu} \cdot \tilde{\nabla} \bsU -\tilde{\nabla} \tilde{p} +\Rey^{-1} \tilde{\nabla}^2 \tilde{\bsu},\quad     \tilde{\nabla} \cdot \tilde{\bsu}=0, 
    \label{LinearizdForm}
    \end{equation}
where the transformed gradient operator is $\tilde{\nabla}:= \text{i}k_x\textbf{i}+\partial_y\textbf{j}+\partial_z\textbf{k}$ and the Laplacian operator is $\tilde{\nabla}^2:=-k_x^2+\partial_y^2+\partial_z^2$. Equation~\eqref{LinearizdForm} can be expressed in a matrix form as
\begin{subequations}
\label{eq:lin}
\begin{equation}
 {
\left[\lambda\mathsfbi{B} \ + \
\begin{pmatrix}
\mathsfbi{L}_{1} & \partial_yU &  \partial_z U & {\text{i}}k_x \\
0 & \mathsfbi{L}_{1} & 0   & \partial_y \\
0 & 0 & \mathsfbi{L}_{1} & \partial_z \\
{\text{i}}k_x & \partial_y & \partial_z & 0 
\end{pmatrix}\right]
\begin{pmatrix}
\bsphi^{u} \\
\bsphi^{v}  \\
\bsphi^{w}  \\
\bsphi^{p} 
\end{pmatrix}
 }=(\lambda \mathsfbi{B} +\mathsfbi{A})
 \begin{pmatrix}
\bsphi^{(u,v,w)} \\
\bsphi^{p} 
 \end{pmatrix}
 =0,
\label{EigenFull}
\end{equation} 
where

\begin{equation}
    \mathsfbi{L}_{1} = \text{i} k_xU -\Rey^{-1}\tilde{\nabla}^2,
\end{equation}
and
\begin{equation}
    \mathsfbi{B}=
    \begin{pmatrix}
    1& 0 & 0 & 0 \\
    0 & 1 & 0 & 0 \\
    0 & 0& 1 &  0 \\
    0 & 0 & 0 & 0 \\
\end{pmatrix}
.
\end{equation}
\end{subequations}
The $\bsphi^{u},\bsphi^{v},$ and $\bsphi^{w}$ correspond to the streamwise and cross-stream components of the eigenmode $\bsphi$, respectively, and the superscript $(\boldsymbol{\cdot})^{(u,v,w)}$ is read as an eigenvector $\bsphi$ with $u,v$, and $w$ components. Here and throughout, \eqref{EigenFull} is completed by the Dirichlet boundary conditions $\bsphi^{(u,v,w)}(y = \pm 1,z) = \bsphi^{(u,v,w)}(y,z = \pm 1) = 0$. 
The differential operator of equation~\eqref{EigenFull} is non-normal; that is $(\mathsfbi{A}\mathsfbi{A}^* \neq \mathsfbi{A}^*\mathsfbi{A})$ where $\mathsfbi{A}^*$ is the adjoint operator. Therefore, its eigenmodes are not orthogonal. 

From this point on, we will only consider streamwise-constant fluctuations, where $k_x = 0$ in \eqref{LinearizdForm}. In this case, \eqref{eq:lin} simplifies to
\begin{equation}
 {
\left[\lambda\mathsfbi{B}\ + \
\begin{pmatrix}
\mathsfbi{L}_{2} & \partial_yU & \partial_z U & 0 \\
0 & \mathsfbi{L}_{2} & 0   & \partial_y \\
0 & 0 & \mathsfbi{L}_{2} & \partial_z \\
0 & \partial_y & \partial_z & 0 
\end{pmatrix}\right]
\begin{pmatrix}
\bsphi^{u} \\
\bsphi^{v}  \\
\bsphi^{w}  \\
\bsphi^{p} 
\end{pmatrix}=0
}, 
\label{EigenNoMean}
\end{equation} 
where $\mathsfbi{L}_{2} = -\Rey^{-1}(\partial_{yy}+\partial_{zz})$. 
In this case, non-normality arises due to the off-diagonal terms, $\partial_yU$ and $\partial_zU$. For this streamwise constant case, the operator is block upper triangular and can be recast in a compact form as
\begin{equation}
\begin{pmatrix}
    A_u &   C_{(v,w) \rightarrow u}\\
    0   &   A_{v,w,p}
\end{pmatrix}
\begin{pmatrix}
    \bsphi^{u}\\
    \bsphi^{v,w,p}
\end{pmatrix}=0,
\label{blockEigenNoMean}
\end{equation}
where $A_u = \lambda + \mathsfbi{L}_{2}$ is the top left block, $A_{v,w,p}$ is the lower right $3\times 3$ block, and $C_{(v,w) \rightarrow u} = \begin{pmatrix}\partial_yU & \partial_z U & 0 \end{pmatrix}$ is the upper off-diagonal block of~\eqref{EigenNoMean}. 
The non-normality of the full operator arises from this off-diagonal block that maps from transverse velocity components to the streamwise velocity component.  Because the operator is
block upper triangular, its generalised eigenvalues are the values of
$\lambda$ for which either diagonal block is singular, yielding the two
separate eigenvalue problems given by

\begin{subequations}
\begin{equation}
\left[\lambda^{u} \ + \ \mathsfbi{L}_2\right]\bsphi^u_\text{Laplace} = 0,
\label{eq:streamwiseeig}
\end{equation}
\begin{equation}
\left[\lambda^{(v,w)}\begin{pmatrix}
  1 & 0   & 0\\
 0 &1 & 0\\
0 & 0 & 0 
\end{pmatrix}\ + \
\begin{pmatrix}
  \mathsfbi{L}_{2} & 0   & \partial_y \\
 0 & \mathsfbi{L}_{2} & \partial_z \\\partial_y & \partial_z & 0 
\end{pmatrix}\right] \begin{pmatrix}
\bsphi^{v}  \\
\bsphi^{w}  \\
\bsphi^{p} 
\end{pmatrix}= 0,
\label{EigenProbvw}
\end{equation}
\end{subequations}
where $\lambda^u$ and $\lambda^{(v,w)}$ are the eigenvalues associated with the streamwise eigenproblem and cross-stream eigenproblem, respectively. Equation \eqref{eq:streamwiseeig}, corresponding to eigenmodes with only a streamwise component, is equivalent to an eigenproblem of the Laplace operator, with solutions given by
\begin{subequations}
    \begin{equation}
           \bsphi^u_{\text{Laplace},(j,k)}= \sin\left( \frac{j\pi}{2} (y+1)\right)\sin\left(\frac{k \pi}{2} (z+1)\right),
           \label{AnalyticUmodes}
    \end{equation}
    \begin{equation}
            \lambda^{u}_{(j,k)} = -\frac{\pi^2}{4Re}\left( j^2+k^2\right),
            \label{EigenValu}
    \end{equation}
\end{subequations}
where the modal indices $j,k=1,2,3,...$, and the eigenmodes $\bsphi^u_\text{Laplace}$ are orthonormal. 

The generalized eigenproblem given in \eqref{EigenProbvw} corresponds to the two-dimensional Stokes operator. The eigenvalues and eigenmodes for this operator must be computed numerically, which will be discussed in the next section. While the eigenvalues $\lambda^{(v,w)}$ only depend on the lower diagonal block of \eqref{EigenNoMean}, the corresponding eigenmodes will also have a streamwise  component, owing to the off-diagonal block $C$. The streamwise component of these eigenmodes is given by 
\begin{equation}
\left[\lambda^{(v,w)} + \mathsfbi{L}_2\right]\bsphi^u_\text{response}=- \left(\partial_yU\; \bsphi^v+\partial_zU\;\bsphi^w\right), \quad \forall \; \lambda^{(v,w)} \;\notin \sigma(\mathsfbi{L}_2)
  \label{streaamwiseeqn}
\end{equation}
where $\sigma (\mathsfbi{L}_2)$ is the spectrum of the operator $\mathsfbi{L}_2$. The equation shows that the cross-stream velocity components of the lower-block eigenmodes on the right-hand side of the equation act as a forcing, triggering a response in the streamwise direction. 

From the previous discussion, for the case $k_x=0$, we note that while the operators in \eqref{eq:streamwiseeig}-\eqref{EigenProbvw} are Hermitian and thus have orthogonal eigenmodes, the eigenmodes of the full operator \eqref{EigenNoMean} are not orthogonal, owing to the streamwise component of the lower-block eigenmodes given in \eqref{streaamwiseeqn}.
The eigenmode of the lower diagonal block, $A_{v,w,p}$, corresponds to the Stokes eigenmodes~\eqref{EigenProbvw} and is self-adjoint with a real spectrum; therefore, the corresponding eigenmodes form a complete orthonormal basis in the space of divergence-free cross-stream velocity fields. Since the present work focuses on modelling the secondary velocity components, the reduced-order models are constructed using the eigenmodes of this cross-stream Stokes problem, rather than the full non-orthogonal eigenmodes of \eqref{EigenNoMean}. Therefore, the basis functions of the ROMs contain only the secondary velocity components and are written as $\bsphi^{(v,w)} (y,z)=\left ( \bsphi^v (y,z), \bsphi^w (y,z)  \right)^T$. The superscript $(\boldsymbol{\cdot})^{(v,w)}$ will be dropped in the following sections for simplicity. The computation and structure of these eigenfunctions will be discussed in \S\ref{sec:computation} and \S\ref{sec:symmetries} respectively.
 
We further note that while the eigenvalues of \eqref{eq:streamwiseeig}-\eqref{EigenProbvw} depend on the Reynolds number, the eigenmodes, and thus the subspace used to define our reduced-order models, will be independent of the Reynolds number. We could eliminate this  Reynolds number dependency altogether by redefining the velocity scale to be $\hat{U}_c=\nu/h$, which would replace $Re$ by unity in \eqref{EigenNoMean}-\eqref{streaamwiseeqn}. 

There are several advantages to using this choice of eigenmodes over others that could be formulated, such as those obtained from resolvent or transient growth analysis, or through balanced POD. The current eigenmode formulation allows for the secondary velocity components to be entirely decoupled from the streamwise components while still yielding orthonormal modes, whereas when applied to the full three-component system, the other methods mentioned generally give modes with coupled velocity components. In practice, this can lead to models that make it difficult to isolate the secondary flow dynamics, since the leading modes are often dominated by streamwise velocity components. Additionally, within the present streamwise-constant formulation, the
cross-stream basis functions are independent of the choice of laminar or
turbulent streamwise base/mean velocity profile. This is advantageous if the mean is not known in advance, and also for showing that the predictions of the ROMs are not sensitive to this choice of linearisation point.

\subsection{Computation of eigenmodes }\label{sec:computation}

We now discuss the approach to numerically compute the eigenvalues and eigenmodes of the Stokes operator described in \eqref{EigenProbvw}. Considerable effort has been directed towards developing efficient numerical solvers for the Stokes operator across both 2-D and 3-D geometries, including square~\citep{leriche2004}, triangular~\citep{chen2016}, and cubic domains~\citep{labrosse2014}. The choice of solution methodology for the Stokes problem depends on the selected variables, whether primitive flow variables~\citep{arrow1958,chorin1968,kleiser1980} or stream-function formulations~\citep{batoul1994,leriche2004}. A common objective of these Stokes solvers is the decoupling between the pressure and velocity fields. However, with modern computational capabilities, direct solutions to large-scale eigenvalue problems are feasible without iterative or splitting schemes. Therefore, we solve the eigenvalue problem of the linear differential operator using the MATLAB function \texttt{eigs} which utilises the implicitly restarted Arnoldi method in the ARPACK library~\citep{ARPACK}. Spatial discretisation is carried out using the Chebyshev collocation spectral method, with an equal number of Chebyshev nodes, denoted as $N_c=96$, in each spatial direction. The no-slip boundary condition is implicitly imposed on the velocity field using the methods described in \citet{trefethen2000spectral}, restricting computations to the internal nodes of the domain.

\subsection{Eigenmode structure and symmetries}
\label{sec:symmetries}
 \begin{figure}
  \centerline{\includegraphics[width=1.03\textwidth]{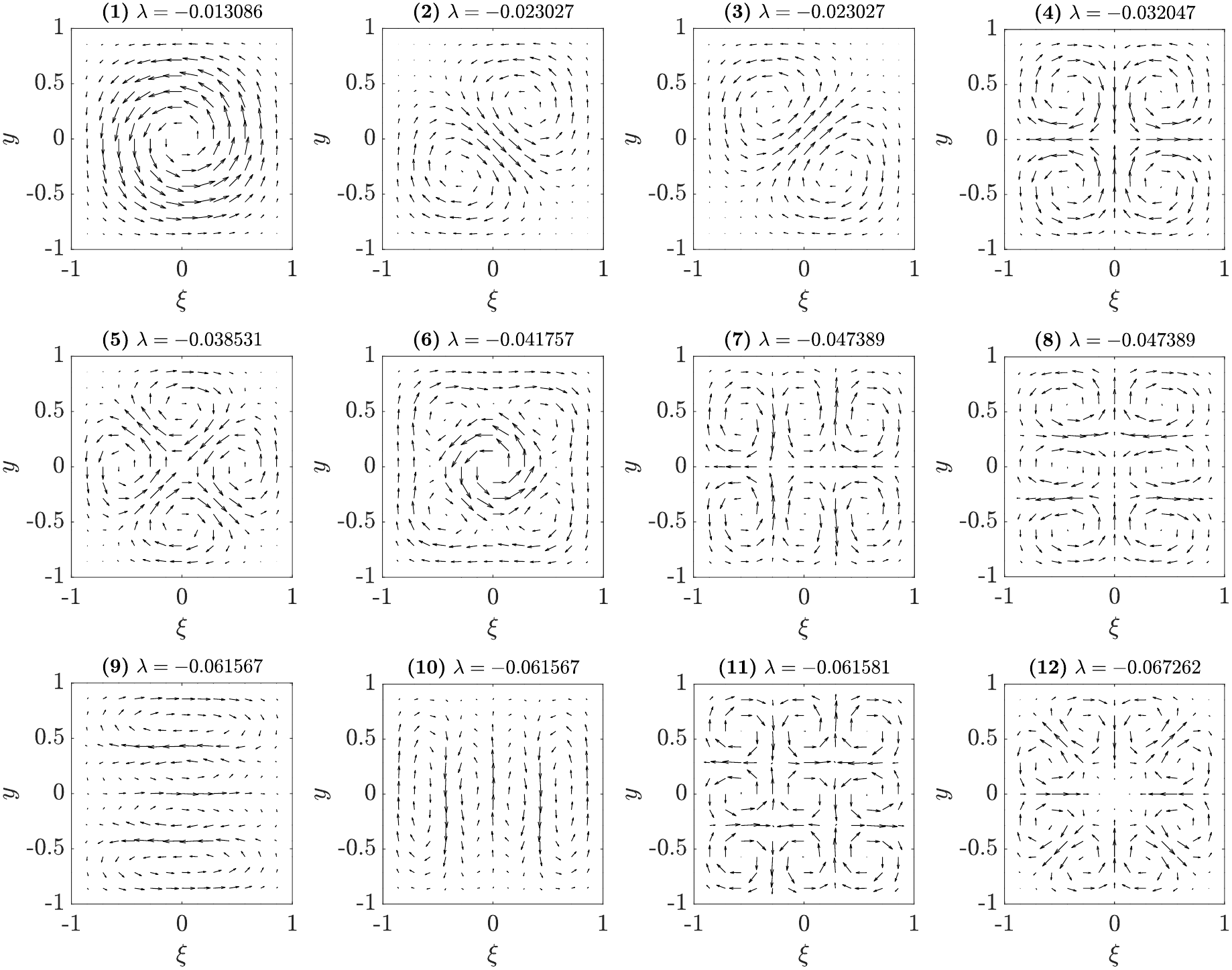}}
  \caption{Vector field plots of the  twelve least stable eigenmodes $\bsphi$ of~\eqref{EigenProbvw}, arranged from least to most stable. A total of 96 Chebyshev nodes are used in both the $y$ and $z$ directions. }
\label{TwelveEigenmodes}
\end{figure}
 \begin{figure}
  \centerline{\includegraphics[width=0.85\textwidth]{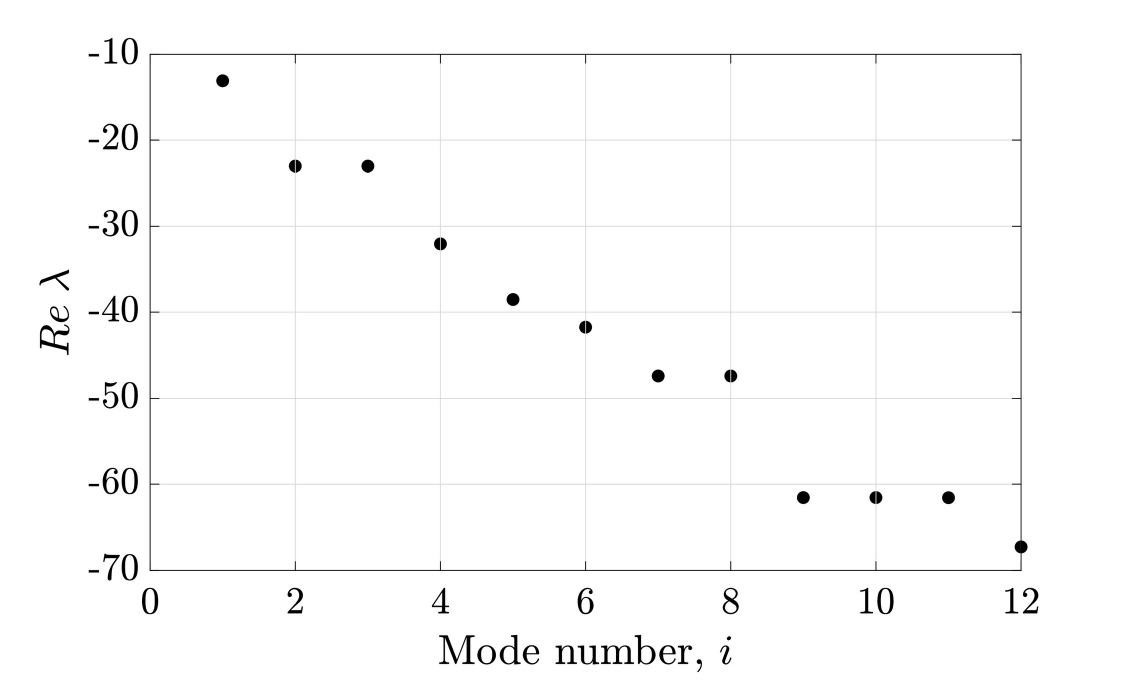}}
  \caption{Spectrum associated with the twelve least-damped cross-stream eigenmodes shown in figure~\ref{TwelveEigenmodes}, obtained from~\eqref{EigenProbvw}. The eigenvalues are real and negative, and are shown after multiplication by the Reynolds number, $Re$.}
\label{TwelveEigevalues}
\end{figure}
The twelve least stable eigenmodes of~\eqref{EigenProbvw} are shown in figure~\ref{TwelveEigenmodes}, arranged in increasing order of stability (damping rate). All eigenmodes exhibit vortical structures, which are largest for the least stable modes. The structure of these modes is consistent with the eigenmodes of the Stokes operator reported in \citet{leriche2004}. The spectrum of these eigenmodes is shown in figure~\ref{TwelveEigevalues} scaled by the Reynolds number, $\Rey$. We find that all eigenvalues are stable and observe that there are several repeated eigenvalues, which arise due to the symmetries present in the square duct geometry. While these cross-stream basis modes are asymptotically stable, the transient secondary flows that they are associated with are a key component of linear amplification mechanisms that can yield large growth, owing to the non-normality of the linear operator. 
\cite{GALLETTI2004} showed that in square ducts, optimally configured cross-stream vortices induce large transient amplification despite being asymptotically stable. They found that, for a turbulent mean profile, large-scale secondary structures can be sustained over long streamwise distances through a linear transient mechanism.

Subsets of these eigenmodes will be used as basis functions for identifying ROMs. The properties and predictions of these ROMs will be dependent on the properties of the eigenmodes used, with the symmetries of the chosen eigenmodes being particularly important for obtaining mean field predictions with expected symmetries.
These eigenmodes are 2D vector fields defined on a square domain, which can feature two fundamental planar isometries: rotational symmetry ($\theta-$rotation) and reflectional symmetry ($\beta-$reflection) \citep{leriche2004}. These symmetries play an important role in shaping the structure of eigenmodes, and the subsequent flow field predicted by ROMs. 
Given a two-dimensional eigenmode $\bsphi (y,z)=\left ( \bsphi^v (y,z), \bsphi^w (y,z)  \right)^T$, its rotational symmetry is expressed as 
\begin{equation}
    \boldsymbol{\phi}'(y',z')=\mathcal{R}(\theta)\bsphi \left(\mathcal{R}(\theta)(y,z)\right)
\end{equation}
where $\mathcal{R}(\theta)$ is a rotational transformation matrix defining a coordinate transformation of the form 
\begin{equation}
\begin{pmatrix}
y' \\
z'  \\
\end{pmatrix}=
\begin{pmatrix}
\cos{\theta} & \sin{\theta} \\
-\sin{\theta}& \cos{\theta} 
\end{pmatrix}
 \begin{pmatrix}
y  \\
z  \\
\end{pmatrix}
\end{equation}

Similarly, a reflection transformation about an angle $\beta$ is accomplished using the reflection transformation matrix $\mathcal{L}(\beta)$, defined as follows:
\begin{equation}  \boldsymbol{\phi}''(y'',z'')=\mathcal{L}(\beta) \bsphi \left( \mathcal{L}(\beta) (y,z)  \right)
\end{equation}
with the coordinate reflection expressed as
\begin{equation}
\begin{pmatrix}
y''  \\
z'' \\
\end{pmatrix}=
\begin{pmatrix}
-\cos{\beta} & \sin{\beta} \\
\sin{\beta}& \cos{\beta} 
\end{pmatrix}
 \begin{pmatrix}
y  \\
z  \\
\end{pmatrix}
\end{equation} 
The $\theta$ and $\beta$ angles denote an anticlockwise rotation by an angle $\theta$ and a reflection around a line with an angle $\beta/2$ with the positive $z$ axis, respectively. Note that the transformation matrices $\mathcal{R}$ and $\mathcal{L}$ are written such that the first row corresponds to the transformation of the vertical coordinate $y$, and the second row corresponds to the horizontal coordinate $z$. The square has four rotational and four reflection symmetries characterised by angles $\theta$ and $\beta$ as multiples of $\pi/2$ and summarised in table~\ref{tab:SymmTable}. 

Of the eigenmodes shown in figure~\ref{TwelveEigenmodes}, only mode 12 satisfies all eight possible geometric symmetries. This will be important when it comes to selecting modes to use for a ROM, since the inclusion of such a fully symmetric mode will be required for the prediction of secondary mean flow which also  satisfies all symmetries. 
Note that while mode 12 is the only such fully-symmetric mode shown in figure~\ref{TwelveEigenmodes}, it is one of a family of such eigenmodes of the square duct geometry, with all others being even more stable.

\begin{table}
  \begin{center}
\def~{\hphantom{0}}
\renewcommand{\arraystretch}{1.7}
\begin{tabular*}{0.9\textwidth}{@{\extracolsep{\fill}} lll}
  \textbf{Angle} & \textbf{Transformed field} & \textbf{Coordinate mapping} \\[3pt]
 $\theta=$ & $\boldsymbol{\phi}'(y',z')=$ &  $(y',z')=$\\[2pt]
$\pi/2$   & $\big(\;\bsphi^w(y',\,z'),\;-\,\bsphi^v(y',\,z')\;\big)^T$ & $(z,-\,y)$ \\
$\pi$     & $\big(-\bsphi^v(y',\,z'),\;-\bsphi^w(y',\,z')\;\big)^T$ & $(-\,y,-\,z)$ \\
$-\pi/2$  & $\big(-\bsphi^w(y',\,z'),\;\bsphi^v(y',\,z')\;\big)^T$ & $(-\,z,\,y)$ \\
$2\pi$    & $\big(\;\bsphi^v(y',\,z'),\;\bsphi^w(y',\,z')\;\big)^T$ & $(y,\,z)$ \\[6pt]
 $\beta/2 =$ & $ \boldsymbol{\phi}''(y'',z'')=$ & $(y'',\,z'')=$ \\[2pt]
$ \pi/4$       & $\big(\;\bsphi^w(y'',\,z''),\;\bsphi^v(y'',\,z'')\;\big)^T$        & $(z,\,y)$ \\
$\pi/2$& $\big(\;\bsphi^v(y'',\,z''),\;-\,\bsphi^w(y'',\,z'')\;\big)^T$      & $(y,\,-z)$ \\
$ 3\pi/4$     & $\big(\;-\bsphi^w(y'',\,z''),\;-\,\bsphi^v(y'',\,z'')\;\big)^T$   & $(-z,\,-y)$ \\
 $ \pi$& $\big(\;-\bsphi^v(y'',\,z''),\;\bsphi^w(y'',\,z'')\;\big)^T$      & $(-y,\,z)$ \\
\end{tabular*}
\caption{Symmetry transformations of the cross-stream eigenmode vector field $\bsphi(y,z)=\left( \bsphi^v,\bsphi^w \right)^T$ under $\mathcal{R}(\theta)$ (rotations) and $\mathcal{L}(\beta)$ (reflections), with the corresponding coordinate mappings.}
\label{tab:SymmTable}
\end{center}
\end{table}

 We finally note that we are not the first to observe the similarity between the secondary flow and eigenmodes of linear operators in the square duct geometry. \citet{pirozzoli2018} found that the Helmholtz equation associated with the secondary flow streamfunction under an inviscid assumption admits solutions (corresponding to eigenmodes of the Laplacian) that resemble the secondary mean. In addition, \citet{wedin2008} used Stokes eigenmodes (equivalent to those shown in figure \ref{TwelveEigenmodes}) to represent streamwise-constant rolls in their model for self-sustaining processes (SSP), though they found that a four-vortex mode (equivalent to mode 4 in figure \ref{TwelveEigenmodes}) was more conducive to self-sustaining dynamics than the eight-vortex configuration.

\section{Galerkin  projection}\label{Galerkin  projection}

In this section, we discuss the Galerkin projection method for approximating the incompressible NSE and the derivation of the evolution equations for internal flow with Dirichlet boundary conditions. Assuming a streamwise-constant flow and neglecting streamwise fluctuations,  we approximate the cross-stream dynamics as
\begin{subequations}
    \begin{equation}
    \bsu (y,z,t) = \begin{pmatrix}
        U(y,z)\\
        0\\
        0
    \end{pmatrix}+
    \begin{pmatrix}
        0\\
        v'(y,z,t)\\
        w'(y,z,t)
    \end{pmatrix}
     \approx \begin{pmatrix}
        U(y,z)\\
        0\\
        0
    \end{pmatrix} +\sum_{i=1}^{N} a_i(t) \bsphi_i(y,z)    \end{equation}
    \begin{equation}
      \frac{\text{d} \boldsymbol{a}}{\text{d} t}=\boldsymbol{F}(\boldsymbol{a}), \label{ODEsfora}   \end{equation}
    \label{LowDimSys}
\end{subequations}
where $\boldsymbol{a}=(a_1,a_2,a_3, \cdots,a_N)^T$ and $\bsphi_i=(0,\bsphi^v_i,\bsphi^w_i)^T$. In this formulation, the total velocity field $\bsu$ is approximated as a laminar velocity profile, $U(y,z),$ and a summation of cross-stream spatial basis functions $\bsphi_i(y,z)$ multiplied by time-evolving magnitudes $a_i(t)$ known as expansion coefficients. The expansion coefficients are governed by a system of ordinary differential equations (ODEs) $\boldsymbol{F}(\boldsymbol{a})$. Equation~\eqref{LowDimSys} is therefore a low-order $N$-dimensional representation of the cross-stream dynamics of the
streamwise-constant Navier--Stokes equations.

The system of ODEs for the expansion coefficients~\eqref{ODEsfora} is derived by solving an optimization problem that involves the projection of the basis functions onto the residual of the Galerkin expansion of the NSE within the same subspace~\citep{cassel2013}. The residual of the NSE is expressed as, 
\begin{equation}
        \mathscr{R}(\bsu)=\partial_t \bsu + \bsu \cdot \nabla \bsu +\nabla p -\Rey^{-1} \;\nabla^2 \bsu
\end{equation}
and the projection would be in the form 
\begin{equation}
        \left\langle  \boldsymbol{\phi}_i, \mathscr{R}\left( \sum_{j=0}^{N}a_j(t)\boldsymbol{\phi}_j(y,z) \right) \right\rangle_{\Upsilon}=0,
\qquad \quad i=1,\ldots,N.
\end{equation}
where $\langle \cdot, \cdot \rangle_\Upsilon $ is the standard inner product on the domain $\Upsilon$, and for simplicity we let $\boldsymbol{\phi_0}=\boldsymbol{U}(y,z)$ and $a_0=1$. Assuming orthonormal basis functions, the resulting quadratically nonlinear system of ODEs can be expressed as 
\begin{equation}
    \dfrac{\text{d}a_i}{\text{d}t}=-\sum_{j=0}^N \sum_{k=0}^N a_j a_k\; \langle \bsphi_j \cdot \nabla \bsphi_k , \bsphi_i \rangle +\Rey^{-1} \sum_{j=0}^N a_j \; \langle \nabla^2 \bsphi_j,\bsphi_i \rangle , \quad i=1,\ldots,N.\label{galerkinmodel}
\end{equation}
where the inner product $\langle \nabla^2 \bsphi_j,\bsphi_i \rangle$ is the projection of the viscous term, $\langle \bsphi_j \cdot \nabla \bsphi_k , \bsphi_i \rangle$ is the nonlinear advection term. The projection of the pressure gradient vanishes identically for internal incompressible flow with Dirichlet and periodic boundary conditions~\citep{holmes2012}, which can be shown by transforming the volume into a surface integral using the divergence theorem. It should be noted that the pressure gradient has no impact on the Galerkin projection of this type of flow field, but for open flows, this term can be crucial and  computationally expensive~\citep{noack2005}. 
 Equation \eqref{galerkinmodel} can also be generalised to include additional closure terms to account for unmodelled dynamics, which will be utilised in the forthcoming analysis. 
To achieve spectral accuracy in solving the system of equations~\eqref{galerkinmodel}, a Chebyshev differentiation scheme with Clenshaw--Curtis integration weights are used \citep{trefethen2000spectral}.

In developing a ROM, two main questions naturally arise: first, how simplified can the model become without losing essential dynamics, and second, which modes must be retained in such a simplified model? A detailed investigation of~\eqref{galerkinmodel} is crucial to answer these fundamental questions. The nonlinear advection term accounts for the modal interactions and energy transfers between different modes, making it essential to the accuracy of any reduced-order model. The divergence-free condition on the velocity field constrains the NSE and the eigenfunctions of their linearised form. Consequently, the projection of any spatial basis function onto the residual of the advection term representing the self-interaction of the same basis function 
vanishes identically for incompressible bounded flows. Using the divergence theorem, this property can be written as
\begin{equation}
\langle \boldsymbol{\phi}_j \cdot \nabla \boldsymbol{\phi}_j ,\boldsymbol{\phi}_j\rangle=0
\label{SingleProjection}
\end{equation}

Since the nonlinearity disappears for individual modes, it becomes clear that at least two basis functions must be included to preserve nonlinear effects in the model. The selection of these modes is guided by the symmetry of the mean secondary flow field. Specifically, as discussed in \S\ref{sec:symmetries}, because the mean secondary flow is fully symmetric (invariant under all the transformations 
described in table \ref{tab:SymmTable}), the reduced-order model must incorporate mode 12 in figure~\ref{TwelveEigenmodes} to reconstruct a secondary mean state with nonzero cross-stream velocity components and also obey all the symmetry properties of the square duct geometry. 

\section{Reduced-order models}\label{results}

This section presents results from applying the methodology described in \S\S\ref{Eigenfunctions and eigenvalues}-\ref{Galerkin  projection} to develop ROMs for the duct flow configuration described in \S\ref{Governing equations}.

\subsection{Galerkin projection ROM}
A model consisting of $N$ 
modes will introduce a system of $N$ ODEs governing the Galerkin expansion coefficients. For the cases considered here, the resulting deterministic system of equations constructed using~\eqref{galerkinmodel} is typically asymptotically stable, with
\begin{equation}
    \lim_{t\rightarrow \infty} a_i (t)=0 \qquad i=1,\cdots, N.
\end{equation}
This is consistent with the stability properties of the linearised equations about the laminar solution. To account for the unresolved scales 
in the NS, a forcing term is introduced into the system of ODEs. Two types of forcing are considered: deterministic forcing and stochastic forcing. The deterministic forcing is chosen to be sinusoidal functions, and the deterministic forced system would be 
\begin{equation}
    \dfrac{\text{d}a_i}{\text{d}t}=-\sum_{j,k=0}^N a_j a_k\; \langle \bsphi_j \cdot \nabla\bsphi_k,\bsphi_i \rangle +\Rey^{-1} \sum_{j=0}^N a_j \; \langle\nabla^2 \bsphi_j,\bsphi_i\rangle +f_i,
    \label{forcedsystem}
\end{equation}
where $f_i =B_i \cos{(\omega_i (t-t_i))}$, $B_i$ denotes the forcing magnitude and $\omega_i$ is the angular frequency of the forcing for the $i$-th equation, normalised by the streamwise centreline velocity and the duct half height. Note that for the case considered here, all terms involving $\phi_0$ disappear, as the laminar solution does not have secondary velocity components. For the stochastic forcing, the stochastic system of ODEs is expressed as~(\cite{sarkka2019,kasdin1995})
\begin{equation}
    \frac{\text{d} \boldsymbol{a} }{\text{d}t}=\boldsymbol{F}(\boldsymbol{a}(t)) +\boldsymbol{G}(\boldsymbol{a}(t))\;\boldsymbol{f}(t).
    \label{StochasticForm}
\end{equation}
In this formulation, $\boldsymbol{F}$ denotes the drift function, corresponding to the right-hand side of~\eqref{galerkinmodel}. The dispersion function $\boldsymbol{G}$ determines how stochastic perturbations are introduced into the system. In this work, we use additive white noise and set $\boldsymbol{G}=\mathsfbi{I}_N\in \mathbb{R}^{N\times N}$, where $\mathsfbi{I}_N$ is the identity matrix. The stochastic forcing $\boldsymbol{f}(t)=(f_1(t),f_2(t),\cdots, f_N(t))^T$ is a Gaussian white noise vector with the properties 
\begin{equation}
    \mathbb{E}[\boldsymbol{f}(t)]=0,\quad \mathbb{E}[\boldsymbol{f}(t)\boldsymbol{f}(\tau)^T]=\mathsfbi{Q}\delta (t-\tau),
\end{equation}
where the spectral density (covariance) matrix is expressed as $\mathsfbi{Q}=q\mathsfbi{I}_N$ with the scalar intensity $q\geq0$.   

The stochastic system of ODEs is simulated using a stochastic fourth-order Runge-Kutta algorithm developed by~\cite{kasdin1995}. Given the absence of a closed-form covariance solution for~\eqref{StochasticForm}, we employ Monte Carlo simulations at various time steps to assess the robustness and numerical accuracy of the computations. A total non-dimensional time $t=2000$ is used for all the following numerical computations. The results are averaged over 4000 realizations with a time step of $5\times 10^{-4}$. Note that while the additive noise assumption is a reasonable approximation for our model, it is worth noting that the multiplicative noise could  lead to a more general representation~\citep{callaham2021,majda2001}.

As discussed above, we are interested in analysing the ability of ROMs to predict the emergence of secondary mean flow components. If we assume that the mean predicted by the model adheres to all of the symmetries described in \S\ref{sec:symmetries}, then only fully symmetric basis functions should have coefficients with nonzero average values. Note that the mode numbering defined in figure~\ref{TwelveEigenmodes} (i.e.~from least to most stable) is retained throughout all subsequent ROMs.  We now consider a hierarchy of models, 
where all models include the fully symmetric mode 12, denoted by $\boldsymbol{\phi}_s:=  \boldsymbol{\phi}_{12}$, with associated modal coefficient $a_s:=a_{12}$. For an $N$-mode model with $N\leq 12$, the remaining basis functions are selected as the $N-1$ least stable modes (e.g., for $N=3$ the ROM includes modes $1,2$, and $12$; for $N=4$ it includes $1,2,3$, and $12$; and so forth). For $N\ge 12$, the ROM basis comprises modes $1$ through $N$. Under these assumptions, and assuming zero-mean forcing, the mean of~\eqref{forcedsystem} with $N$ ODEs simplifies to
\begin{equation}
\overline{a_{i}}=\Rey\sum_{j=1}^N \sum_{k=1}^N\frac{ \langle \bsphi_j \cdot \nabla \bsphi_k ,\bsphi_{i} \rangle}{\langle \nabla^2 \bsphi_{i}, \bsphi_{i} \rangle }\;\overline{a_{j}a_{k}},\label{multimode}
\end{equation} 
The omitted terms in deriving~\eqref{multimode} are discussed in appendix~\ref{appA}.

\subsubsection{Minimal two-mode model} 
\label{sec:2modeModel}
 It is apparent from~\eqref{SingleProjection} and the surrounding discussion that at least two modes are required to form a Galerkin model with nonzero nonlinear terms. Here, the properties of such a minimal two-mode model are investigated. Utilising the properties discussed in appendices~\ref{appA} and~\ref{AppendixB}  alongside \eqref{SingleProjection}, a system of ODEs including mode $\bsphi_i$ and the symmetric mode ($\bsphi_s$) with general form given by~\eqref{galerkinmodel}   simplifies to 
\begin{subequations}
\begin{equation}
        \frac{\text{d}a_i}{\text{d}t}=-a_i a_s \langle \bsphi_i \cdot \nabla \bsphi_s,\bsphi_i \rangle +\frac{1}{\Rey} \langle \nabla^2 \bsphi_i,\bsphi_i \rangle a_i +f_i(t)   \label{antisymsingle}
    \end{equation}
\begin{equation}
        \frac{\text{d}a_s}{\text{d}t}=-a_i^2 \langle \bsphi_i \cdot \nabla \bsphi_i,\bsphi_s \rangle +\frac{1}{\Rey} \langle \nabla^2 \bsphi_s,\bsphi_s \rangle a_s +f_s(t)
        \label{symmetricsingle}
    \end{equation}  \label{TwoModeModel}
\end{subequations}
where $\bsphi_i$ is any eigenmode that does not satisfy all of the rotation and reflection symmetry relations. The temporal average of~\eqref{symmetricsingle} reduces to
\begin{equation}\overline{a_{s}}=\frac{\Rey \langle \bsphi_i \cdot \nabla \bsphi_i ,\bsphi_{s} \rangle \overline{a_{i}^2}}{\langle \nabla^2 \bsphi_{s}, \bsphi_{s} \rangle }.
 \label{OneModeMean}
\end{equation}
\begin{figure}
  \centerline{\includegraphics[width=0.77\textwidth]{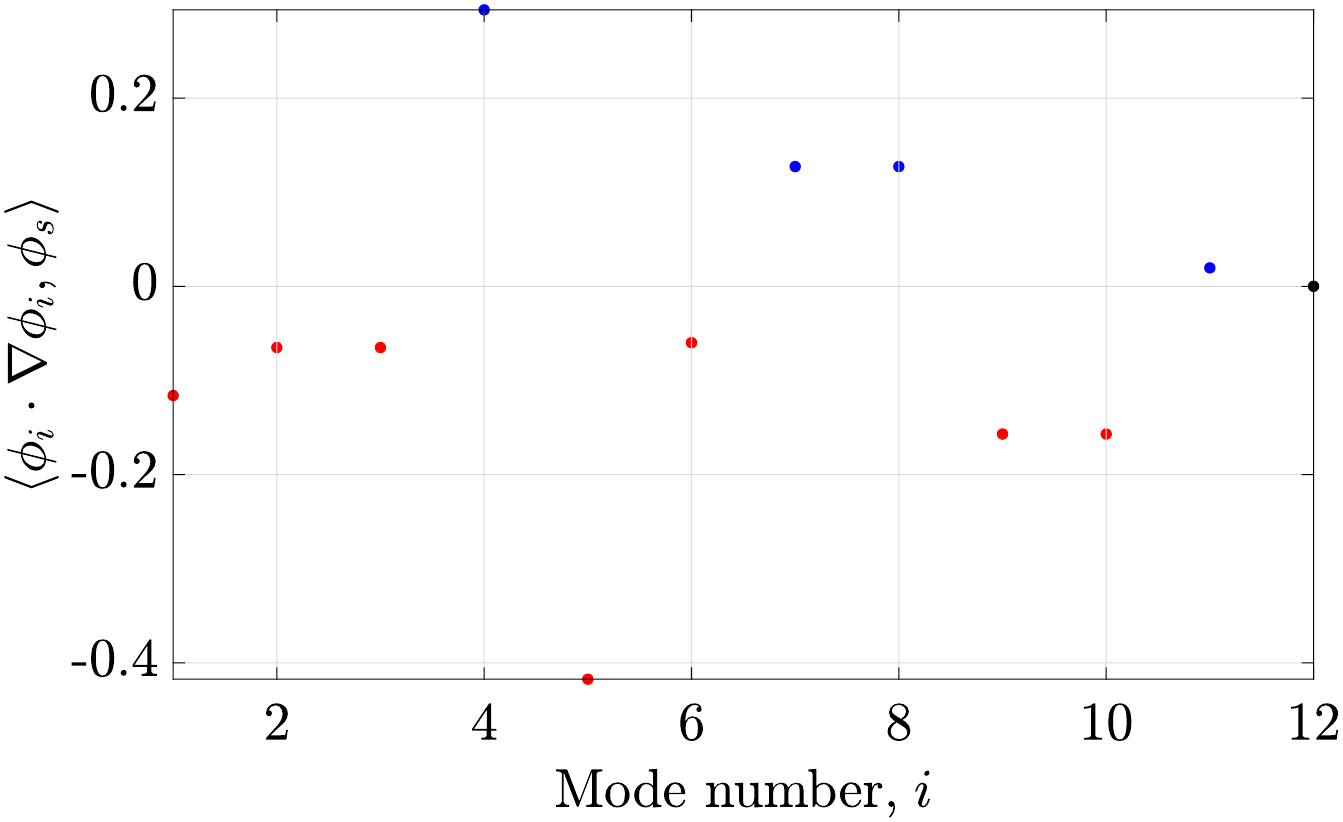}}
\caption{ 
The diagonal nonlinear terms \(\langle \boldsymbol{\phi}_i \cdot \nabla \boldsymbol{\phi}_i , \boldsymbol{\phi}_s \rangle \) in the Galerkin model equation for the symmetric mode coefficient $a_s$, 
across different mode numbers \( i = 1,2, \dots, 12 \). Positive and negative values of the advection term are indicated with blue and red colours, respectively. Because $\langle \nabla^2 \bsphi_s, \bsphi_s\rangle < 0$, negative values correspond to positive $a_s$ in the two-mode model prediction.}
\label{NonlinearProjection}
\end{figure}
\begin{figure}
    \centering
    \begin{subfigure}{0.9\textwidth} 
        \centering
        \includegraphics[width=0.78\textwidth]{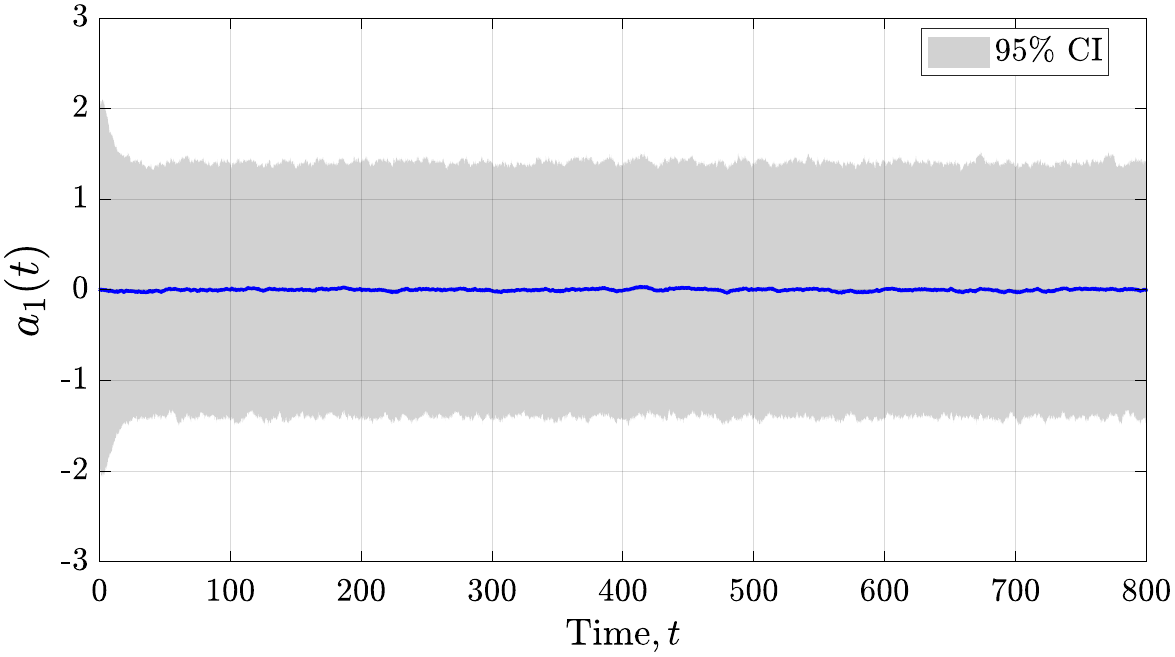}
         \caption{ }
         \label{TimeHistorya1}
    \end{subfigure}
        \begin{subfigure}{0.9\textwidth} 
        \centering
        \includegraphics[width=0.78\textwidth]{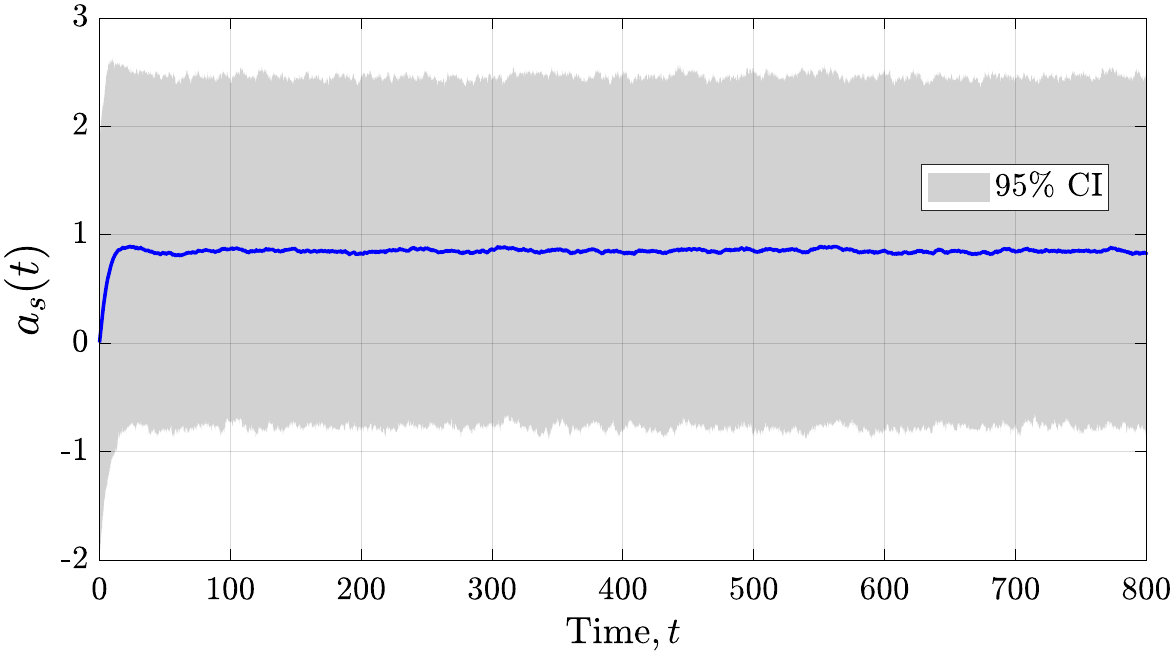}
         \caption{ }
         \label{TimeHistoryas}
    \end{subfigure}
    \caption{The time evolution of the two-mode model coefficients (a) $a_1(t)$ and (b) $a_s(t)$ across 4000 realisations. The solid blue line is the ensemble average, and the grey shaded area is the $95\%$ confidence interval. The coefficients are computed over $0 \le t \le 2000$; for clarity, only the interval $0 \le t \le 800$ is shown.}
    \label{TimeHistory}
\end{figure}
The denominator of~\eqref{OneModeMean} is always negative, since the Laplacian operator is negative definite. Therefore, the sign of $\overline{a_{s}}$ is opposite to the sign of the nonlinear projection term. From figure \ref{NonlinearProjection}, five out of the leading six modes (and seven out of the leading twelve) give a negative value of this projection, corresponding to a secondary flow in agreement with observations from the full nonlinear system (e.g.,~as shown in figure \ref{LamAndTurb}(b)). 

We present results from simulating this two-mode model with the least stable eigenmode, $i=1$. The corresponding system of ODEs is expressed as
\begin{subequations}
\begin{equation}
        \frac{\text{d}a_1}{\text{d}t}=-0.1160\;a_1 a_s -\frac{13.0913}{\Rey}  a_1 +f_1(t),   
    \end{equation}
\begin{equation}
        \frac{\text{d}a_s}{\text{d}t}=0.1160\;a_1^2  -\frac{67.3394}{\Rey}  a_s +f_s(t).
    \end{equation}  \label{TwomodeDetailed}
\end{subequations}

\begin{figure}
    \centerline{\includegraphics[width=0.8\textwidth]{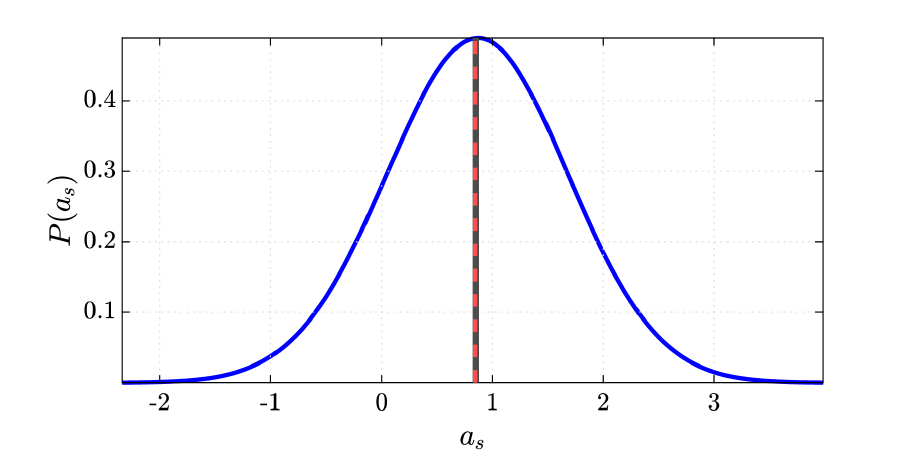}}
    \caption{Probability density function of the symmetric coefficient $a_s(t)$ in a two-mode model, evaluated over all realizations and time steps. The figure shows a Gaussian distribution around the coefficient mean value. The solid black vertical line indicates the mean of the coefficient over all realizations and time $\overline{a_s}$, whereas the dashed red line represents the value of $\overline{a_s}$ computed using~\eqref{OneModeMean}. }
    \label{probabilitydensity}
\end{figure}

\begin{figure}
    \centerline{\includegraphics[width=0.8\textwidth]{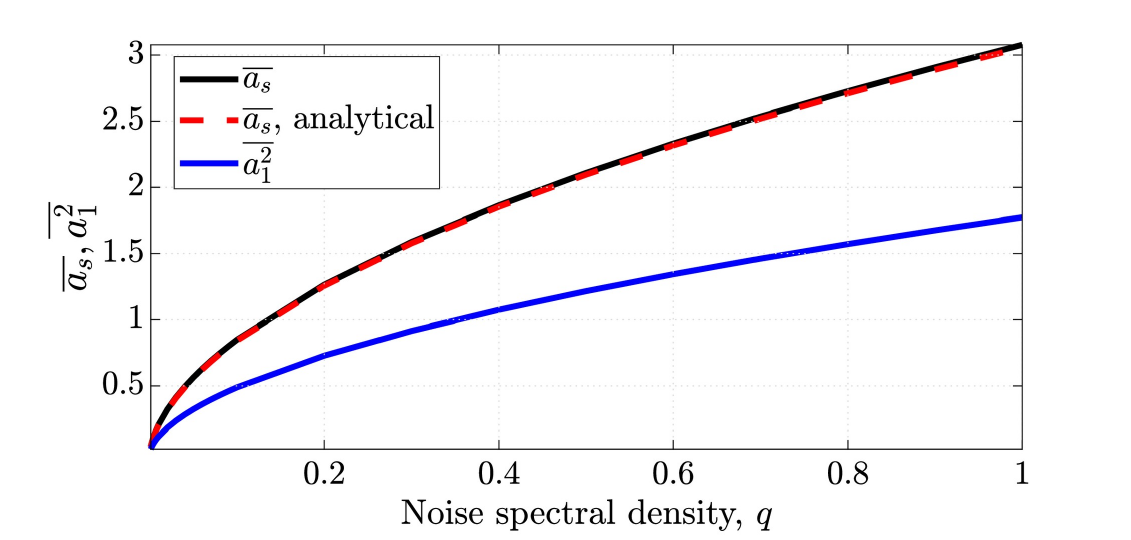}}
    \caption{Mean value of the coefficient of the symmetric mode, ($\overline{a_s}$, black) and variance of the non-symmetric mode ($\overline{a_1^2}$ blue) for the two-mode model as a function of noise spectral density, $q$. The red dashed line shows the predicted mean state from~\eqref{OneModeMean} using this variance.
    }
    \label{MeanVsNoise}
\end{figure}

We first consider a stochastically forced system with noise intensity $q = 0.1$ applied to both model equations, with $Re = 1000$. The ensemble-averaged time histories of the mode coefficients are presented in figure~\ref{TimeHistory}, along with $95\%$ confidence intervals computed over the ensembles. The coefficient $a_1$ fluctuates around a zero mean state, whereas the symmetric mode coefficient $a_s$ clearly exhibits a nonzero mean, as expected from our prior analysis. Figure~\ref{TimeHistoryas} shows that after a transient period of approximately 100 time units, the ensemble-averaged solution and its corresponding confidence interval converge to a statistically stationary behaviour. To further characterise this behaviour, we show the probability density function (PDF) of $a_s$ across all post-transient times and all realizations in figure~\ref{probabilitydensity}. The distribution of the PDF appears to be Gaussian, with the mean in close agreement with the value predicted from \eqref{OneModeMean}. To further verify numerical consistency of \eqref{OneModeMean},  figure~\ref{MeanVsNoise} shows the numerical and analytical values of the symmetric mode coefficient, $\overline{a_s}$, and the variance of the first mode coefficient, $\overline{a_1^2}$, across a range of noise spectral densities. The excellent agreement between the analytical and numerical values of $\overline{a_s}$ increases confidence in the numerical analysis, and the constant ratio between $\overline{a_s}$ and $\overline{a_1^2}$ for the noise spectral density range shown can be inferred from~\eqref{OneModeMean} to be the ratio $\Rey \;\langle \bsphi_1 \cdot \nabla \bsphi_1 ,\bsphi_{s} \rangle /\langle \nabla^2 \bsphi_{s}, \bsphi_{s} \rangle$.

\begin{figure}
\centerline{\includegraphics[width=0.6\textwidth]{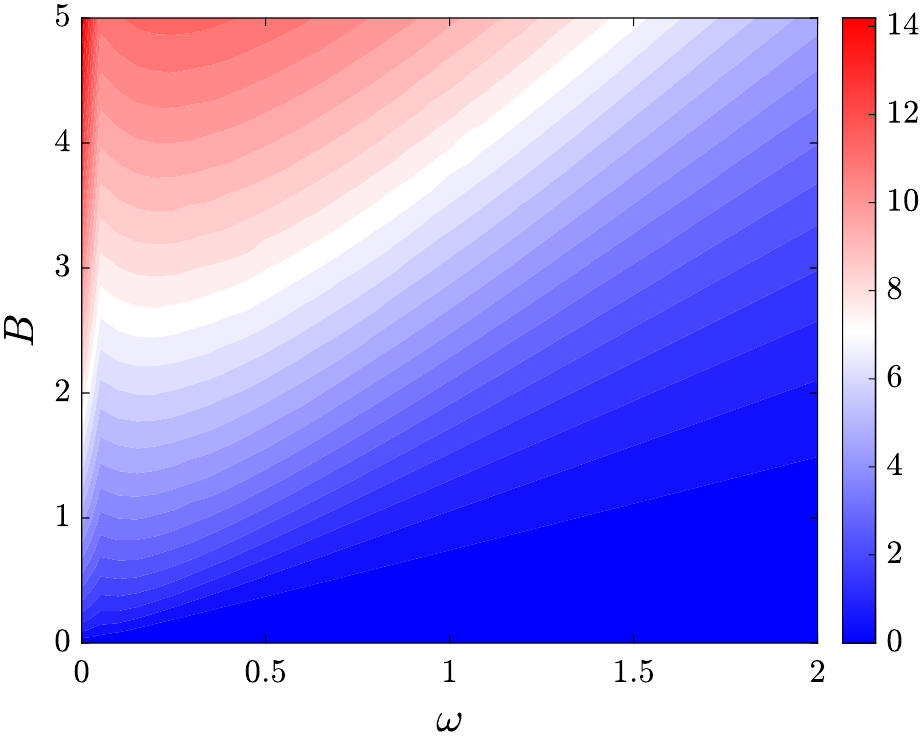}}
  \caption{Contour plot of the symmetric mode coefficient mean $\overline{a_s}$ for a two-mode model as a function of forcing magnitude $B$ and frequency $\omega$. The system is driven by a sinusoidal forcing applied to the first mode, given by $f_1=B \cos{(\omega t)}$. }
\label{DetForcingcontours}
\end{figure}

To demonstrate that these qualitative findings hold for inputs other than white noise, we now simulate~\eqref{TwoModeModel} with a deterministic sinusoidal forcing of the form $f_1=B\cos(\omega t)$, while keeping $f_s=0$. Figure~\ref{DetForcingcontours} shows a contour plot of the mean symmetric mode coefficient $\overline{a_s}$ as a function of the forcing amplitude ($B$) and frequency ($\omega$). We  observe that $\overline{a_s}$ again has a positive value across all forcing parameters. The coefficient magnitude increases with $B$, and also generally decreases as $\omega$ increases. This behaviour arises from the decomposition of the system~\eqref{TwoModeModel} into slow and fast components. The fast oscillatory term varies inversely with forcing frequency, whereas the slow component (governing the long-term system response) scales proportionally to $B^2/\omega^2$~\citep{yao2013,oxlade2015,rallu2018}. For the two-mode model forced through $\bsphi_1$, $a_s$ is positive for both stochastic and deterministic forcing across all amplitudes and frequencies considered, showing that the predicted direction is robust for this particular modal forcing.

\begin{figure}
    \centerline{\includegraphics[width=0.6\textwidth]{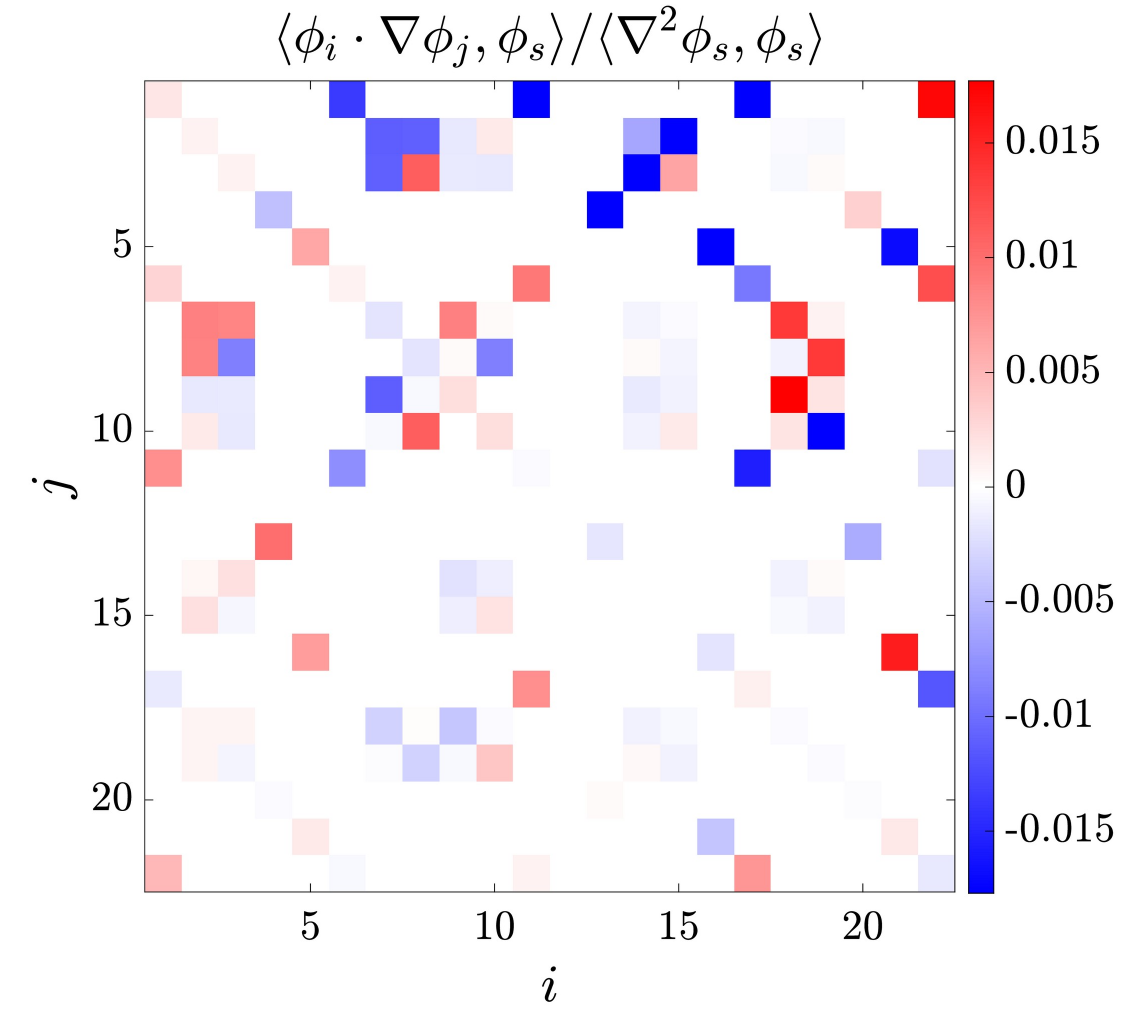}}
    \caption{ The projection of the normalised advection term  $\langle \bsphi_i \cdot \nabla \bsphi_j,\bsphi_s\rangle$  onto the fully symmetric eigenmode, for different mode $i$ and $j$ combinations.}
\label{fig:advectionMode22}
\end{figure}

\begin{figure}
\centering

\begin{subfigure}{0.47\textwidth}
    \centering
    \includegraphics[width=\textwidth]{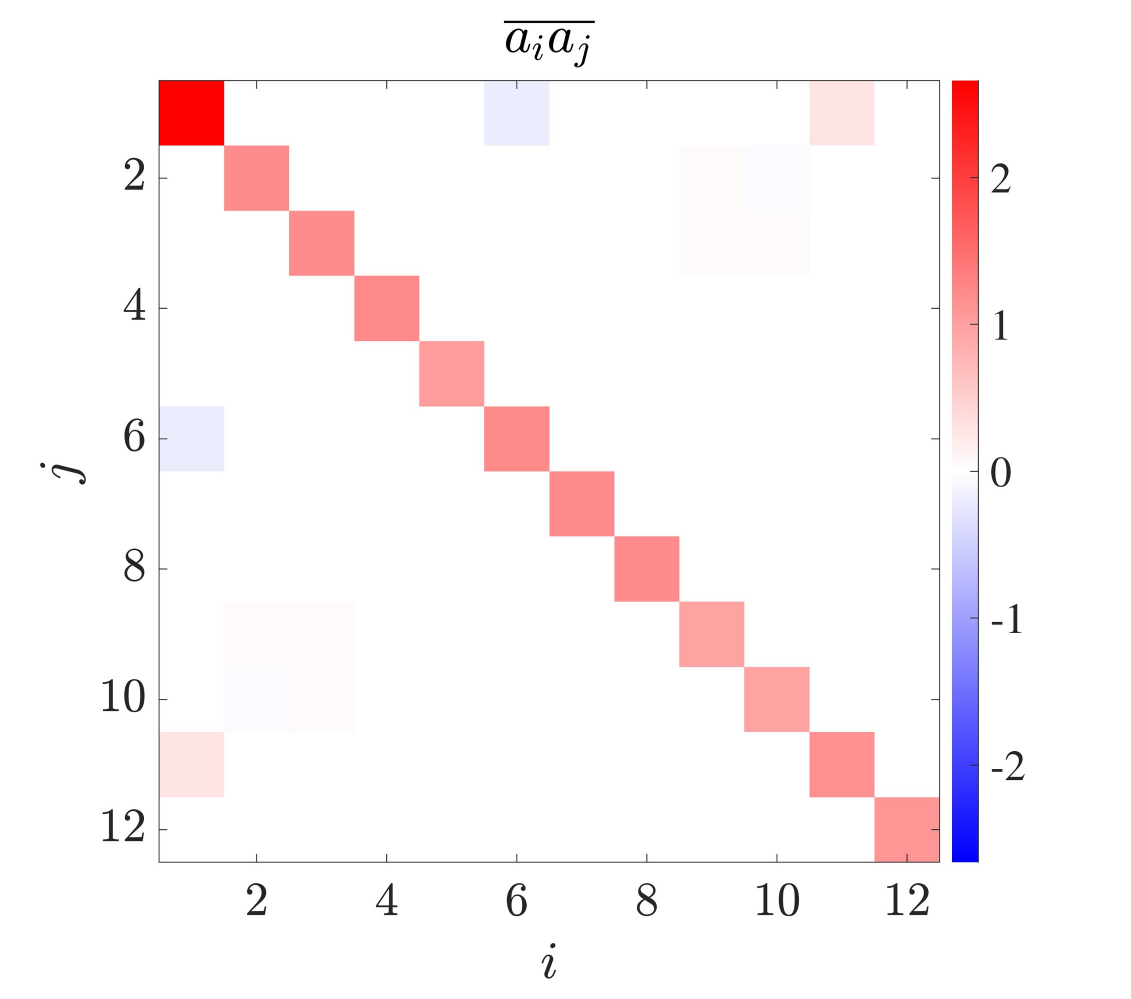}
    \caption{}
    \label{fig:Covariance12Stochastic}
\end{subfigure}
\begin{subfigure}{0.47\textwidth}
    \centering
    \includegraphics[width=\textwidth]{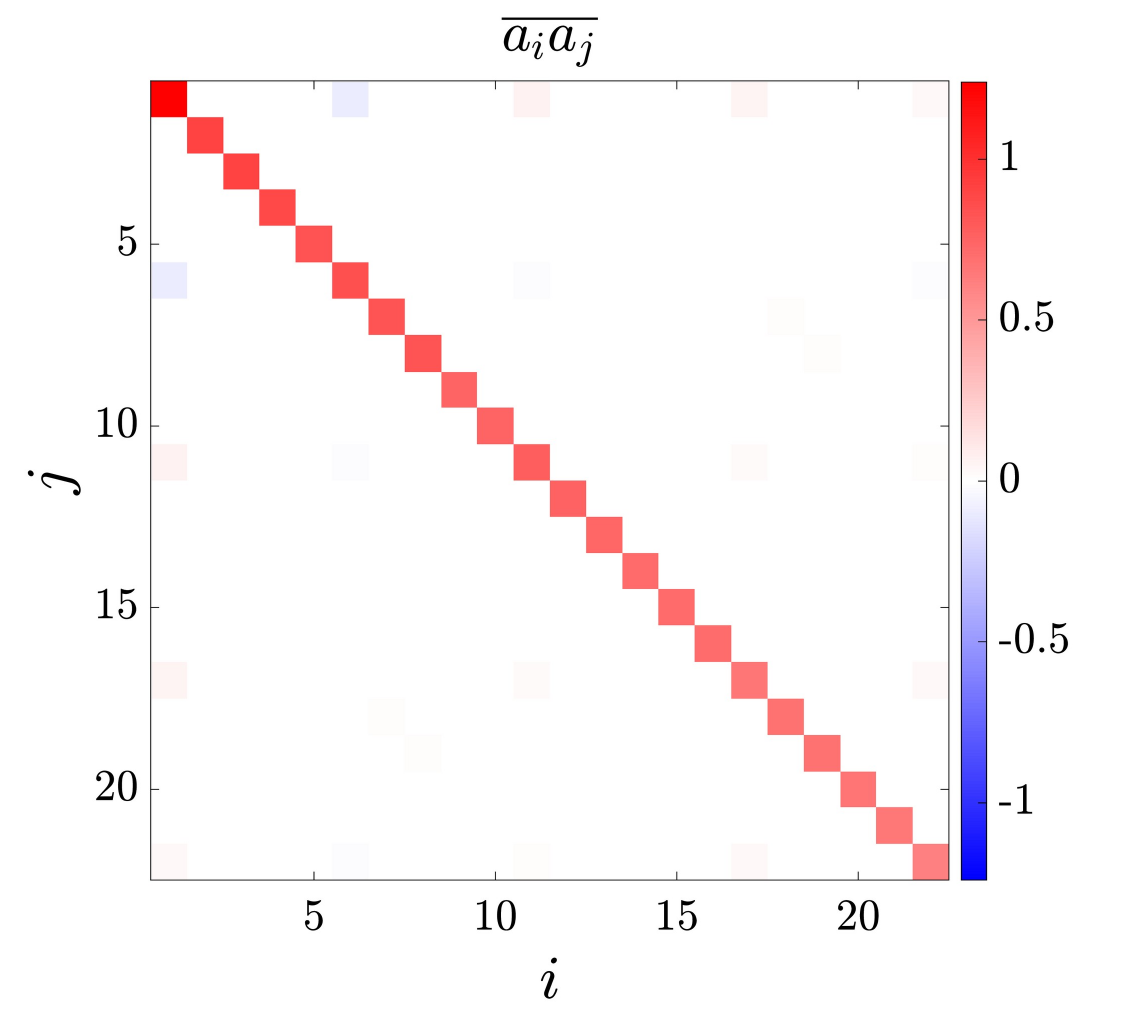}
    \caption{}
    \label{fig:Covariance22Stochastic}
\end{subfigure}
\begin{subfigure}{0.47\textwidth}
    \centering
    \includegraphics[width=\textwidth]{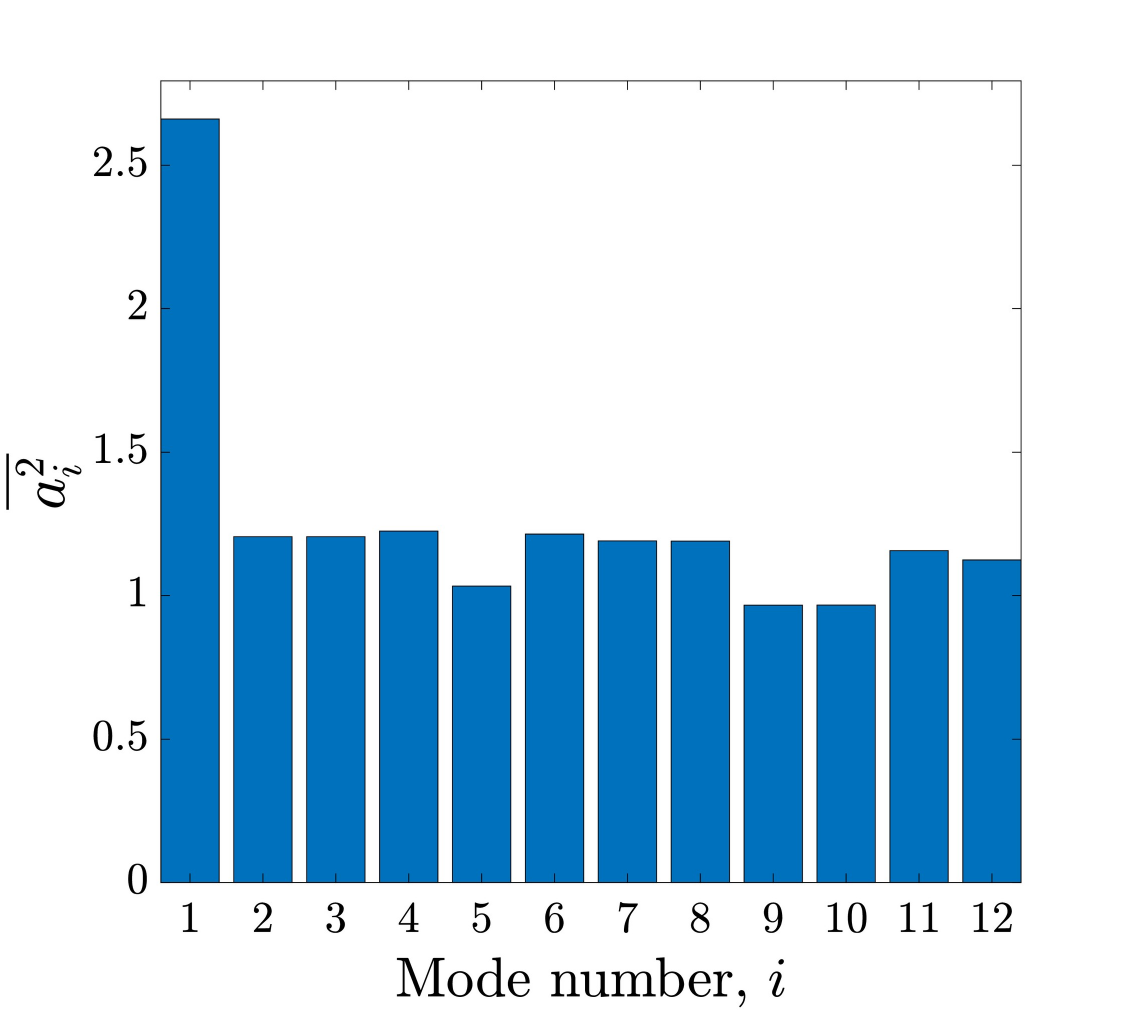}
    \caption{}
    \label{fig:Variance12Stochastic}
\end{subfigure}
\begin{subfigure}{0.47\textwidth}
    \centering
    \includegraphics[width=\textwidth]{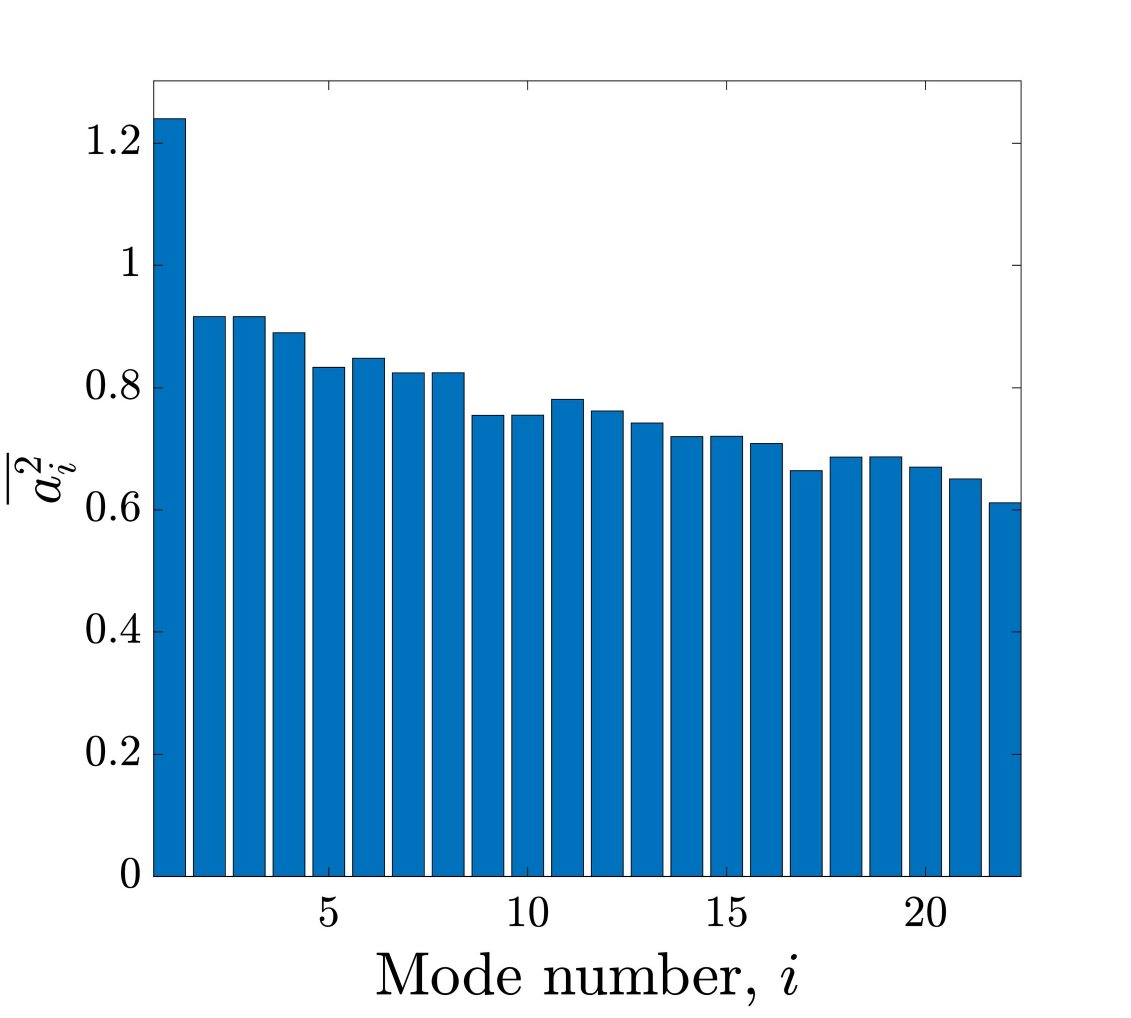}
    \caption{}
    \label{fig:Variance22Stochastic}
\end{subfigure}
\begin{subfigure}{0.47\textwidth}
    \centering
    \includegraphics[width=\textwidth]{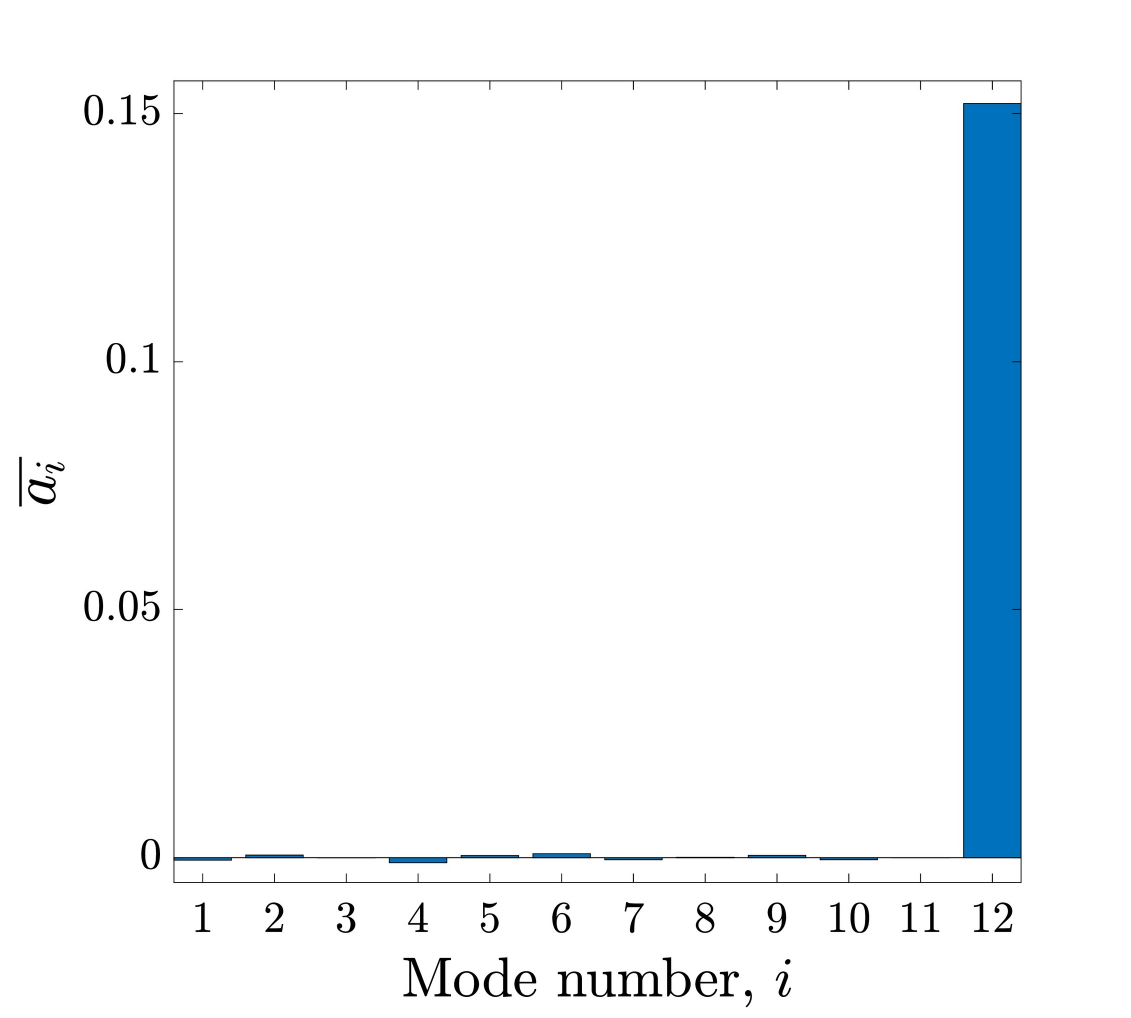}
    \caption{}
    \label{fig:Mean12Stochastic}
\end{subfigure}
\begin{subfigure}{0.47\textwidth}
    \centering
    \includegraphics[width=\textwidth]{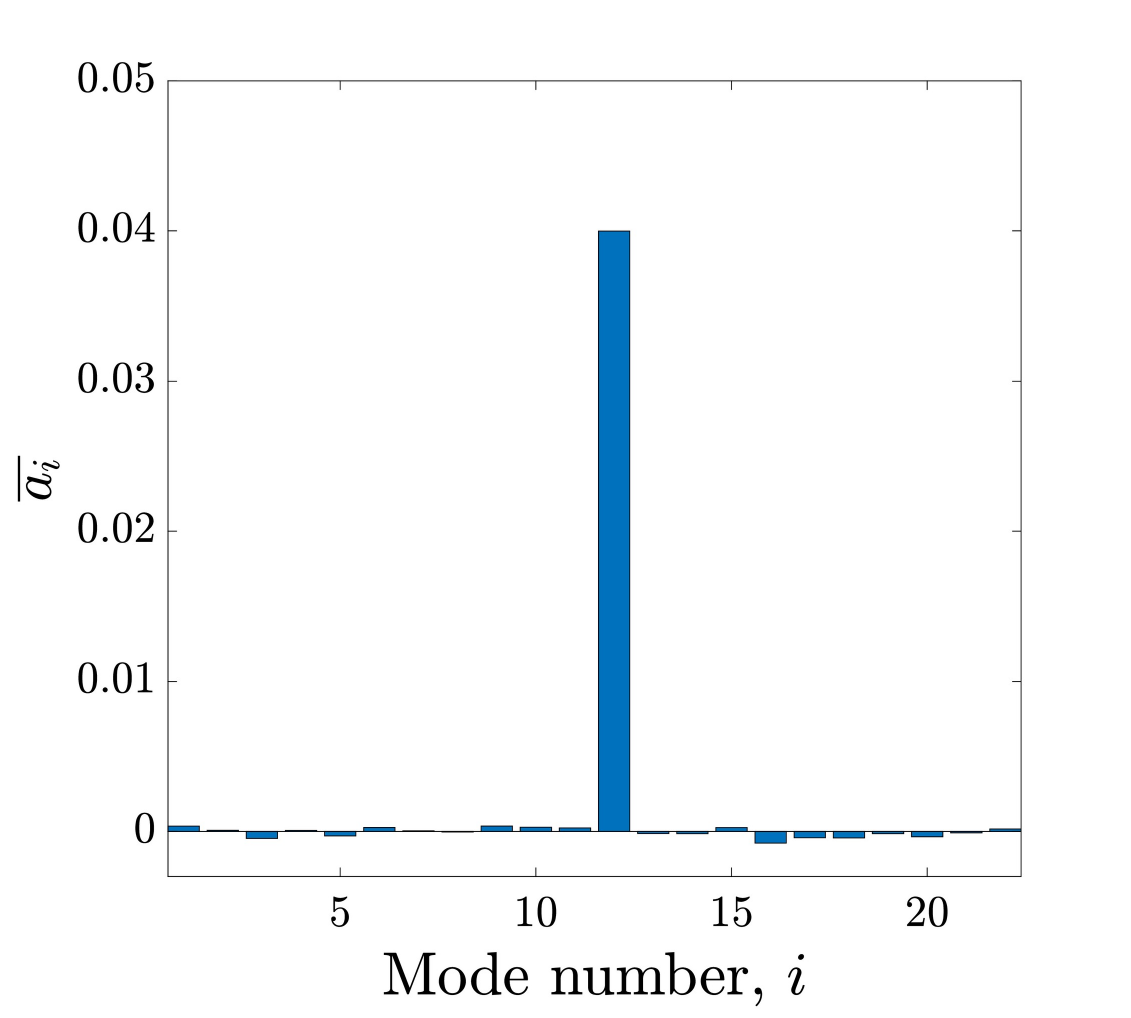}
    \caption{}
    \label{fig:Mean22Stochastic}
\end{subfigure}
\caption{
(a,c,e) Twelve-mode ROM: (a) second-moment colour map $\overline{a_i a_j}$, (c) diagonal second moments $\overline{a_i^2}$, and (e) temporal mean coefficients $\overline{a_i}$. (b,d,f) Corresponding quantities for the twenty-two-mode ROM. The models are stochastically forced with white-noise forcing of intensity $q=0.1$ applied equally to all the modes.
In both models, the second-moment matrices are dominated by their diagonal entries, while the temporal mean is dominated by the fully symmetric coefficient.
}
\label{twelve-TwentyTwomodeMeanCovariance}
\end{figure}

\subsubsection{Higher-order ROMs and modal interactions}\label{GeneralROM}

In \S\ref{sec:2modeModel}, we demonstrated how the self-interaction of a single non-symmetric eigenmode can generate a nonzero mean of the fully-symmetric mode. Here, we extend this analysis to higher-order ROMs, where we investigate the effects of additional modal interactions on the mean symmetric coefficient. A colour map of the normalised projection term $\langle \bsphi_i \cdot \nabla \bsphi_j ,\bsphi_{s} \rangle/\langle \nabla^2 \bsphi_{s}, \bsphi_{s} \rangle $ is shown in figure~\ref{fig:advectionMode22} for all mode combinations. The colour map shows nonzero projection coefficients between $\bsphi_s$ and the advection terms. The nonzero off-diagonal terms correspond to modal interactions between modes sharing the same symmetry family feeding into the fully symmetric eigenmode. For example, $\bsphi_1$ ,$\bsphi_6$ and $\bsphi_{11}$ are invariant under rotation, so the corresponding joint second moment $\overline{a_1a_6}$ and $\overline{a_1a_{11}}$ can contribute to the mean symmetric coefficient through~\eqref{multimode}.

To further understand the contributing terms in~\eqref{multimode}, stochastically forced ROMs are introduced. For these models, a stochastic white noise forcing with noise intensity $q=0.1$ is applied to all equations. The modal statistics of twelve- and twenty-two-mode ROMs are shown in figure~\ref{twelve-TwentyTwomodeMeanCovariance}. The mean-coefficient bar plot shows that the non-symmetric modal means are negligible compared with $\overline{a_s}$. The second-moment colour map shows that the autocorrelations of the coefficients dominate the system behaviour, and the coefficient $a_s$ is approximately uncorrelated with all the other model coefficients. In contrast, small but non-zero correlations are observed between $a_1$, $a_6$, and $a_{11}$. These non-vanishing correlations arise from the shared symmetries between these modes, as also suggested by the corresponding nonzero projection terms in figure~\ref{fig:advectionMode22}. Moreover, their relatively larger magnitudes compared with the other cross-correlations are attributed to the fact that mode 1 is the least damped eigenmode and has the largest diagonal second moment as shown in figures~\ref{fig:Variance12Stochastic} and \ref{fig:Variance22Stochastic}. From these figures, we notice that, although all modes are forced with the same noise intensity, the diagonal second moments decrease with increasing mode number, reflecting the increasing damping rate of the higher-order eigenmodes, which suppresses their fluctuation amplitudes relative to the least-damped modes and limits their contribution to the mean symmetric coefficient.

Equation~\eqref{multimode} and figure~\ref{twelve-TwentyTwomodeMeanCovariance} provide insight into the system dynamics and the contributors to the mean state of the symmetric mode. Although the non-symmetric modes do not directly contribute to the mean of the reduced-order model, their covariance plays a crucial role in determining the mean of the fully symmetric mode coefficient. In particular, figure \ref{twelve-TwentyTwomodeMeanCovariance} suggests that the diagonal terms where $i=j$ in~\eqref{multimode} are likely to be dominant contributors to $\overline{a_s}$.
\begin{figure}
    \centerline{\includegraphics[width=0.7\textwidth]{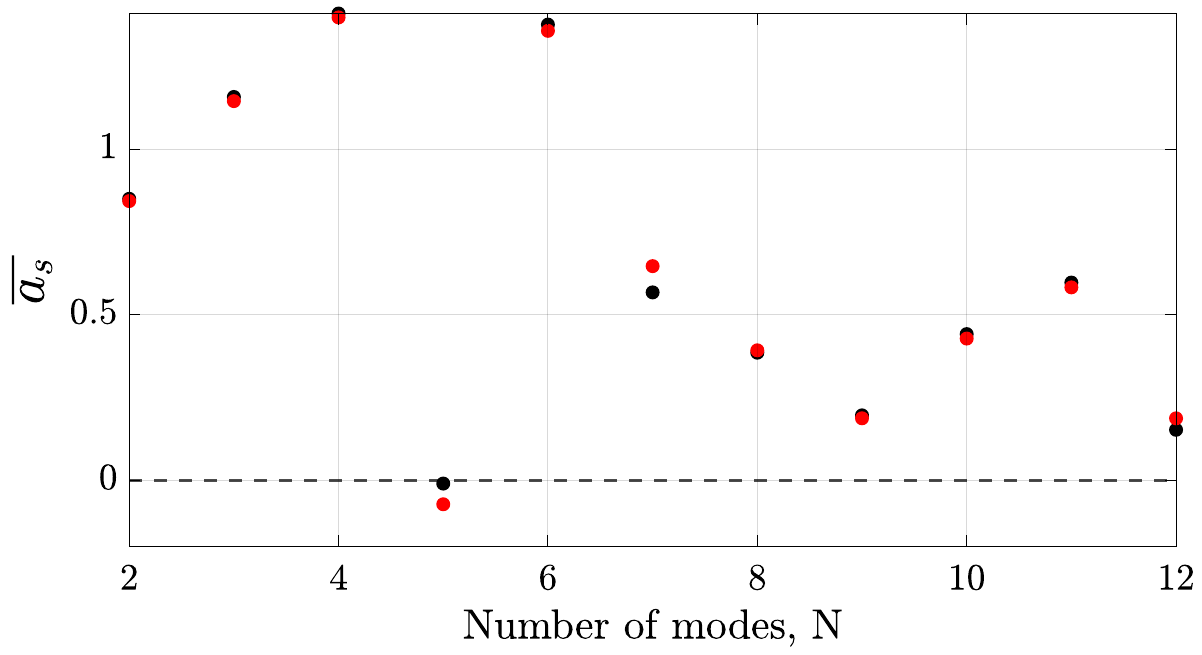}}
    \caption{Variation of the mean symmetric mode coefficient $\overline{a_s}$ with the number of model modes $N$. A white noise forcing with an intensity $q=0.1$ is applied to all modes. The numerical results using the stochastic RK4 algorithm (black dots) are compared to the values computed using~\eqref{multimode} (red dots).  }
    \label{fig:meanVsnumModes}
\end{figure}

 To verify the validity of~\eqref{multimode}, figure~\ref{fig:meanVsnumModes} shows a comparison between a numerical computation of the fully symmetric mode coefficient from simulating the ROMs and the values computed analytically using~\eqref{multimode}. A white noise forcing with an intensity $q=0.1$ is applied to all the model modes.  Close agreement is observed between these two quantities for different model orders. While the analytical computation of the coefficient mean relies on the second moments $\overline{a_ia_j}$ obtained from simulations, the agreement between both methods still demonstrates the validity of~\eqref{multimode}. We also observe from figure~\ref{fig:meanVsnumModes} that the symmetric mode mean is positive for all models except for the five-mode model, which exhibits a small negative mean. This suggests a clear preferred direction to the secondary mean flow predicted by the ROMs. From the shape of mode twelve in figure \ref{TwelveEigenmodes}, positive $\overline{a_s}$ corresponds to a mean secondary flow directed towards the wall along the corner bisectors, which qualitatively agrees with the secondary flow pattern observed in full DNS simulations (shown in figure \ref{TurbulentProfile}). 

It is only the five-mode model that yields a small negative $\overline{a_s}$, leading to a  reversed flow direction of the secondary mean. This behaviour can be explained by Figure~\ref{NonlinearProjection}, where $\bsphi_s$ is projected onto the single-mode advection term $\left ( \bsphi_i \cdot \nabla \bsphi_i \right)$ for each of the first twelve eigenmodes. Given the fact that the dissipation term $\langle \nabla^2 \bsphi_s,\bsphi_s \rangle$ is strictly negative and $\overline{a_i^2}>0$, the nonlinear projection provides intuition about the energy exchange mechanisms contributing to the direction of the secondary flow structure from the diagonal terms in~\eqref{multimode}. 
The modes associated with negative projection (red) contribute towards a secondary flow in the ``correct" direction that is observed in turbulent duct flows, while those with positive projections (blue) contribute towards a  secondary flow in the reverse direction to what is observed in the full nonlinear system.

For the five-mode model, the projection terms associated with modes $1-3$ are negative, while the term associated with mode four is positive and significantly larger. The effect of mode four thus dominates the contribution to $\overline{a_s}$ in the five-mode model, leading to its negative value. While several other modes also have positive projection values, they do not outweigh the contributions of the modes with negative values for all other model orders, due in part to the large negative value of the mode five projection. 
  \begin{figure}
 \centering
\begin{subfigure}{0.45\textwidth} 
    \centerline{\includegraphics[width=\textwidth]{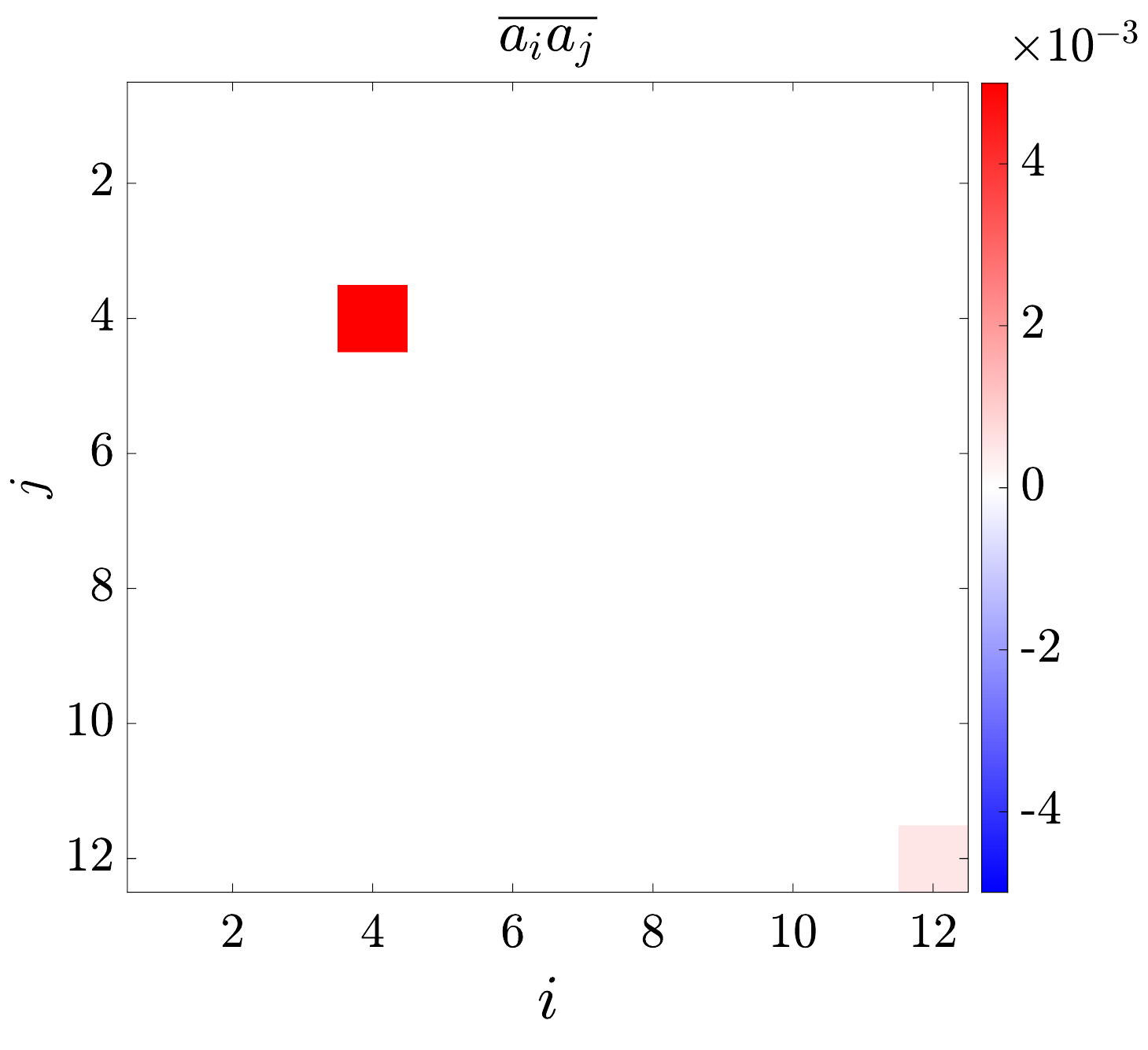}}
        \caption{\label{fig:CovarianceMode4q0_1}}
\end{subfigure}
\begin{subfigure}{0.45\textwidth} 
        \centerline{\includegraphics[width=\textwidth]{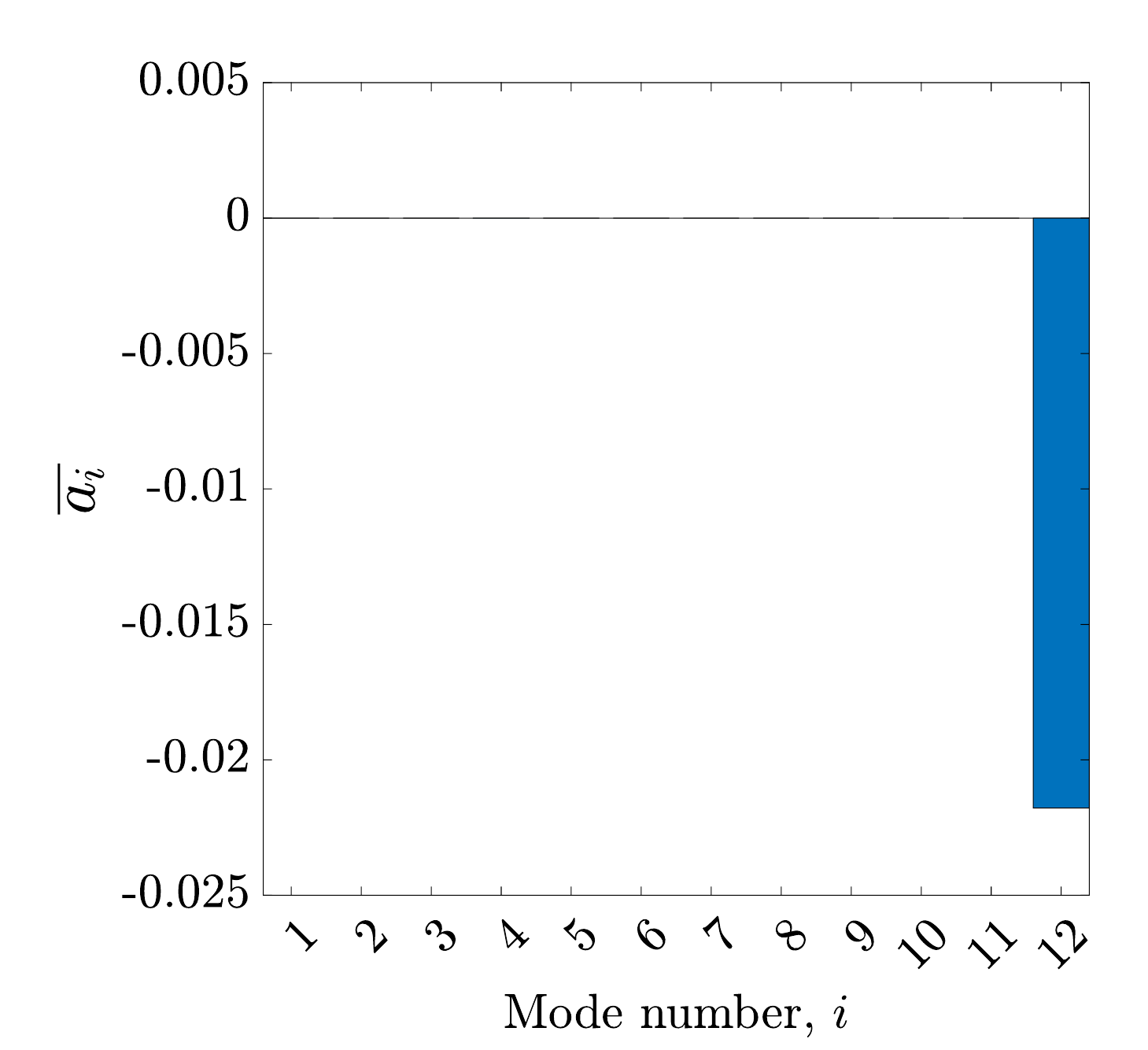}}
        \caption{ \label{fig:MeanMode4q0_1}}
    \end{subfigure}
    \centering
    \begin{subfigure}{0.45\textwidth} 
        \centerline{\includegraphics[width=\textwidth]{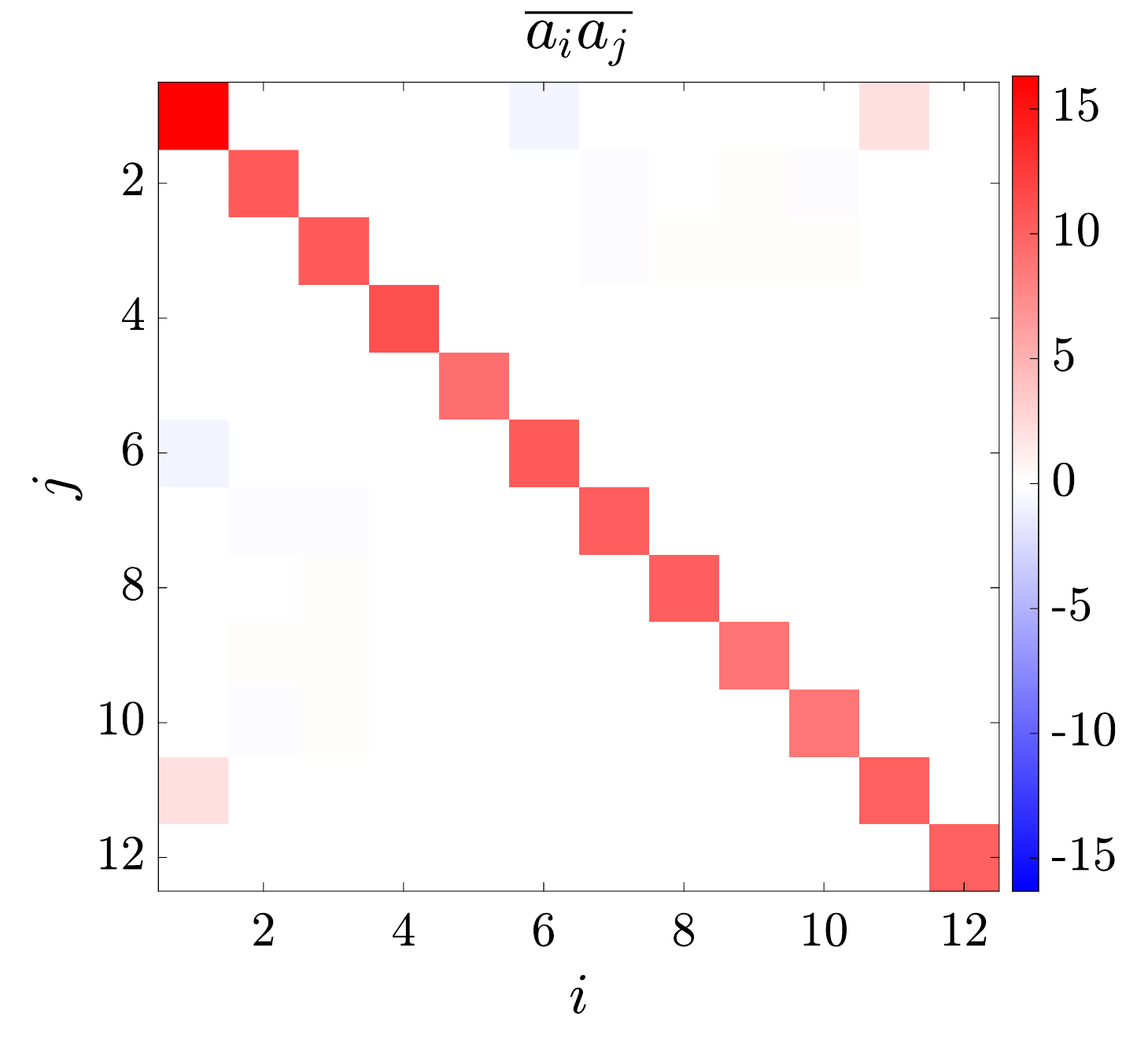}}
        \caption{ \label{fig:CovarianceMode4q2}}
    \end{subfigure}
\begin{subfigure}{0.45\textwidth} 
        \centerline{\includegraphics[width=\textwidth]{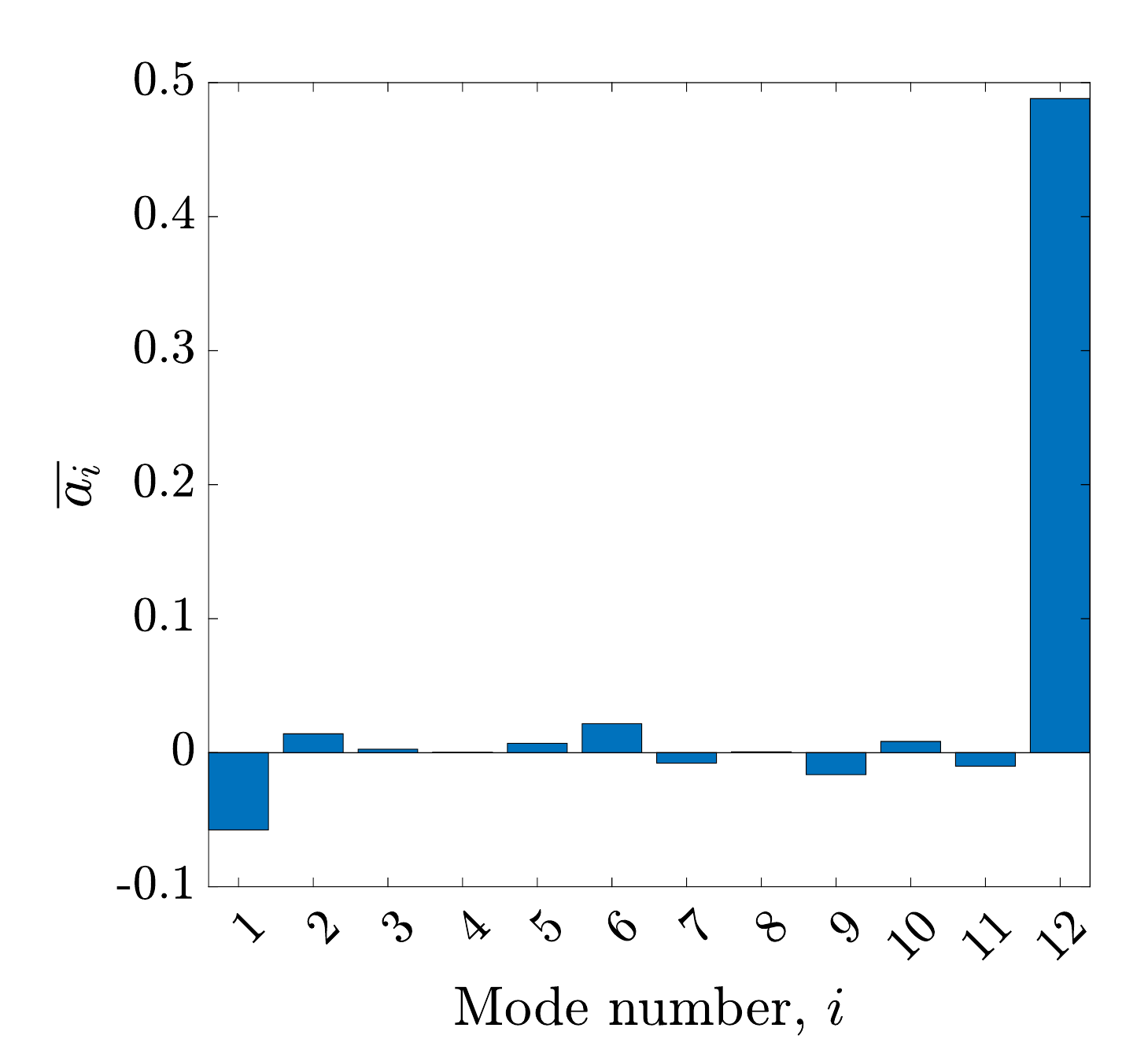}}
        \caption{ \label{fig:MeanMode4q2}}
    \end{subfigure}
  \caption{Modal statistics of the twelve-mode ROM with sinusoidal forcing applied only to $\bsphi_4$, with forcing $f_4(t)=B\cos(t)$. (a) Second-moment colour map $\overline{a_i a_j}$ and (b) temporal mean coefficients $\overline{a_i}$ for forcing amplitude $B=0.1$. (c), (d) show corresponding quantities for forcing amplitude $B=10$.
    }
    \label{twelvemodeDetforcingM4}
\end{figure}

These results motivate further examination of how the modal interactions depend on the forcing structure. We consider a twelve-mode ROM in which an external deterministic forcing is applied to $\bsphi_4$ only. Although only $\bsphi_4$ is directly forced, the nonlinear terms can transfer activity to other modal coefficients through quadratic interactions. Figure~\ref{twelvemodeDetforcingM4} shows the modal second moments and temporal mean coefficients for the twelve-mode ROM forced by $f_4(t)=B\cos(t)$. Figures~\ref{fig:CovarianceMode4q0_1} and~\ref{fig:MeanMode4q0_1} correspond to $B=0.1$, while figures~\ref{fig:CovarianceMode4q2} and~\ref{fig:MeanMode4q2} correspond to $B=10$.

Figure~\ref{twelvemodeDetforcingM4} demonstrates how the twelve-mode ROM response changes when only $\bsphi_4$ is externally forced. At the smaller forcing amplitude, the response is dominated by the forced-mode second moment, $\overline{a_4^2}$, while the remaining modal second moments are negligible. In this limit, the twelve-mode ROM behaves as the minimal two-mode model discussed in \S\ref{sec:2modeModel}, and the mean symmetric coefficient is controlled by the self-interaction of the forced mode. The sign of $\overline{a_s}$ is therefore determined by the nonlinear projection $\langle\bsphi_4\cdot\nabla\bsphi_4,\bsphi_{12}\rangle$ (see figure~\ref{NonlinearProjection}), leading to a negative mean symmetric coefficient and a reversed secondary mean flow.

As the forcing amplitude increases, additional modes are excited through nonlinear interactions, producing a distribution of diagonal second moments. In this regime, the two-mode approximation is no longer sufficient, and $\overline{a_s}$ must be interpreted using~\eqref{multimode}. These additional interactions modify the net nonlinear forcing of the fully symmetric mode and change $\overline{a_s}$ from negative to positive. For the forcing amplitudes considered here, the temporal mean remains dominated by the fully symmetric coefficient, while the non-symmetric modal means remain comparatively small.

\begin{figure}
    \centerline{\includegraphics[width=0.85\textwidth]{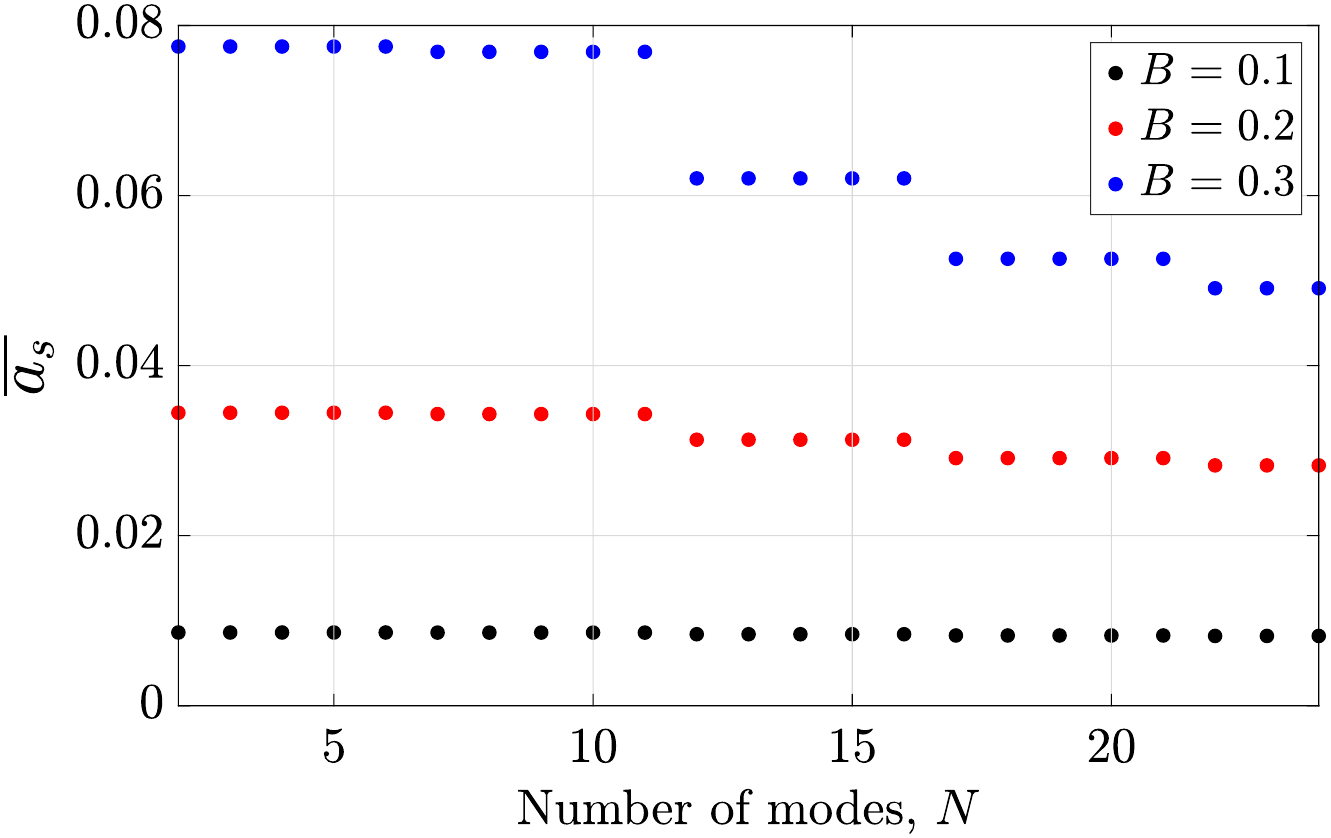}}
    \caption{Variation of the mean symmetric mode coefficient $\overline{a_s}$ with the number of model modes $N$. A sinusoidal forcing of the form $f_1=B\cos{t}$ is applied to mode one, $\bsphi_1$, with three forcing amplitudes $B=0.1$ (black dots),  $B=0.2$ (red dots) and $B=0.3$ (blue dots).  }
    \label{fig:meanVsnumModesDetForcing}
\end{figure}

These results show that the two-mode model captures the leading weak-forcing mechanism, whereas higher-order modal interactions become important at larger forcing amplitudes. For sufficiently large forcing, however, the dominance of the fully symmetric mean coefficient is not guaranteed a priori, and non-symmetric modal means or additional modes may become important.

Figure~\ref{fig:meanVsnumModesDetForcing} shows the variation of the symmetric mean coefficient $\overline{a_s}$ with the number of model modes for three sinusoidal forcing amplitudes, $B=0.1$, $B=0.2$ and $B=0.3$. In all considered cases, the forcing is applied only to the first mode, $\bsphi_1$. For the weaker forcing amplitude, $B=0.1$, the value of $\overline{a_s}$ is approximately independent of model order, suggesting that the two-mode mechanism captures the dominant contribution to the mean symmetric response. When the forcing amplitude is increased to $B=0.3$, the magnitude of $\overline{a_s}$ increases and the symmetric mean coefficient becomes dependent on the number of ROM modes. Thus, a larger number of modes is required before the symmetric mean coefficient appears to converge over the range of model orders considered. This indicates that, although the sign of $\overline{a_s}$ is robust, higher forcing amplitudes activate additional modal interactions that affect the quantitative value of the mean symmetric coefficient.

\section{Forced direct numerical simulations}\label{sec:dns}

In order to assess the validity and predictive capacity of the reduced-order modelling procedure described in \S\ref{results}, we now consider direct numerical simulations of incompressible flow through a square duct. Since there has been substantial prior work performing three-dimensional turbulent duct flow simulations as detailed in \S\ref{Introduction} and reporting on the structure of the secondary mean (e.g.,~as shown in figure \ref{LamAndTurb}(b)), we choose instead to run streamwise-constant simulations in which transverse spatial eigenmodes are imposed onto the laminar streamwise mean velocity and forced with zero-mean deterministic forcing in the same manner as the ROMs considered in \S\ref{results}. The stochastic forcing introduced in~\S\ref{results} is limited to the ROM analysis, where independent stochastic forcing can be prescribed directly in modal space. In the DNS calculations, we restrict the comparisons to deterministic harmonic forcing of selected modes. This provides a direct test of the nonlinear mechanism identified by the ROM, without introducing the additional statistical convergence requirements and computational expense associated with stochastic forcing in DNS.

These simulations are not intended to reproduce fully turbulent square duct flow or its streamwise velocity profile. Instead, they are designed to simulate the nonlinear mechanisms determined by the ROM to examine whether a zero-mean time periodic forcing of selected eigenmodes can generate a nonzero mean secondary flow resembling the leading fully symmetric eigenmode.

To run these simulations, we utilise the spectral-element code Semtex \citep{blackburn2019}. The two-dimensional $(y,z)$ domain is decomposed into quadrilateral spectral elements, with each element consisting of Gauss–Lobatto–Legendre shape functions in both spatial directions. Time integration uses a fractional step velocity-correction method \citep{guermond2003velocity}. This code has been utilised across a range of applications and extensions \citep[e.g.]{blackburn2004formulation,blackburn2025semtex}; refer to~\cite{blackburn2019} for further details about the numerical method.

The simulations are performed in the square duct geometry as shown in figure \ref{fig:schematic}. The flow is assumed to be constant in the streamwise direction but is solved for three components of the velocity field (i.e.~$2D/3C$). The computational domain consists of $31 \times 31 $ spectral elements, using 7th-order polynomials within each element in both spatial directions. The element boundaries are chosen to conform to Chebyshev collocation points to give near-wall refinement. To ensure computational convergence, it was verified that doubling the number of spectral elements in each direction did not have a significant effect on the results. 

The flow is forced by applying a time-varying body force of the form
\begin{equation}
\label{eq:DNSf}
    {\bsf} = B\sin(\omega t)\bsphi_j,
\end{equation}
where $\bsphi_j$ is one of the eigenmodes shown in figure \ref{TwelveEigenmodes}. To maintain a fixed centreline Reynolds number when applying this forcing, an iterative procedure is employed, where the pressure gradient is adjusted to maintain a fixed mean centreline velocity. We use a nominal Reynolds number based on this centreline velocity of \( \Rey = 1000 \). Note that the laminar flow is stable at this Reynolds number, so if the forcing is turned off, the flow reverts to the steady laminar velocity profile. This Reynolds number is intentionally selected to isolate the effect of
the applied zero-mean forcing and to test whether the mechanisms predicted
by the ROM can be reproduced by the full streamwise-constant
Navier--Stokes equations. Thus, the secondary mean-flow structure can be
attributed to the applied forcing rather than to transition to turbulence. To ensure statistical convergence, an initial transient period of \( T_i = 200 \) convective time units after the forcing is turned on is discarded, followed by an averaging period of \( T_A = 2000 \) convective units.

\begin{figure}
   \centering
    \begin{subfigure}{0.49\textwidth} 
        \centering
        \includegraphics[width=\textwidth]{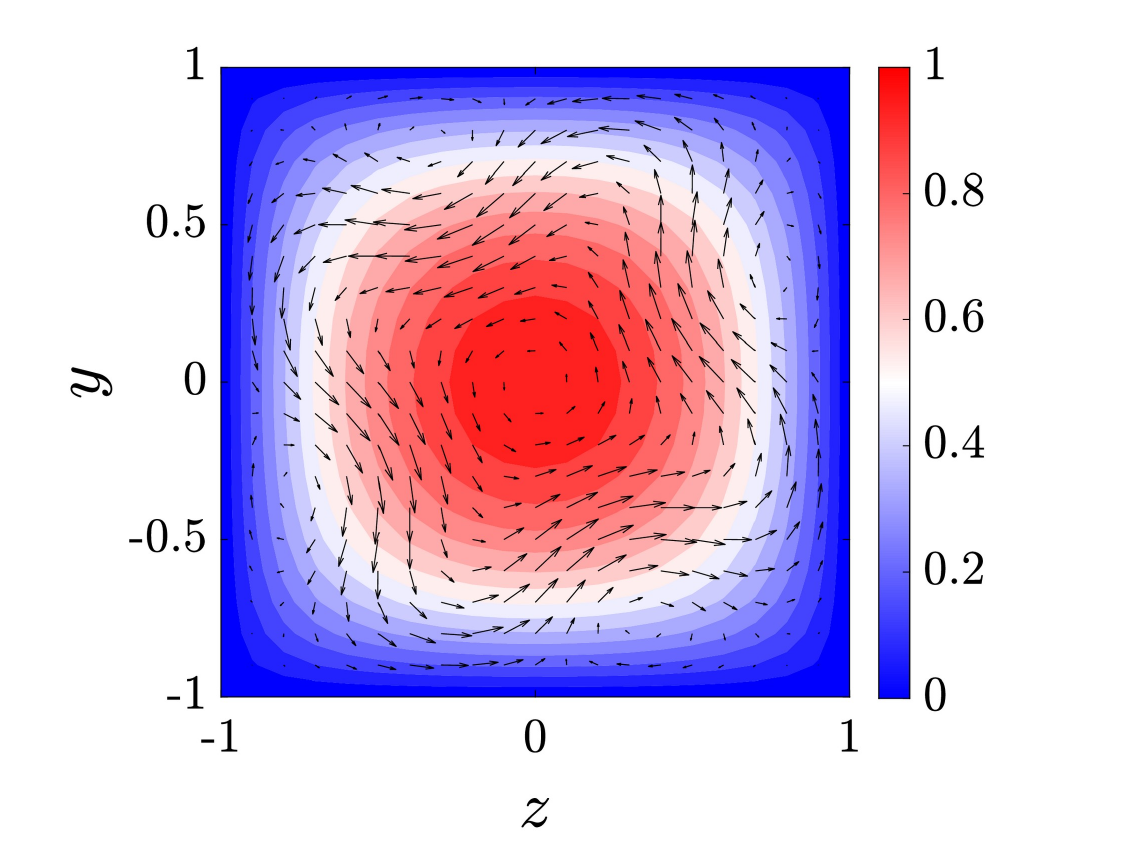}
        \caption{        \label{MeanDNST350}}
    \end{subfigure}
        \begin{subfigure}{0.49\textwidth} 
        \centering
        \includegraphics[width=\textwidth]{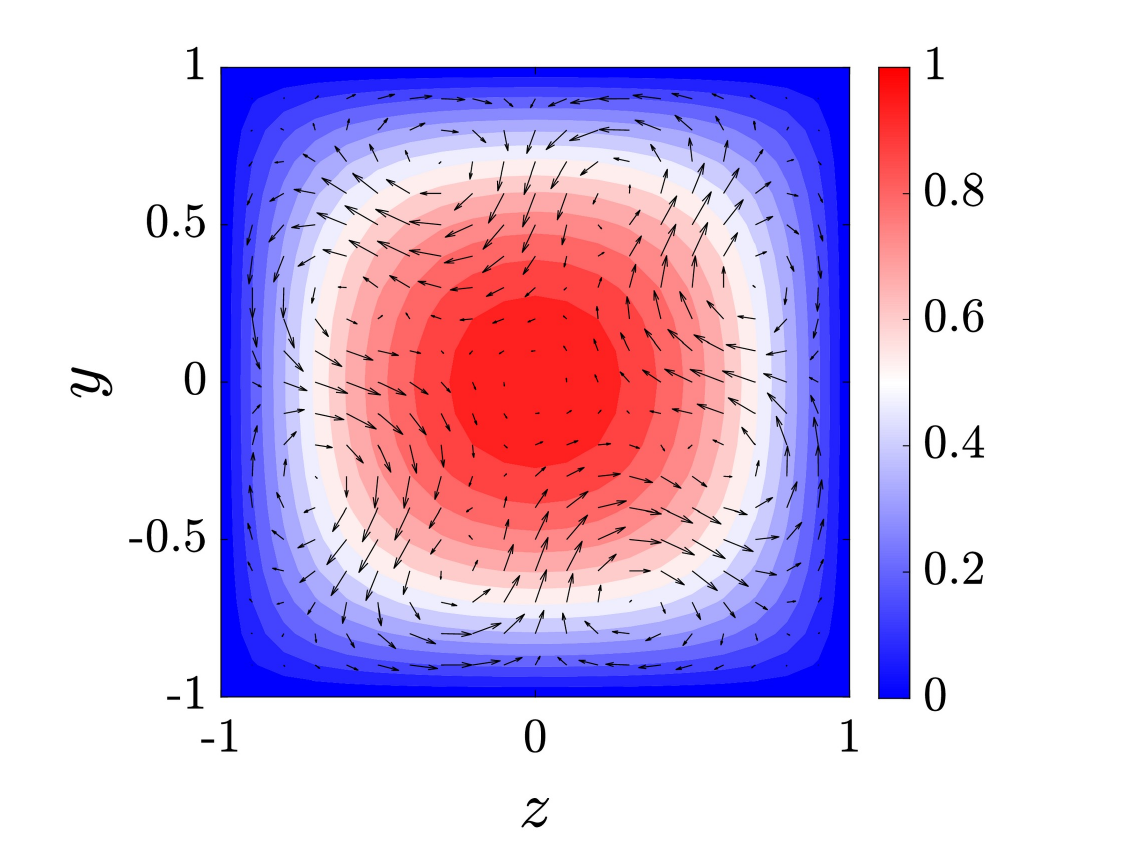}
        \caption{        \label{MeanDNST500}}
    \end{subfigure}
        \begin{subfigure}{0.49\textwidth} 
        \centering
        \includegraphics[width=\textwidth]{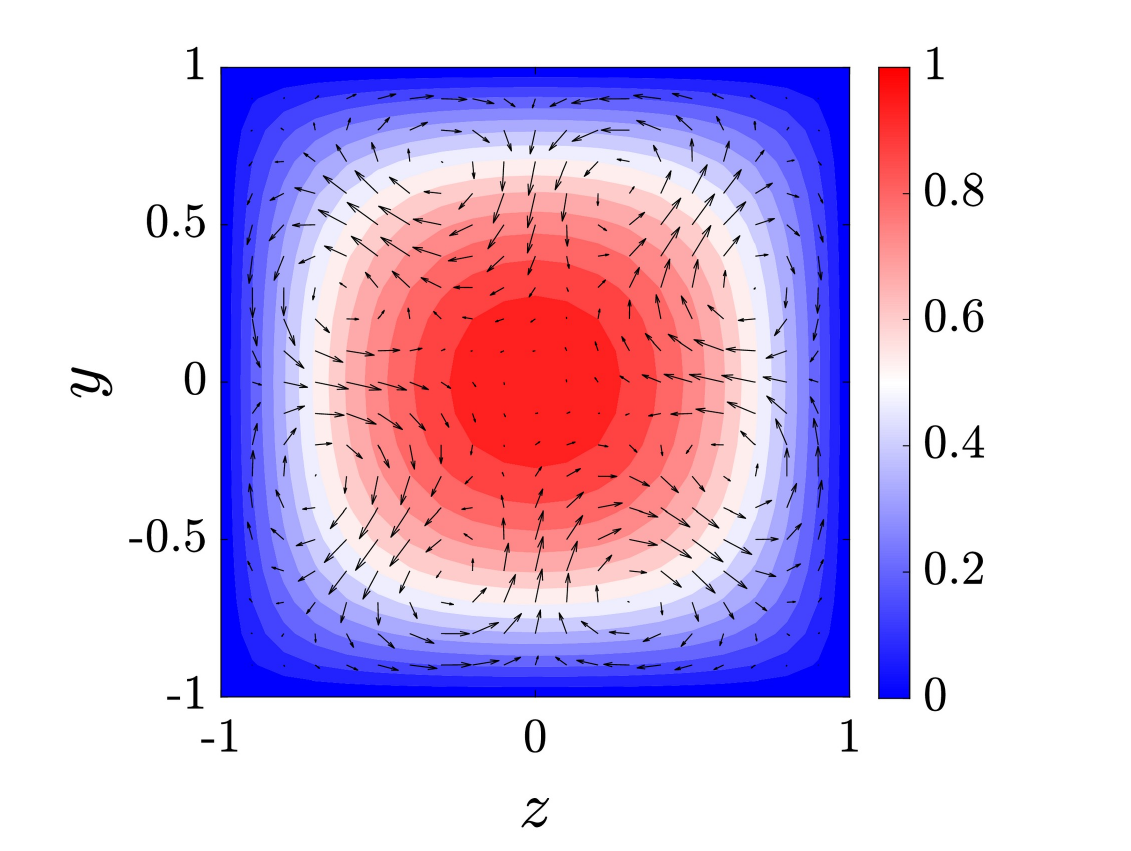}
        \caption{\label{MeanDNST1000}}
    \end{subfigure}
\caption{Contours of mean streamwise velocity with superimposed vector fields of mean cross-stream velocities from 2D/3C DNS simulations averaged over times $T_A=350$ (a), $T_A=500\;(b)$, and $T_A=1000\; (c)$.}
    \label{DifferentTimeaverages}
\end{figure}

Results showing the necessity of this lengthy averaging period are illustrated in figure~\ref{DifferentTimeaverages}, which shows the mean streamwise and cross-stream velocities for three different averaging periods, $T_A=350,500$ and $1000$. Here, forcing is applied as specified in \eqref{eq:DNSf}, using $B=0.01$, $\omega = 1$, in the direction of the least-stable eigenmode ($j=1$). For the shorter averaging time shown in figure~\ref{MeanDNST350}, a central vortex somewhat reminiscent of the forcing mode (mode 1 in figure~\ref{TwelveEigenmodes}) is discernible, along with secondary counterrotating vortices along each sidewall. For the larger averaging times (figures~\ref{MeanDNST500}~and~\ref{MeanDNST1000}), the central vortex has been distorted into four separate vortices, which align with the counterrotating vortices to give an approximately symmetric secondary flow structure that closely resembles the leading fully symmetric eigenmode (mode 12 in figure \ref{TwelveEigenmodes}). These results suggest that the two-mode model formulated in \S\ref{sec:2modeModel} captures the same essential physics observed in these simulations, with a zero-mean forcing resulting in a mean secondary flow field with counter-rotating vortex pairs in each corner.

\begin{figure}
   \centering
    \begin{subfigure}{0.49\textwidth} 
        \includegraphics[width=\textwidth]{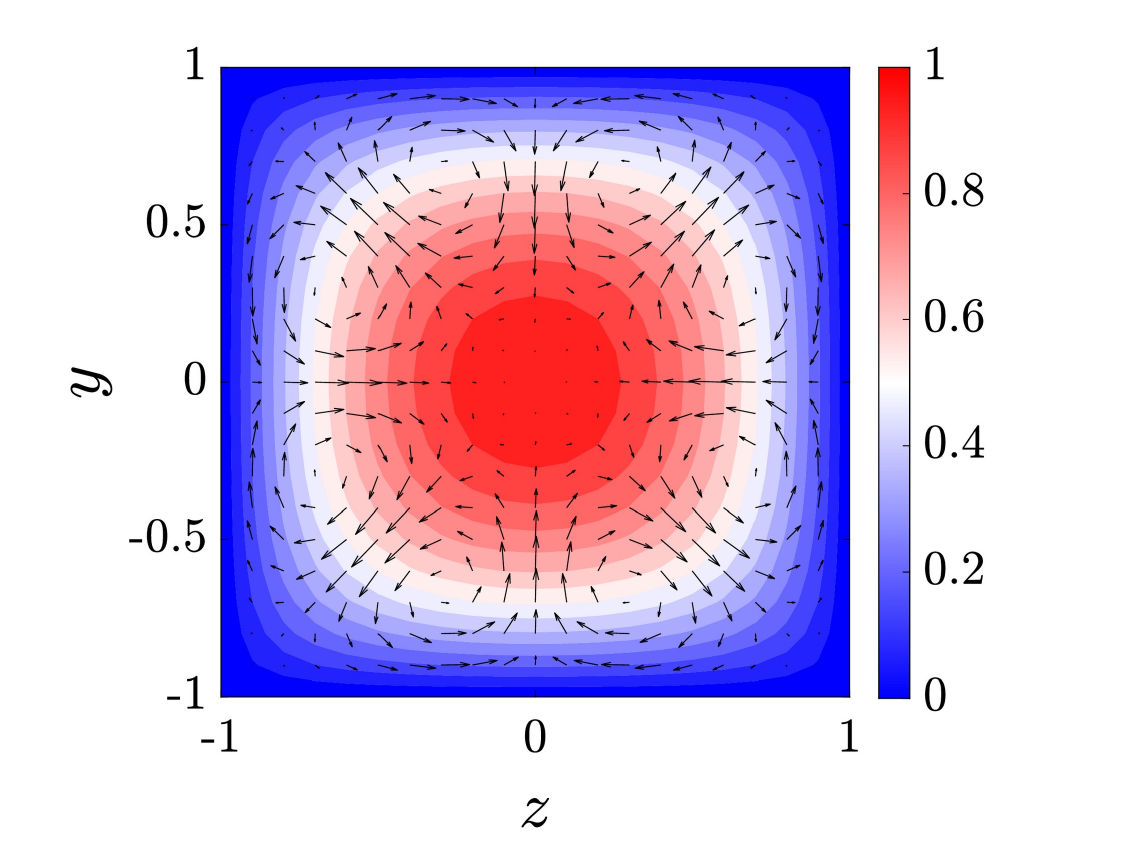}
        \caption{        \label{StreamandTransversecontoursDNSsub1}}
    \end{subfigure}
        \begin{subfigure}{0.49\textwidth} 
        \includegraphics[width=\textwidth]{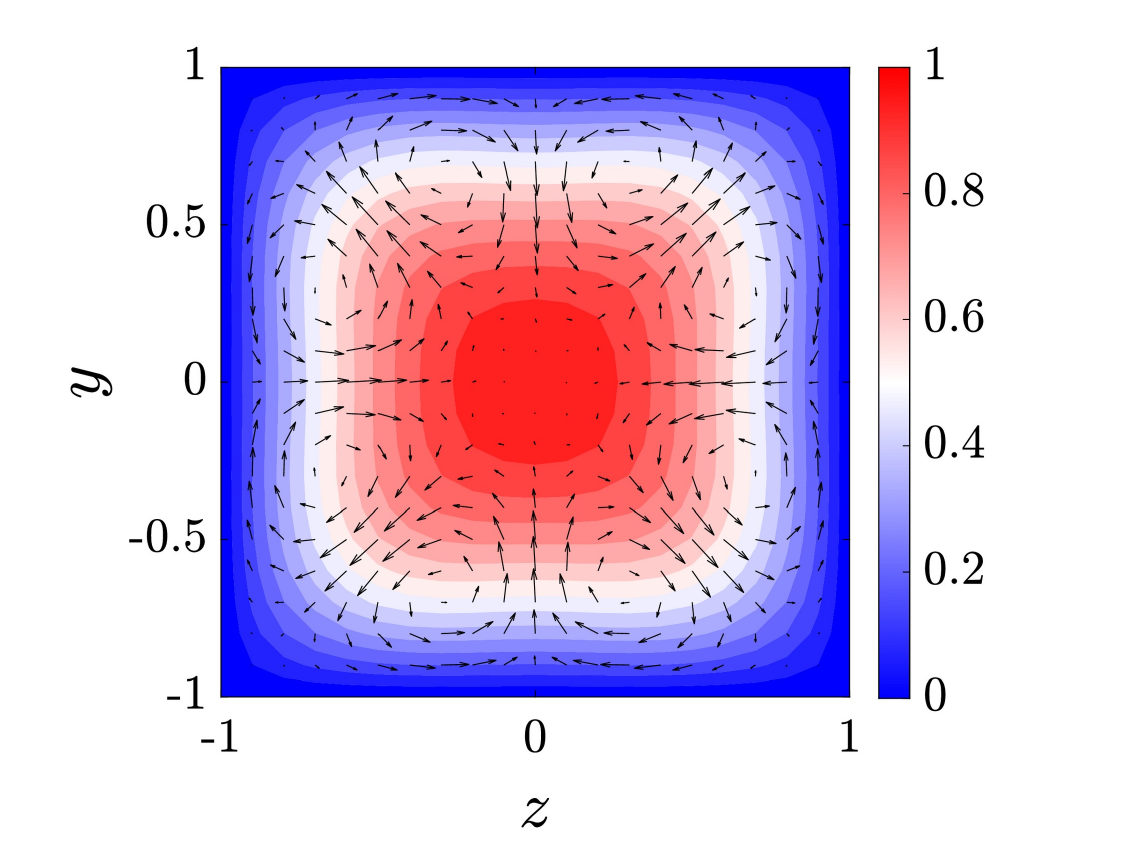}
        \caption{        \label{StreamandTransversecontoursDNSsub2}}
    \end{subfigure}
        \begin{subfigure}{0.49\textwidth} 
        \includegraphics[width=\textwidth]{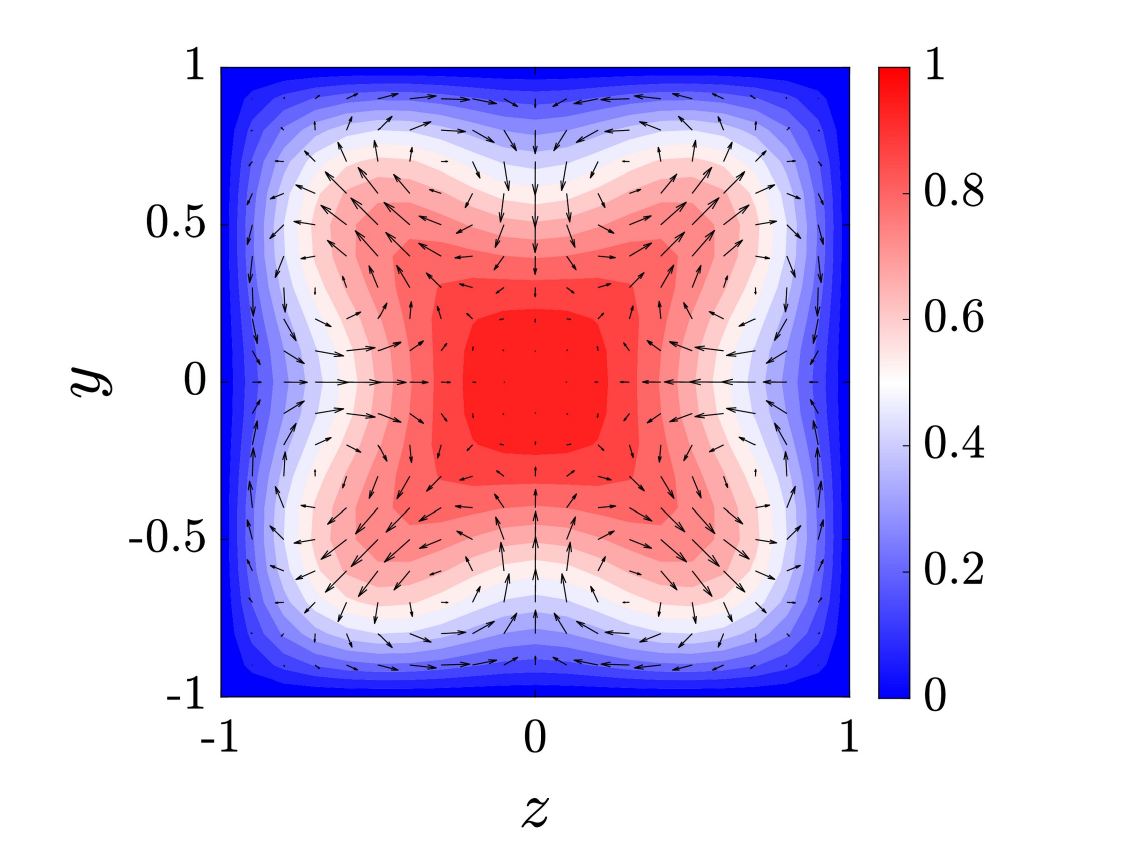}
        \caption{\label{StreamandTransversecontoursDNSsub3}}
    \end{subfigure}
\caption{Contours of the mean streamwise velocity overlaid with vector fields of the mean cross-stream velocities from forced 2D/3C DNS simulations. The results are obtained using sinusoidal forcing of the least stable eigenmode, with forcing magnitudes (a) $B=0.01$, (b) $B=0.05$, and (c) $B=0.1$.}
    \label{StreamandTransversecontoursDNS}
\end{figure}

\begin{figure}
    \begin{subfigure}{0.322\textwidth} 
        \centering
        \includegraphics[width=1.15\textwidth]{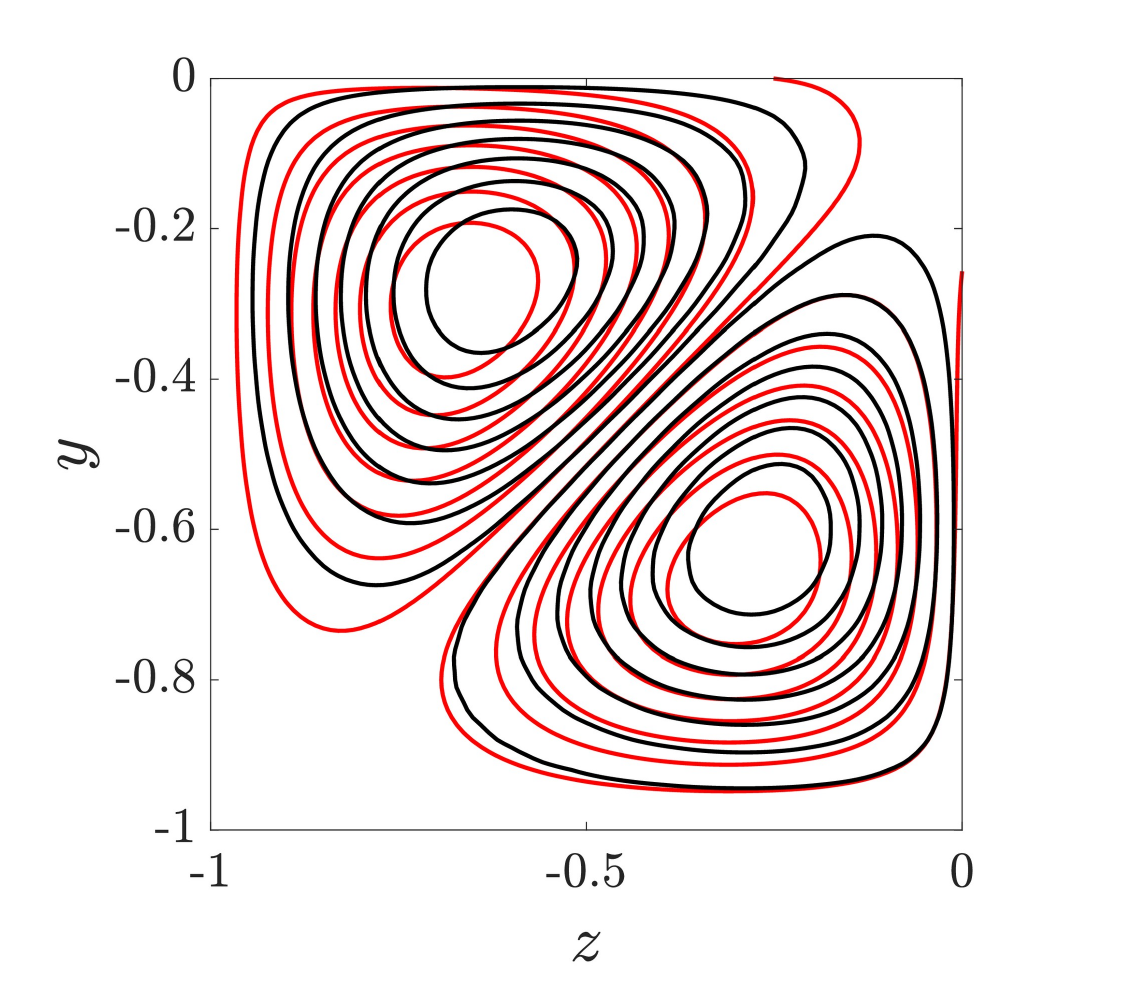}
        \caption{        \label{StreamLinessub1}}
    \end{subfigure}
        \begin{subfigure}{0.322\textwidth} 
        \centering
        \includegraphics[width=1.15\textwidth]{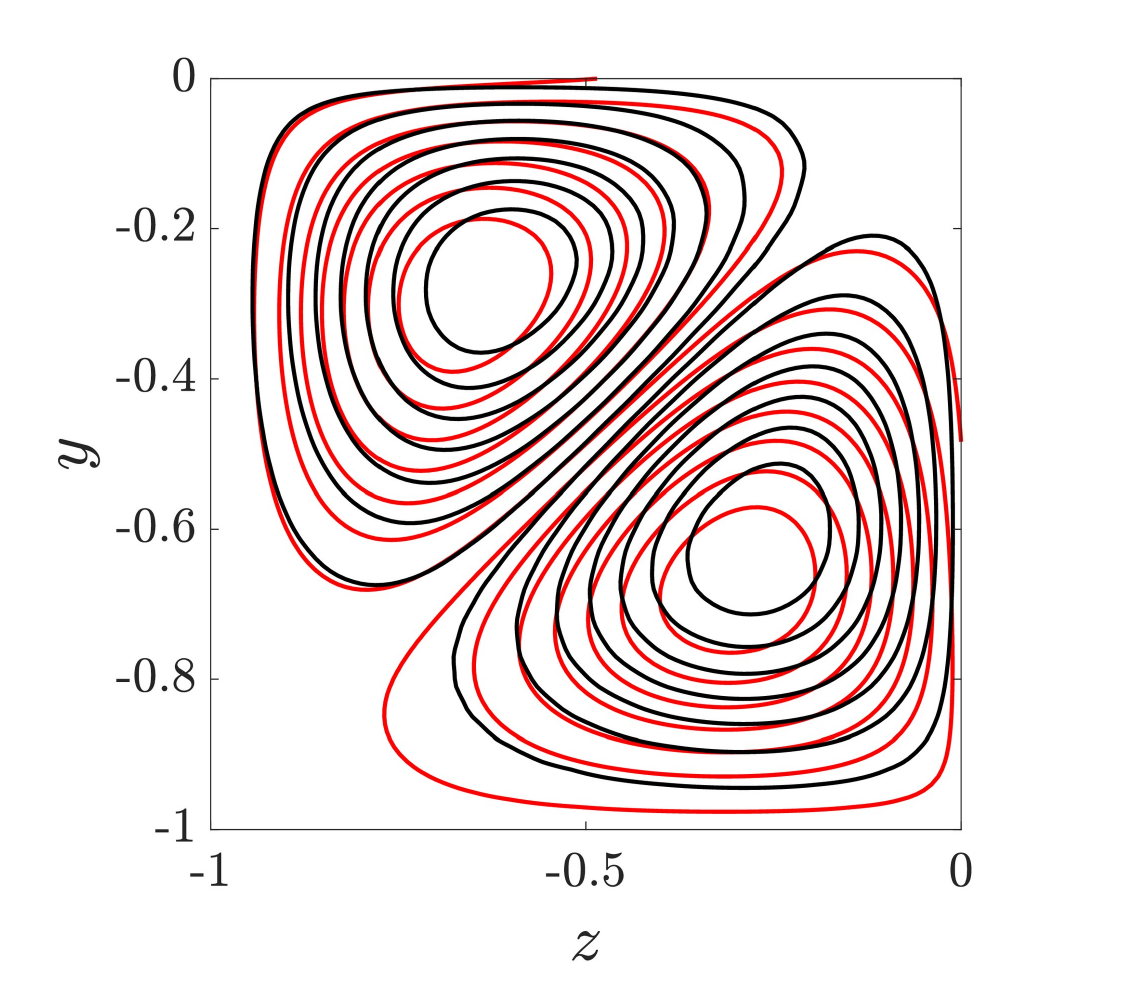}
        \caption{        \label{StreamLinessub2}}
    \end{subfigure}
        \begin{subfigure}{0.322\textwidth} 
        \centering
        \includegraphics[width=1.15\textwidth]{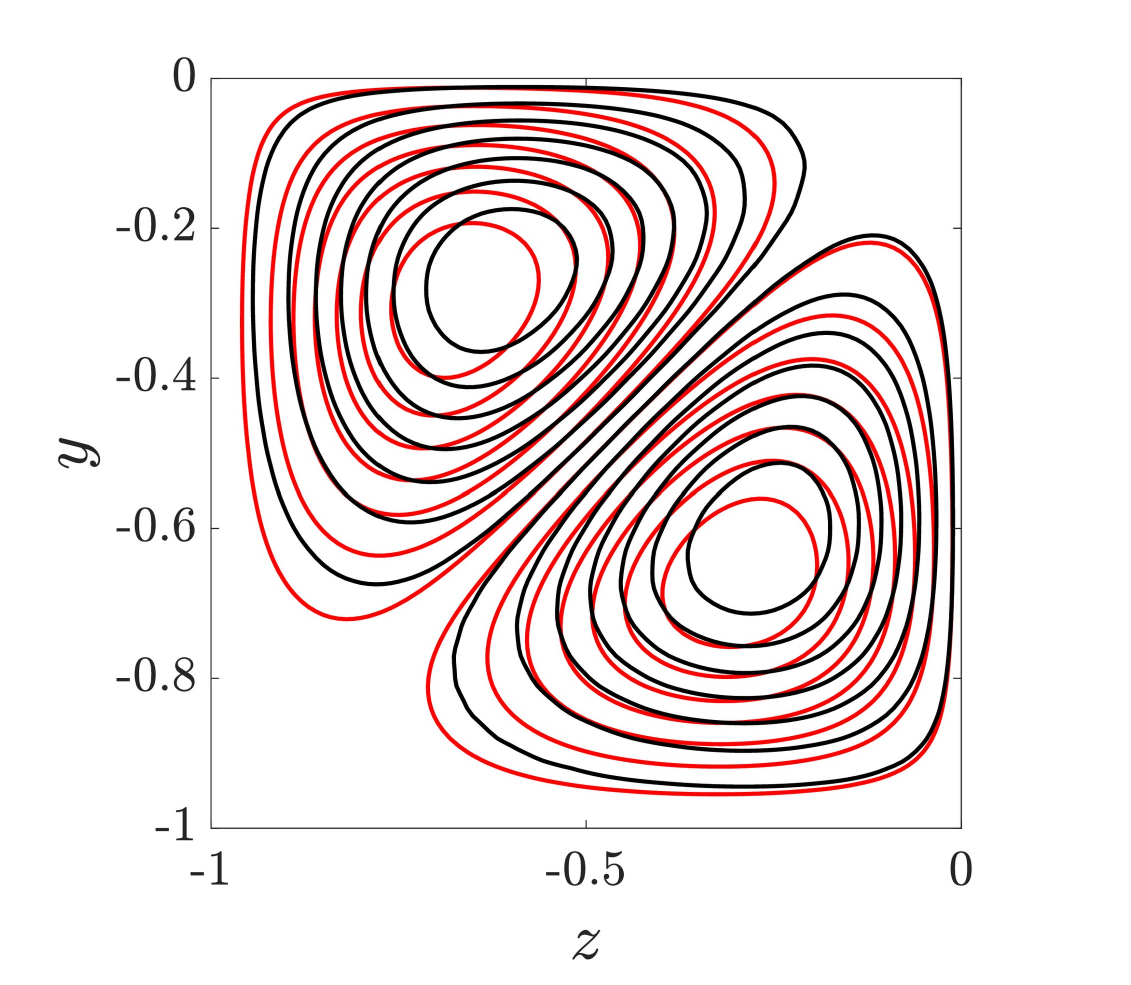}
        \caption{        \label{StreamLinessub3}}
    \end{subfigure}
\caption{Comparison of secondary mean flow streamlines $\;\Psi$ (red) and the twelfth mode reconstruction (black) in the lower-left corner of the duct. The flow is driven by sinusoidal forcing of the least stable eigenmode $(\bsphi_1)$, with forcing magnitudes (a) $B=0.01$, (b) $B=0.05$, and (c) $B=0.1$.}
    \label{StreamLines}
\end{figure}

To determine the influence of the forcing amplitude, figure~\ref{StreamandTransversecontoursDNS} shows the average flow fields for $B=0.01, 0.05$ and $0.1$. In all cases, the mean secondary flow converges toward the same structure. To quantify the effect of forcing amplitude on the mean flow field, we use the cross-stream velocity norm,
\begin{equation}
    \|\overline{\boldsymbol{u}}_\perp\|_2=\sqrt{\int_\Omega\left(\overline{V}^2+\overline{W}^2\right) d\Omega},
\end{equation}
where $\overline{\boldsymbol{u}}_\perp=(\overline{V},\overline{W})$ denotes the mean cross-stream velocity field. As the forcing magnitude increases, the secondary flow strengthens, leading to a distortion of the mean streamwise velocity contours. For $B=0.01$, $0.05$ and $0.1$, the corresponding cross-stream velocity norms are $8.8\times10^{-5}$, $2.2\times10^{-3}$ and $8.5\times10^{-3}$, respectively. This effect is particularly noticeable at higher forcing values ($B=0.1$) in figure~\ref{StreamandTransversecontoursDNSsub3}, where the increased secondary circulation redistributes streamwise momentum to a larger extent, altering the mean streamwise velocity distribution in a manner similar to what is observed in marginally turbulent duct flows \citep{pinelli2010}. To quantitatively compare the secondary mean obtained from ROMs and these simulations, figure~\ref{StreamLines} plots the secondary mean streamfunction $\Psi$ (where  $W=-\partial \Psi /\partial y$ and $V=\partial \Psi /\partial z$) in the lower left corner of the duct for both the ROMs and DNS. We observe close agreement across all forcing amplitudes, with the centre of the vortices slightly closer to the corner in each of the DNS cases. We emphasize here that while the mean secondary flow in the ROMs is essentially constrained to follow that of the symmetric eigenmode, the DNS cases have no such constraint, with this structure emerging naturally. As an aside, we note that more generally this phenomenon, where a zero-mean time-varying input results in a change in the mean flow field is related to the phenomenon of viscous streaming~\citep{riley2001}.

From the ROM discussion in \S\ref{results}, in the weak-forcing regime and when forcing a single non-symmetric mode,  the sign of the mean coefficient of the symmetric mode depends on the sign of its corresponding advection term (see \eqref{OneModeMean}). This suggests that forcing a non-symmetric eigenmode whose self-interaction projects with the opposite sign onto the fully symmetric mode generates a reversed secondary mean flow. The results in \S\ref{GeneralROM} further suggest that, as the forcing amplitude increases, additional modal interactions are activated and other diagonal second moments become non-negligible. Thus, the mean symmetric coefficient recovers the sign associated with the physically expected secondary flow.
\begin{figure} 
    \begin{subfigure}{0.5\textwidth}
        \centerline{\includegraphics[width=\textwidth]{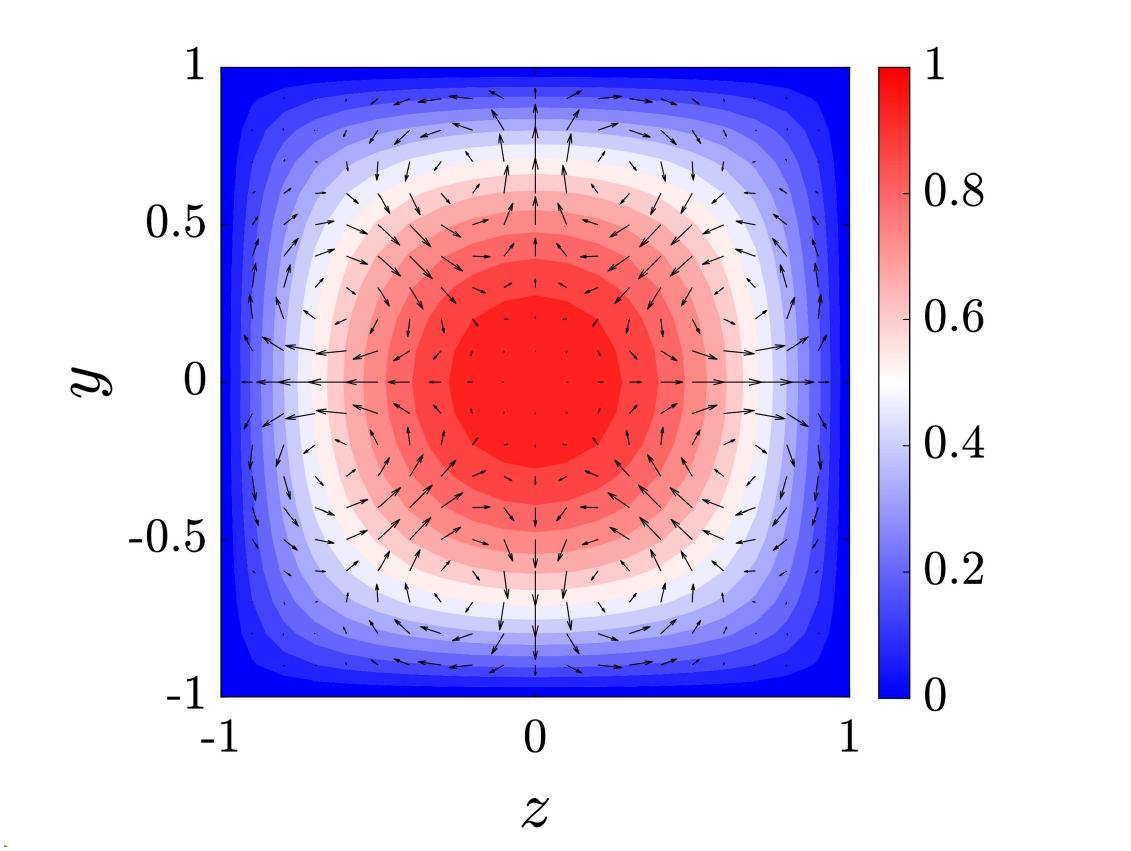}} 
        \caption{} 
        \label{FourthModeDNSforcingA} 
    \end{subfigure} 
    \begin{subfigure}{0.5\textwidth} 
        \centerline{\includegraphics[width=\textwidth]{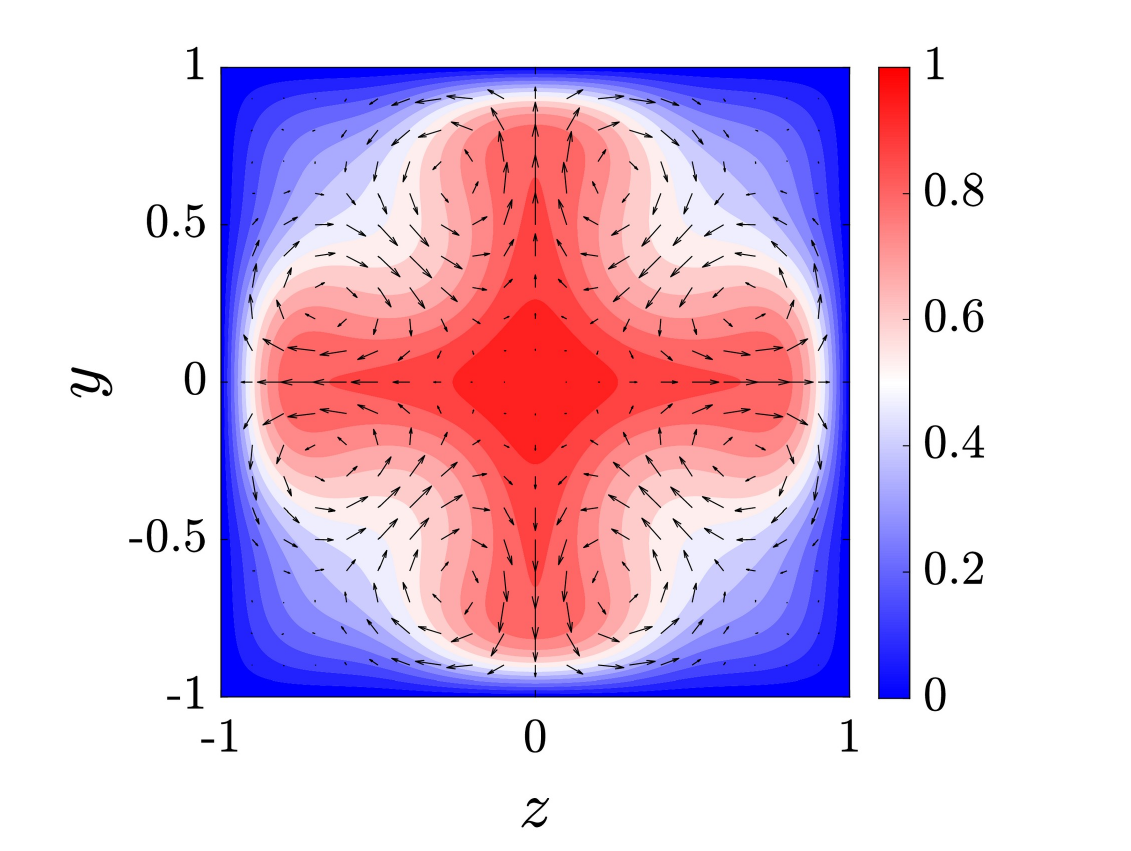}} 
        \caption{} 
        \label{FourthModeDNSforcingB} 
    \end{subfigure} 
    \caption{Contours of the mean streamwise velocity overlaid with vector fields of the mean cross-stream velocities from 2D/3C DNS simulations using sinusoidal forcing of mode four with forcing magnitudes (a) $B=0.01$ and (b) $B=10$.}
    \label{FourthModeDNSforcing}
\end{figure}

To test this reasoning, we now run simulations where the spatial structure of the forcing is given by the fourth eigenmode $(\bsphi_4)$ using sinusoidal forcing magnitudes $B=0.01$ and $B=10$ as shown in figure~\ref{FourthModeDNSforcing}. The corresponding cross-stream velocity norms are $2.24\times 10^{-4}$ and $2.58\times 10^{-2}$, respectively. The resulting mean streamwise velocity distribution reveals that the secondary mean flow vector field is directed towards the duct core along the corner bisector and towards the walls along the wall bisector, opposite to the physically expected flow pattern. The secondary-flow magnitude for $B=10$ is significantly larger than that for $B=0.01$, leading to a significant distortion of the streamwise velocity contours. In contrast to the ROM prediction, however, the secondary mean flow does not reverse back to the physically expected direction in the DNS simulations over the range of forcing amplitudes considered. This difference may be attributed to the fact that, unlike the finite-dimensional ROM, the DNS is not restricted to the retained ROM subspace. Nonlinear interactions in the DNS can redistribute activity over a broader set of spatial structures, whereas modal truncation confines this redistribution to the retained modes and may amplify the influence of the included modal interactions. It remains possible that the physically expected direction could be recovered at forcing amplitudes beyond the range considered here, but such cases would require substantially higher computational cost. This observation confirms the weak-forcing ROM prediction that forcing of $\bsphi_4$ produces an opposite secondary-flow structure. Based on this result, along with the findings in figure~\ref{NonlinearProjection}, we conclude that in a two-mode model, only an eigenmode with a negative nonlinear projection term should be selected for a model with the correct secondary flow structure.

\section{Secondary flow production mechanisms}\label{SecFlwProd}

In \S\ref{Introduction}, we briefly introduced the efforts to explain the mechanisms responsible for secondary flow generation. This section provides a more in-depth discussion of the origin of secondary flows, along with predictions made using the present ROM. The derivation of the streamwise vorticity equation in turbulent flow serves as the foundation for understanding the existence of secondary flows. The mean streamwise vorticity equation is expressed as~\citep{einstein1958}
\begin{equation}
    \underbrace{V\frac{\partial\Omega_x}{\partial y}+W \frac{\partial \Omega_x}{\partial z}}_{\text{Convection}}
    -
    \underbrace{\frac{1}{\Rey} \nabla^2 \Omega_x}_{\text{Viscous diffusion}}
    =
    \underbrace{\left(\frac{\partial^2}{\partial y^2} -\frac{\partial^2}{\partial z^2} \right)  (-\overline{v'w'})}_{\text{Production, $P_1$}}
    +
    \underbrace{ \frac{\partial^2}{\partial y\partial z} (\overline{v'^2}-\overline{w'^2})}_{\text{Production, $P_2$}}.
    \label{MeanVorticity}
\end{equation}

Note that in this equation the fluctuations are based on a turbulent mean with secondary flow components, in contrast to the decomposition used previously.
 The equation is temporally averaged and independent of the streamwise direction. The mean vorticity vector 
 $\bm{\Omega}=\nabla\times\bsu=\Omega_x \textbf{i}+\Omega_y \textbf{j}+\Omega_z \textbf{k}$, has a streamwise component given by $\Omega_x=\partial W/\partial y-\partial V/\partial z$. The terms on the left-hand side of \eqref{MeanVorticity} are the convection of vorticity along the flow streamlines and the viscous diffusion, respectively. The terms on the right-hand side are the spatial derivatives of the Reynolds stresses and are responsible for the production of secondary flows (note that without these terms \eqref{MeanVorticity} is satisfied by $\Omega_x=0$).

 Previous studies have demonstrated that non-vanishing Reynolds stresses are essential for the existence of streamwise vorticity and secondary flows. The location of vorticity production within the duct cross-section was predicted by \cite{brundrett1964}, who suggested that a pair of positive and negative vorticity structures emerge at each duct corner, with zero production along the symmetry lines. They also indicated that the normal Reynolds stress production term $P_2$ plays the dominant role over the shear-stress contribution,$P_1$ in the generation of secondary flows. However, more sophisticated experimental investigations~\citep{gessner1965,demuren1984} have challenged this notion, revealing that both $P_1$ and $P_2$ contribute comparably, though with opposite signs. Additionally, the viscous term remains small except in the vicinity of the corners.  For the ROMs developed in \S\ref{results} where the state variables are coefficients $a_i$ of modes $\bsphi_i$ with components $\bsphi^{v}_i$ and $\bsphi^{w}_i$, the
  quantities in~\eqref{MeanVorticity} can be computed by 
\begin{equation}
    \begin{aligned}
        V &= \sum_{i=1}^{N} \overline{a_i} \bsphi^{v}_i, \quad 
        W = \sum_{i=1}^{N} \overline{a_i} \bsphi^{w}_i, \\
        \overline{v'^2} &= \sum_{i=1}^{N}\sum_{j=1}^{N} 
        \text{covar}(a_i,a_j)
        (\bsphi^{v}_i\odot \bsphi^{v}_j), \quad
        \overline{w'^2} =\sum_{i=1}^{N} \sum_{j=1}^{N} 
        \text{covar}(a_i,a_j)
        (\bsphi^{w}_i\odot \bsphi^{w}_j), \\
        \overline{v'w'} &=  \sum_{i=1}^{N} \sum_{j=1}^{N}  
        \text{covar}(a_i,a_j)
        (\bsphi^{v}_i \odot \bsphi^{w}_j),
    \end{aligned}
    \label{eq:summation}
\end{equation}
where the symbol $\odot$ denotes the Hadamard (element-wise) product. Note that all of these quantities are functions of $y$ and $z$.
\begin{figure}
    \begin{subfigure}{\textwidth} 
        \centering
        \includegraphics[width=\textwidth]{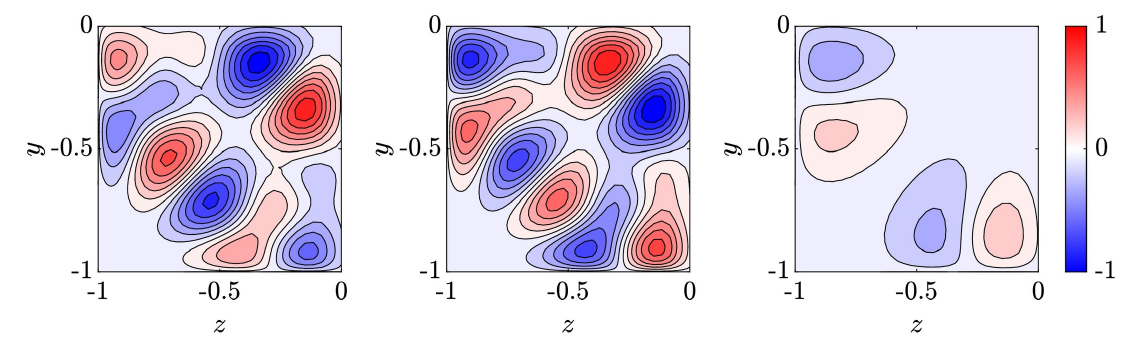}
    \end{subfigure}
    \vspace{-0.7cm}
    \caption*{$(a)\;N=2$}

    \begin{subfigure}{\textwidth} 
        \centering
        \includegraphics[width=1\textwidth]{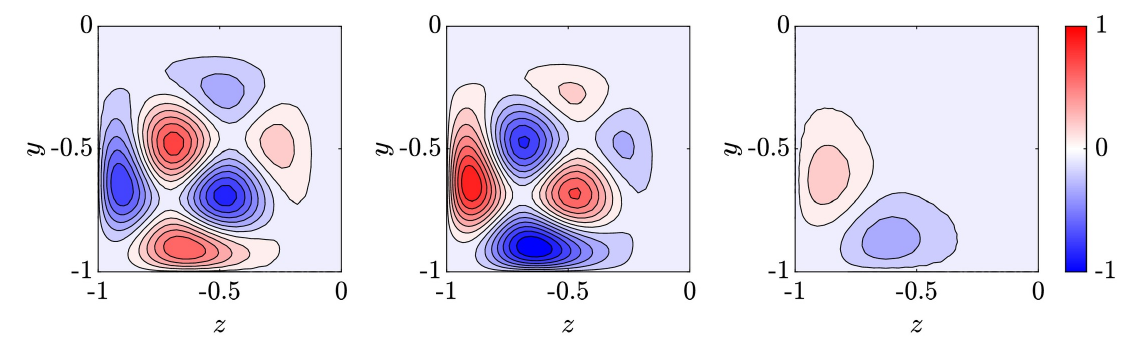}
    \end{subfigure}
    \vspace{-0.7cm}
    \caption*{$(b)\;N=12$}
    
\begin{subfigure}{\textwidth} 
        \centering
        \includegraphics[width=\textwidth]{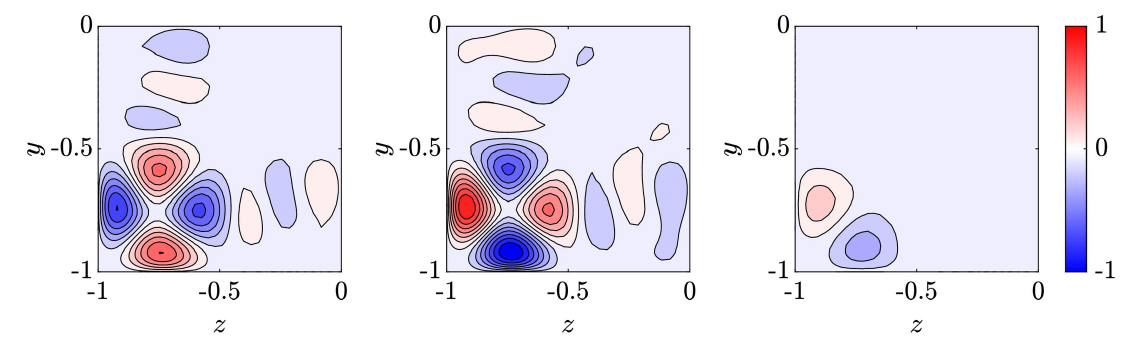}
    \end{subfigure}
        \vspace{-0.7cm}
    \caption*{$(c)\;N=22$}
    \vspace{-0.1cm}
\begin{subfigure}{\textwidth} 
        \centering
        \includegraphics[width=\textwidth]{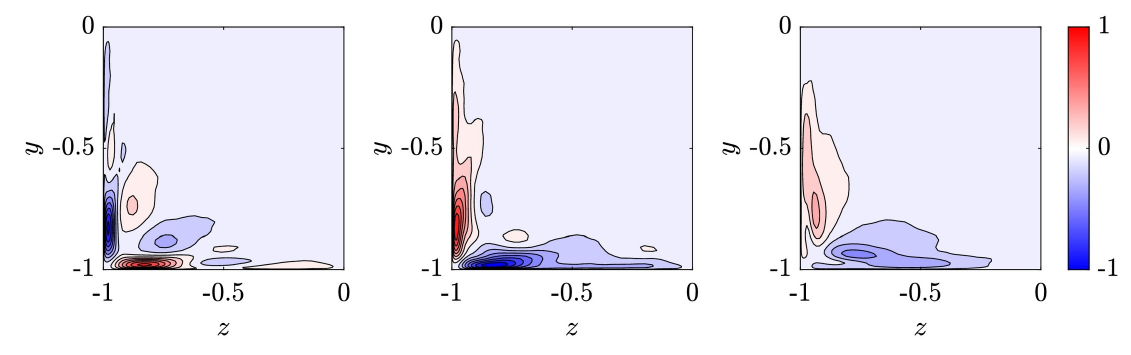}
        \vspace{-0.5cm}
\caption*{%
\makebox[0pt][l]{\hspace{0.4cm}$P_1$}\hspace{4.7cm}$P_2$ \hspace{3.7cm}$P_1+P_2$%
}
    \end{subfigure}
\caption*{$(d)$ DNS simulations}
 \caption{Contours of the secondary flow production terms, $P_1$, $P_2$ and the total production $P_1+P_2$, from the reduced-order model (a), (b), and (c) for model orders 2, 12, and 22, respectively, compared to (d)  DNS simulations of a square duct at $\Rey=2796$~\citep{vinuesa2014}. Production terms are normalised by the maximum absolute value of the corresponding $P_2$ field, and a linear colour scale ranging from $-1$ to $1$ is used for all the plots. Modes are chosen as described in \S \ref{results}, with the $N=22$ case obtained using the leading 22 modes.}
    \label{TotalProd}
\end{figure}

All models shown in this section are forced with white-noise forcing with a spectral density $q=0.1$. Figure~\ref{TotalProd}$(a-c)$ illustrate the vorticity production terms for several ROM orders, alongside a comparison to DNS simulations by~\cite{vinuesa2014} in (d). These DNS simulations were conducted at a Reynolds number $Re = 2796$ (based on the centreline velocity) and averaged over a duct length of $25h$. The mean flow from these simulations was previously shown in figure \ref{TurbulentProfile}. The DNS production terms $P_1$ and $P_2$ exhibit antisymmetric patterns around the corner bisector, with peak values occurring approximately at coordinates $(-0.82,-0.98)$ and $(-0.98,-0.82)$. Notably, normal stress contributions display slightly higher magnitudes compared to shear stresses, which influence the spatial distribution of the total production term $P = P_1 + P_2$. This combined distribution also exhibits an antisymmetric pattern around the corner bisector.

It should be noted that the goal here is not to match the exact magnitude of the DNS simulations (since these depend on the forcing magnitude) but rather to evaluate the qualitative structural accuracy of the ROM predictions. Therefore, the production terms are normalised by the maximum absolute value of their corresponding $P_2$ field. ROMs of all orders capture certain features of the DNS data. In particular, in all cases, the nearest peak to the corner along the vertical (horizontal) wall is negative (positive) for $P_1$, and positive (negative) for both $P_2$ and the sum $P_1 + P_2$. The ROMs also capture the antisymmetry of all fields along the corner bisector, and the property that the contributions of $P_1$ and $P_2$ to the total production tend to oppose each other, with $P_2$  typically having a slightly larger amplitude. 
 
 These similarities notwithstanding, the $N=2$ model in particular exhibits clear differences in the spatial distribution of these production terms in comparison to the DNS, with peaks observed much further from the corner and large production observed nearer to the duct centre. However, as the model order increases, the maxima in all terms move towards the corner, appearing to be converging towards the structure observed in the DNS results. This suggests that increasing the model order allows for a closer representation of aspects of the flow physics observed in DNS. 

\section{Conclusion}\label{sec:conc}

This investigation has shown that secondary mean flow direction and qualitative spatial distribution in a square duct can be predicted using a ROM with as few as two states, with one state being a mode that possesses all of the symmetry properties that the secondary mean is expected to satisfy. However, a higher number of states are necessary to accurately predict the spatial distribution of the secondary flow production terms near the corner region. The family of ROMs considered in this work were obtained by projecting the incompressible Navier--Stokes equations onto a set of eigenmodes of the linearised equations. By only considering streamwise-constant eigenmodes, the eigendecomposition decouples into streamwise-only and secondary modes, all of which are linearly stable. These findings demonstrated that the qualitative structure of the secondary mean flow can be predicted from a streamwise-constant model, justifying this simplifying assumption.

The shape of the secondary mean flow is governed by the shape of the  fully-symmetric mode included in the model, while its direction is determined by the sign of the average symmetric mode coefficient. In turn, the sign of this mean coefficient is determined by the projection of the nonlinear convective terms involving other modes onto the symmetric mode. The majority of the leading (least-stable) modes give a projection that agrees with the observed direction of the secondary flow (as shown in figure~\ref{NonlinearProjection}), and the resulting ROM means are generally consistent with this direction (figures \ref{fig:meanVsnumModes} and~\ref{fig:meanVsnumModesDetForcing}). This naturally leads to a hypothesis that the overall structure and direction of turbulent secondary mean flow can be predicted only from the structure of these leading eigenmodes. Note that these streamwise-constant cross-stream basis functions are independent of both the streamwise base flow and the Reynolds number. If (as in this work) only a single symmetric mode is used, then the models will predict the same secondary mean flow streamlines (as well as related quantities such as mean streamwise vorticity) regardless of Reynolds number. Future work will analyse whether the inclusion of additional fully-symmetric eigenmodes can capture the scaling properties of these quantities with Reynolds number \citep{zhang2015direct,pirozzoli2018}.

It was determined that the minimal model capable of producing a secondary mean flow consists of only two modes, corresponding to a set of two coupled nonlinear differential equations, forced with a zero-mean disturbance. This is simpler than the models proposed by \citet{wedin2008} when seeking SSP in a square duct, which also included streamwise streaks and streamwise-varying waves. While the focus of \citet{wedin2008} was to search for SSP that did not require external forcing (necessitating the additional modes), they also reported a nonzero secondary mean resembling the expected pattern (though not fully symmetric when averaged over the time window considered).

The relevance of the family of ROMs identified in the current work to the physics of the duct geometry was investigated in several ways. To begin with, we performed streamwise-constant DNS subject to zero-mean sinusoidal forcing with the same spatial structure as the eigenmodes used in the ROMs. It was found that these simulations produced secondary mean flow fields with very similar structure to the fully symmetric eigenmode used in the ROMs. Moreover, the direction of the secondary mean was found to be dependent on the eigenmode used for the spatial forcing, with the direction generally consistent with the predictions from the ROMs. One exception is in the case of high amplitude sinusoidal forcing in the direction of the fourth mode, where increasing the forcing amplitude was observed to swap the (initially reversed) direction of the mean in the ROMs, without such a reversal being observed in the DNS at equivalent forcing amplitudes. These simulations confirmed the ROM prediction that secondary mean flow can arise from a forced streamwise-constant flow and further corroborated the relevance of the leading eigenmodes and their interactions in predicting the structure and direction of this secondary mean.

As secondary mean flow can be related to the mean streamwise vorticity equation \eqref{MeanVorticity}, we next investigated the extent to which the ROMs capture the production terms in this equation required to support a secondary mean. It was found that the ROMs captured overall trends of both the individual and summed production terms, with closer spatial agreement with DNS observed as more modes were included in the ROM. Note that while the production terms predicted by our ROMs are dependent on the number of modes used, the mean streamwise vorticity field is independent of the number of modes and is constrained by the structure of the fully-symmetric mode. As mentioned above, the inclusion of additional fully-symmetric modes could allow for more quantitatively accurate prediction of the mean streamwise vorticity.

This work has provided evidence that low-dimensional ROMs can be used to predict and explain the presence and structure of the secondary mean field in turbulent flow through a square duct. There are several extensions that remain to be explored that could further enhance the fidelity of this modelling approach. This includes the inclusion of modes with nonzero streamwise wavenumber and the use of streamwise velocity (as well as secondary velocity) modes to predict the shape of the streamwise turbulent profile. The inclusion of streamwise velocity modes could also 
enable the study of transient bursting events, which have been shown to be associated with the secondary mean in corner flows~\citep{marin2016characterization}. 
Note that when streamwise modes are included, the eigenmodes of the linearised system no longer  decouple into streamwise and secondary directions and are also no longer orthogonal, potentially complicating the methodology. Further work could also investigate the application of these methods to other geometries featuring secondary flows, such as rectangular ducts and corner geometries in open domains.

\begin{bmhead}[Acknowledgements]
The authors thank B.~Lopez-Doriga for insightful discussions related to the physics and linear analysis of duct flows.
\end{bmhead}

\begin{bmhead}[Funding]
A.I.E-N.~and S.T.M.D.~acknowledge support from U.S.~National Science Foundation grant CBET-2238770.
\end{bmhead}

\begin{bmhead}[Declaration of Interests]
The authors report no conflict of interest.
\end{bmhead}

\begin{appen}

\section{}
\label{appA}
This appendix shows how the temporally averaged form of~\eqref{forcedsystem} simplifies to~\eqref{multimode} after applying the incompressibility, boundary-condition, and orthogonality properties of the cross-stream Stokes eigenmodes.\\

The temporal average of~\eqref{forcedsystem} or~\eqref{StochasticForm} can be expressed as 
\begin{equation}
Re^{-1}
\sum_{j=1}^N
\left\langle\nabla^2\boldsymbol{\phi}_j,\boldsymbol{\phi}_i\right\rangle
\overline{a_j}
=
\sum_{j=1}^{N} \overline{a_j^2}
\left\langle \boldsymbol{\phi}_j\cdot\nabla\boldsymbol{\phi}_j, \boldsymbol{\phi}_i \right\rangle 
+
\sum_{j=1}^{N-1} \sum_{k=j+1}^{N}
\overline{a_ja_k}
\left[
\left\langle\boldsymbol{\phi}_j\cdot\nabla\boldsymbol{\phi}_k,\boldsymbol{\phi}_i\right\rangle
+
\left\langle\boldsymbol{\phi}_k\cdot\nabla\boldsymbol{\phi}_j,\boldsymbol{\phi}_i\right\rangle
\right]
\label{MeanofAllModes}
\end{equation}
where $i=1,\dots, N$. 

For an eigenmode $\bsphi_j$ that is mapped to either $\bsphi_j$ or
$-\bsphi_j$ under each of the rotation and reflection transformations
listed in table~\ref{tab:SymmTable}, the corresponding self-advection term
$\bsphi_j\cdot\nabla\bsphi_j$ satisfies all eight symmetries. Any change
in sign of $\bsphi_j$ appears twice in this quadratic term and therefore
does not change the transformed self-advection term. Its projection onto
a mode that does not satisfy all eight symmetries consequently vanishes:
\begin{equation}
\left\langle
\bsphi_j\cdot\nabla\bsphi_j,\bsphi_i
\right\rangle
=0,
\qquad
\bsphi_i\neq\bsphi_s.
\label{SelfAdvectionNonSymmetric}
\end{equation}
This argument applies when each rotation or reflection maps
$\bsphi_j$ to either itself or its negative. When a transformation instead maps $\bsphi_j$ to another eigenmode, as can occur for modes associated with a repeated eigenvalue, the self-advection term of an individual mode does not necessarily satisfy all eight symmetries. In that case, fully symmetric contributions may instead arise from combinations of the self-advection and cross-advection terms or from the temporal averaging of the corresponding ROM coefficients.\\

Because the cross-stream modes are orthonormal eigenmodes of the
self-adjoint Stokes operator,
    \begin{equation}
    \left\langle \nabla^2\bsphi_i,\bsphi_j\right\rangle=0
    \quad \text{for} \quad i\neq j, 
    \label{Viscousprojection}
    \end{equation}
    after projecting onto the divergence-free cross-stream eigenspace. This can be shown by expressing the Stokes eigenproblem~\eqref{EigenProbvw} in a compact form as
    \begin{equation}
        \lambda_i^{(v,w)}\boldsymbol{\phi}_i+\mathsfbi{L}\boldsymbol{\phi}_i+\nabla p=0.
        \label{stokes expansion}
    \end{equation}
    Taking the inner product of this equation with $\phi_j$ as
    \begin{equation}
        \lambda_i^{(v,w)}\langle \boldsymbol{\phi}_i, \bsphi_j\rangle+\langle\mathsfbi{L}\boldsymbol{\phi}_i,\bsphi_j \rangle+\langle\nabla p,\bsphi_j\rangle=0,\label{ProjectLaplace}
    \end{equation}
    the projection of the pressure gradient vanishes for bounded domains with Dirichlet boundary conditions, and~\eqref{ProjectLaplace} reduces to
    \begin{equation}
        \langle\mathsfbi{L}\bsphi_i,\bsphi_j\rangle=-\lambda_i^{(v,w)}\langle\bsphi_i,\bsphi_j\rangle=-\lambda_i^{(v,w)}\delta_{ij}\label{projectionOnStokes}
    \end{equation}
    assuming orthonormal eigenmodes. Note that orthogonality is guaranteed in this context except in the case of repeating eigenvalues. For repeating eigenvalues, the selected basis functions for the eigenspace need not be orthogonal, but can be chosen to be. Applying these properties to~\eqref{MeanofAllModes}, the ROM equations reduce to~\eqref{multimode}.

\section{}
\label{AppendixB}

This appendix shows why the advective projection term $\langle\boldsymbol{\phi}_s\cdot\nabla\boldsymbol{\phi}_i,\boldsymbol{\phi}_i\rangle$ vanishes as a consequence of incompressibility and the no-penetration boundary condition. Although this cancellation also applies to the general ROM~\eqref{multimode}, it appears most explicitly in the two-mode reduction and is therefore discussed separately here.

In deriving~\eqref{TwoModeModel}, the projection term $\langle\bsphi_s \cdot \nabla\bsphi_i,\bsphi_i \rangle$ vanishes due to the incompressibility of the eigenmodes and the no-penetration boundary conditions. This can be shown as follows:
    \begin{equation}
        \langle\bsphi_s \cdot \nabla\bsphi_i,\bsphi_i \rangle=\int_\Omega \left[
(\boldsymbol{\phi}_s\cdot\nabla)\boldsymbol{\phi}_i
\right]\cdot\boldsymbol{\phi}_i\,\mathrm{d}\Omega
        =\frac{1}{2}\int_\Omega\bsphi_s\cdot\nabla|\bsphi_i|^2 \; \mathrm{d}\Omega
    \end{equation}
    Given that
    \begin{equation}
        \nabla\cdot(|\bsphi_i|^2\bsphi_s)=|\bsphi_i|^2\nabla\cdot\bsphi_s+\bsphi_s\cdot\nabla|\bsphi_i|^2,
    \end{equation}
    the integral can be expressed as 
    \begin{equation}
         \langle\bsphi_s \cdot \nabla\bsphi_i,\bsphi_i \rangle=\frac{1}{2} \left[\int_\Omega \nabla\cdot(|\bsphi_i|^2\bsphi_s) \;d\Omega-\int_\Omega |\bsphi_i|^2\nabla\cdot\bsphi_s \;d\Omega  \right]
    \end{equation}
    The second term vanishes due to the incompressibility conditions ($\nabla \cdot\bsphi_s=0$). Applying the divergence theorem to the first term on the right-hand side, we get
    \begin{equation}
        \langle\bsphi_s \cdot \nabla\bsphi_i,\bsphi_i \rangle=\frac{1}{2}\int_{\partial\Omega}|\bsphi_i|^2\bsphi_s\cdot\boldsymbol{n} \; \mathrm{d}S=0
        \end{equation}
    as the eigenmodes satisfy the no-penetration boundary condition at the duct boundaries.\\

    For the fully symmetric mode, the self-advection term does not contribute to any of the retained modal equations in the present ROMs:
    \begin{equation}
    \left\langle \bsphi_s \cdot\nabla\bsphi_s,\bsphi_i \right\rangle=0,
    \qquad i=1,\ldots,N .
    \end{equation}
    For $i=s$, this self-projection vanishes by the incompressibility of the eigenmodes, as stated in~\eqref{SingleProjection}. For $i\neq s$, the projection vanishes by symmetry. Since $\bsphi_s$ is fully symmetric, the nonlinear term $\bsphi_s\cdot\nabla\bsphi_s$ is invariant under all eight transformations listed in table~\ref{tab:SymmTable}. Therefore, its inner product with any retained eigenmode that is not fully symmetric is zero.

\section{}\label{Appendixc}

\noindent Here, we present the expanded forms of the four-mode and five-mode ROMs. The modal numbering follows the notation introduced in figure~\ref{TwelveEigenmodes}.\\

The four-mode ROM is:
\begin{subequations}
\begin{flalign}
&\dot{a}_1
= -1.160\mathrm{E}{-1}\;a_1a_s
+\Rey^{-1}\left(-1.309\mathrm{E}{1}\;a_1\right) +f_1(t), &&\\
&\dot{a}_2
= -3.422\mathrm{E}{-1}\;a_1a_3 -6.502\mathrm{E}{-2}\;a_2a_s +5.0\mathrm{E}{-5}\;a_3a_s +\Rey^{-1}\left(-2.306\mathrm{E}{1}\;a_2 
+1.790\mathrm{E}{-2}\;a_3
\right) +f_2(t), &&\\
&\dot{a}_3
= 3.422\mathrm{E}{-1}\;a_1a_2 +5.0\mathrm{E}{-5}\;a_2a_s -6.502\mathrm{E}{-2}\;a_3a_s +\Rey^{-1}\left(
1.790\mathrm{E}{-2}\;a_2 
-2.306\mathrm{E}{1}\;a_3\right) +f_3(t), &&\\
&\dot{a}_s
= 1.160\mathrm{E}{-1}\;a_1^2 + 6.502\mathrm{E}{-2}\;a_2^2 - 1.0\mathrm{E}{-4}\;a_2a_3 + 6.502\mathrm{E}{-2}\;a_3^2 +\Rey^{-1}\left(-6.733\mathrm{E}{1}\;a_s\right) + f_s(t). &&
\end{flalign}
\end{subequations}

The five-mode ROM is:
\begin{subequations}
\begin{flalign}
&\dot{a}_1
= -1.160\mathrm{E}{-1}\;a_1a_s +\Rey^{-1}\left(-1.309\mathrm{E}{1}\;a_1\right) +f_1(t), &&\\
&\dot{a}_2
= -3.422\mathrm{E}{-1}\;a_1a_3 + 1.87\mathrm{E}{-3}\;a_2a_4 - 6.502\mathrm{E}{-2}\;a_2a_s - 3.241\mathrm{E}{-1}\;a_3a_4 + 5.0\mathrm{E}{-5}\;a_3a_s \notag\\
&\qquad
+\Rey^{-1}\left(-2.306\mathrm{E}{1}\;a_2 
+1.790\mathrm{E}{-2}\;a_3
\right) +f_2(t), &&\\
&\dot{a}_3
= 3.422\mathrm{E}{-1}\;a_1a_2 - 3.241\mathrm{E}{-1}\;a_2a_4 + 5.0\mathrm{E}{-5}\;a_2a_s - 1.37\mathrm{E}{-3}\;a_3a_4 - 6.502\mathrm{E}{-2}\;a_3a_s \notag\\
&\qquad
+\Rey^{-1}\left(
1.790\mathrm{E}{-2}\;a_2 
-2.306\mathrm{E}{1}\;a_3\right) +f_3(t), &&\\
&\dot{a}_4
= -1.87\mathrm{E}{-3}\;a_2^2 +6.482\mathrm{E}{-1}\;a_2a_3 +1.37\mathrm{E}{-3}\;a_3^2 +2.936\mathrm{E}{-1}\;a_4a_s
+\Rey^{-1}\left(-3.207\mathrm{E}{1}\;a_4\right) +f_4(t), &&\\
&\dot{a}_s
= 1.160\mathrm{E}{-1}\;a_1^2 + 6.502\mathrm{E}{-2}\;a_2^2 - 1.0\mathrm{E}{-4}\;a_2a_3 + 6.502\mathrm{E}{-2}\;a_3^2 - 2.936\mathrm{E}{-1}\;a_4^2 \notag\\
&\qquad
+\Rey^{-1}\left(-6.733\mathrm{E}{1}\;a_s\right) +f_s(t). &&
\end{flalign}
\end{subequations}
Note that the small off-diagonal entries in the viscous terms are due to the eigenmodes for repeating eigenvalues not being perfectly orthogonal.

\end{appen}

\bibliographystyle{jfm}
\bibliography{jfm}

\end{document}